\documentclass[a4paper, 12pt]{iopart}
 \expandafter\let\csname equation*\endcsname\relax
 \expandafter\let\csname endequation*\endcsname\relax
\usepackage{amsmath}
\usepackage{iopams}
\usepackage{graphicx,colordvi,gnuplotcolors}
\usepackage{subfigure}
\usepackage[dvips]{color}
\usepackage[bottom]{footmisc} 
\usepackage{rotating}
\usepackage{hyperref}
\usepackage{url}
\usepackage{mciteplus}
\usepackage{tabularx}
\usepackage{alltt}
\usepackage{soul}
\usepackage{accents}
\usepackage{fancyhdr}
\usepackage{cite}

\newcounter{bla}

\usepackage[margin=1in]{geometry}
\usepackage[framemethod=TikZ]{mdframed}

\mdfdefinestyle{MyFrame}{%
    linecolor=gray,
    outerlinewidth=0.4pt,
    roundcorner=4pt,
    innertopmargin=0.5\baselineskip,
    innerbottommargin=0.5\baselineskip,
    innerrightmargin=\parindent,
    innerleftmargin=\parindent,
    backgroundcolor=gray!05!white}

\input{paperdef.tex}

\newcommand{\HBclassic}{\emph{classic}}
\newcommand{\HBfull}{\emph{full}}

\newcommand{\nHplus}{n_{H^{\pm}}}
\newcommand{\nHneut}{n_{H^0}}
\newcommand{\cbox}[1]{\\[.3em]\centerline{\tt #1}\\[.3em]}
\newcommand{\subroutine}[2]{\\ \begin{mdframed}[style=MyFrame]\small\begin{tabularx}{\linewidth}{r@{}X} \texttt{#1} & \texttt{#2} \end{tabularx}\end{mdframed}\vspace{0.2em}}

\newcommand{\HBv}[1]{{\tt Higgs\-Bounds-{#1}}}

\definecolor{gnuplotorange}{rgb}{1,0.647059,0}
\definecolor{gnuplotyellow}{rgb}{1,1,0}
\definecolor{gnuplotred}{rgb}{1,0,0}
\definecolor{gnuplotgreen}{rgb}{0,1,0}
\definecolor{gnuplotdarkblue}{rgb}{0,0,0.545098}
\definecolor{gnuplotlightblue}{rgb}{0.678431, 0.847059, 0.901961}
\definecolor{gnuplotblue}{rgb}{0,0,1}
\definecolor{myblue}{rgb}{0,0,1}

\begin{document}

\hfill \texttt{BONN-TH-2013-21}

\hfill \texttt{DESY 13-110}

\title[\HBv{4} User Manual]{\HBv{4}: Improved Tests of Extended\\ Higgs Sectors against Exclusion Bounds\\ from LEP, the Tevatron and the LHC}

\author{Philip Bechtle$^1$, Oliver Brein$^2$, Sven Heinemeyer$^3$, Oscar St{\r{a}}l$^4$,\\ Tim Stefaniak$^{1,5}$, Georg Weiglein$^6$ and Karina E. Williams$^{1,5,*}$}
\address{$^1$ Physikalisches Institut der Universit\"at Bonn, Nu{\ss}allee 12, 53115 Bonn, Germany}
\address{$^2$ Grosskarlbacher Stra{\ss}e 10, D-67256 Weisenheim am Sand, Germany}
\address{$^3$ Instituto de F\'isica de Cantabria (CSIC-UC), Santander, Spain}
\address{$^4$ The Oskar Klein Centre, Department of Physics, Stockholm University, SE-106 91 Stockholm, Sweden}
\address{$^5$ Bethe Center for Theoretical Physics, University of Bonn, Nu{\ss}allee 12, 53115 Bonn, Germany}
\address{$^6$ Deutsches Elektronen-Synchrotron DESY, Notkestra{\ss}e 85, D-22607 Hamburg, Germany}
\address{$^*$ Former affiliations}
\eads{\mailto{bechtle@physik.uni-bonn.de}, \mailto{Sven.Heinemeyer@cern.ch}, \mailto{oscar.stal@fysik.su.se}, \mailto{tim@th.physik.uni-bonn.de}, \mailto{Georg.Weiglein@desy.de}}

\begin{abstract}
We describe the new developments in version {\tt 4} of the public computer code \HB. \HB\ is a tool to test models with arbitrary Higgs sectors, containing both neutral and charged Higgs bosons, against the published exclusion bounds from Higgs searches at the LEP, Tevatron and LHC experiments. From the model predictions for the Higgs masses, branching ratios, production cross sections and total decay widths --- which are specified by the user in the input for the program --- the code calculates the predicted signal rates for the search channels considered in the experimental data. The signal rates are compared to the expected and observed cross section limits from the Higgs searches to determine whether a point in the model parameter space is excluded at 95\% confidence level. In this paper we present a modification of the \HB\ main algorithm that extends the exclusion test in order to ensure that it provides useful results in the presence of one or more significant excesses in the data, corresponding to potential Higgs signals. We also describe a new method to test whether the limits from an experimental search performed under certain model assumptions can be applied to a different theoretical model. Further developments discussed here include a framework to take into account theoretical uncertainties on the Higgs mass predictions, and the possibility to obtain the $\chi^2$ likelihood of Higgs exclusion limits from LEP. Extensions to the user subroutines from earlier versions of \HB\ are described. The new features are demonstrated by  additional example programs.
\end{abstract}


\maketitle

\tableofcontents

\clearpage

\section{Introduction}
\label{Sec:Intro}
The search for Higgs bosons~\cite{Englert:1964et,*Higgs:1964ia,*Higgs:1964pj,*Guralnik:1964eu,*Higgs:1966ev,*Kibble:1967sv} is, and has been, a major cornerstone of the physics programmes
of past, present and future high energy colliders. This has become even more important in view of the recent discovery of a Higgs signal by ATLAS \cite{ATLASdiscovery,*ATLAS2013034} and CMS \cite{CMSdiscovery,*Chatrchyan:2013lba,*CMS13005}. Determining the properties of this newly observed state and comparing the measurements to explicit theories beyond the Standard Model (SM) is one of the present challenges. These theories often contain enlarged Higgs sectors with multiple Higgs bosons. Even in the presence of a signal, it is therefore important that the LEP, Tevatron and LHC experiments present exclusion limits from the non-observation of Higgs bosons in various channels. These are very useful for constraining the available parameter space of those models which are able to fit correctly the observed Higgs signal. Such constraints will need to be taken into account also in the future interpretation of the Higgs results in the context of models of new physics.

In this paper we describe new developments in version 4 of the publicly available Fortran code \HB\ \cite{arXiv:0811.4169,*arXiv:1102.1898,OnlineManual}, which has been designed for exactly this purpose. For the complementary approach, to test whether a model is compatible with the observed LHC Higgs signal (and possible future signals of additional Higgs bosons), we have also developed the sister program \HS, which has been described in Ref.~\cite{Bechtle:2013xfa}. It is highly recommended to use these two programs in parallel to obtain the most complete test for extensions of the SM Higgs sector.

The experimental analyses implemented in \HB\ usually take one of two forms. Dedicated analyses have been
carried out in order to constrain some of the most popular models, such as
the SM~\cite{ATLASdiscovery,CMSdiscovery,*Chatrchyan:2013lba,hep-ex/0306033} and various benchmark scenarios in the Minimal Supersymmetric Standard Model
(MSSM)~\cite{hep-ex/0602042,Abbiendi:2013hk,Chatrchyan:2011nx,Aad:2012cfr}. In  addition, model-independent limits on the cross sections of individual
 signal topologies (such as $e^+e^-\to h_iZ\to b\bar{b}Z$) have been
 published. In the former type of analyses several search channels (or signal topologies) are typically combined in order to maximize the discovery / exclusion reach. However, the re-interpretation of these results in the context of different models than those already investigated by the search analysis requires detailed knowledge of the individual efficiencies (or signal contaminations) of the investigated search channels. In contrast, the latter type of analysis can be used easily to test a wider class of models.
  
\HBv{4} has been designed to facilitate the task of comparing Higgs sector predictions with existing exclusion limits, thus allowing the user to quickly and conveniently check a wide variety of models against the state-of-the-art results from Higgs searches. Version 4 differs significantly from previous versions of the code (described in \cite{arXiv:0811.4169,*arXiv:1102.1898,Bechtle:2013gu}) in several respects. The code now fully supports testing models against exclusion limits from the LHC, which are implemented for analyses performed at center-of-mass energies of both $\sqrt{s}=7 $ and 8 TeV. The main algorithm of \HB\ has been extended to ensure a reliable application of exclusion limits in the presence of a signal (as is now observed in the LHC data). The model-likeness test, which tests whether a given model fulfills the assumptions of a particular Higgs search to a sufficient degree, has been fully rewritten to enable in particular the limits from SM Higgs searches at the LHC to be applied in a wider context. We introduce an option to take into account theory uncertainties on the Higgs mass predictions, which are relevant, for instance, for the lightest Higgs boson mass in the MSSM. An alternative statistical treatment for the LEP constraints (in the form of a $\chi^2$ output) is provided. Finally, we describe an improved input/output for supersymmetric (SUSY) models that can now be given in the SLHA format \cite{arXiv:0801.0045,*hep-ph/0311123}. The main focus of this updated documentation is to provide a detailed description of these new developments, to show relevant physics examples of where improvements can be expected, and to introduce the user to how the improved \HB\ code can be used in practice. 

This paper is structured as follows. In Section \ref{Sec:General} we give a general introduction to the statistical approach employed in the \HB\ code, and describe, in particular, the way in which this approach has been extended in \HBv{4}. Section \ref{Sec:Input} gives a thorough description of the different methods of providing theory input for \HB, and their extension to LHC7/8 predictions. This is followed by Section \ref{Sec:New}, which contains a discussion of the major new developments, including numerical examples. Finally, Section \ref{Sec:Manual} contains the technical details on how these new features can be used in practice, extending the original \HB\ manual \cite{arXiv:0811.4169,*arXiv:1102.1898} with a description of the new subroutines, data files and example programs that are provided. In an Appendix we list and provide references to all the experimental analyses that provide results implemented in the code, including the analyses added in the latest public version (\HBv{4.1}).

\section{General Approach of \HBv{4}}
\label{Sec:General}
The general concept of \HB, including details on the treatment of limits from LEP and the Tevatron, has already been published in \cite{arXiv:0811.4169,*arXiv:1102.1898,OnlineManual} (see also Ref.~\cite{Bechtle:2013gu}). From a conceptual point of view, the extension of \HB\ to include LHC limits is straightforward. The technicalities of this implementation, and how it modifies the user input, is discussed in Section~\ref{Sec:Input}. Our aim here is to give a brief introduction to the purpose of the code and the methods it uses. We also introduce one conceptual change with respect to previous versions, which has been prompted by the application of \HB\ to models which feature a Higgs boson with a mass close to the observed LHC Higgs signal.

The basic input for \HB\ (which the user has to provide) are the relevant physical quantities predicted for the Higgs sector of the model under consideration. The necessary predictions for each Higgs boson $H_i \;  (i=1,\ldots, \nHneut + \nHplus)$  in the model are, schematically, %
\begin{equation*}
\begin{aligned}
\label{basic input}
& M_{H_i},\quad 
\Gamma_{\TOT}(H_i),\quad  
\BR_\MOD(H_i\to ...),\quad
 \frac{\sigma_\MOD(P(H_i))}{\sigma_{\REF}(P(H))},
\end{aligned}
\end{equation*}
i.e.~the Higgs boson mass, its total decay width (it is assumed that the narrow width approximation holds), its decay branching ratios, and the production cross sections, normalized to a particular reference value. Here, $P$ denotes a specific Higgs production process. If $P$ exists in the SM, its cross section, $\sigma^\SM(P(H))$, evaluated at the same mass value, $M_H = M_{H_i}$, is typically used as the reference cross section, $\sigma_{\REF}$. In some cases it can also be necessary to supply additional predictions, such as the $\BR(t\to bH^+)$, or the $\cp$ properties of the neutral Higgs bosons. Variations on the input format are offered, which allow the user to specify a simpler set of input quantities when certain basic approximations are valid.
A complete list of the options for giving model input is given in \refse{Sec:Input}.
 
In addition to the model predictions, the other important ingredient of \HB\ is the experimental data. Exclusion limits from (negative) Higgs searches are collected from the experimental publications, with the aim of keeping the code as up-to-date as possible with the latest developments. Currently the code includes results from LEP, the Tevatron and the LHC experiments. More information on which analyses are available in \HB\ is provided in Appendix~\ref{Appendix:Data}. The data for these analyses is contained in tables of expected exclusion limits at 95\% C.L. in the absence of a signal (based on Monte Carlo simulations), and the corresponding observed limits, as a function of the Higgs boson mass. 
The list consists both of analyses for which model-independent limits were
published, and of dedicated analyses carried out specifically under the assumption of the SM (like most LHC searches to date), or for Higgs bosons with certain $\cp$ properties. These limits can be applied to models with Higgs bosons which show these characteristics \emph{to a sufficient degree}\footnote{This statement will be quantified in Sect.~\ref{Sec:SMtest}.} at the parameter point being considered.  

We call the application of the limit from a particular Higgs search to one of the Higgs bosons 
of the model under study (or to two of the Higgs bosons, for searches involving two Higgs bosons) 
an ``analysis application'', which we denote by $X$ in the following.\footnote{As an example,
suppose that a model with three neutral Higgs bosons ($h_1$, $h_2$, $h_3$) 
should be checked against the limits from two neutral Higgs searches,
$A_1$ and $A_2$. Then there are six possible analysis applications, $X\in \{A_1(h_1), A_1(h_2), A_1(h_3), A_2(h_1), A_2(h_2), A_2(h_3)\}$,
for this model.} Each analysis application has a corresponding signal cross section 
prediction $\sigma(X)$, which \HB\ uses to calculate the relevant quantity $Q_\MOD(X)$ for which the experimental limit is given; typically this is a conveniently normalized cross section times a branching ratio. The corresponding experimental quantities are denoted $Q_{\EXPEC} (X)$ and $Q_\OBS(X)$ for the expected and observed limits, respectively. If two Higgs bosons have a narrow mass separation, $\delta M=M_{h_i}-M_{h_j}$, then their predicted cross sections are added for certain analyses where the mass resolution is limited and interference effects are expected to be negligible. The settings for the maximal $\delta M_h$ can be varied by the user separately for LEP, Tevatron, and LHC analyses (the default values are $0$~GeV for LEP and $10$~GeV for Tevatron/LHC).

\HB\ operates by considering, for each analysis application, the ratio of the model predictions, $Q_\MOD(X)$, to the experimental limits. To ensure that the result can be interpreted as an exclusion at $95\%$ C.L.~(which is the same confidence level as adopted by the individual analyses), it is crucial that the model prediction is only compared to the experimentally \emph{observed} limit for \emph{one} particular analysis application. In a first step, \HB\ therefore uses the \emph{expected} experimental limits to determine the analysis application $X_0$ with the highest statistical sensitivity to exclude the model point under consideration,
\begin{equation}
X_0=X:\max \frac{Q_\MOD(X)}{Q_\EXPEC(X)}.
\label{eq:modvsexp}
\end{equation} 

In the second step, \HB\ then performs the exclusion test for 
the Higgs boson and analysis combination represented by $X_0$,
by computing the ratio to the \emph{observed} limit
\begin{equation}
k_0=\frac{Q_\MOD(X_0)}{Q_\OBS(X_0)}.
\label{eq:modvsobs}
\end{equation} 
If $k_0>1$, \HB\ concludes
that this parameter point of the tested model is excluded
at $95 \%$~C.L.\footnote{If we had instead compared the predicted cross sections
directly to the experimentally observed limits for {\em all} available search
channels and considered the model excluded if at least one of them gave exclusion at 95~\% C.L., the result would in general \emph{not} correspond to an exclusion at $95 \%$~C.L. The combined probability of yielding a false exclusion from any of the individual comparisons of $Q_\MOD$ to $Q_\OBS$ would also yield an overall probability for false exclusion higher than that from applying a single limit.} 

The statistical method as described here (in the following referred to as the \HBclassic\ method) has been the only mode of operation available in previous {\tt HiggsBounds} versions. For \HBv{4}, we have extended this method to perform better in situations where a Higgs boson signal is  present (as is now the case in the LHC data). The problem of the \HBclassic\ method arises for models with multiple Higgs bosons. If one of these has a mass close to that of the observed signal (which is likely, since any reasonable model should also explain this signal), its analysis applications will test the model predictions against limits (for various channels) in the signal region. In this region, the \emph{expected} limits (based on the \emph{background-only} hypothesis) will continue to improve with more experimental data and optimized analysis methods, whereas the \emph{observed} limits can never be expected to reach exclusion at the SM level (provided a true signal of near-SM strength is what is observed). For model points where the most sensitive analysis application $X_0$ is a test of the signal-like Higgs boson, the \HBclassic\ \HB\ method would therefore never yield exclusion. Moreover, constraints on the remaining Higgs spectrum (with less expected sensitivity) are not applied. Even if the exclusion remains formally valid at $95\%$~C.L., it could be anticipated that this problem would eventually become serious enough to limit the usability of the code.
 
Among the several possible ways that the \HB\ algorithm could be extended to address this problem, all involving different compromises, we have opted for a solution which involves a slight violation of the strict testing of only one experimental limit. We call this the \HBfull\ \HB\ method. In summary, this method performs the original \HB\ test separately for each Higgs boson in the model. In the full \HB\ method, the first step is to evaluate the most sensitive analysis application $X_i$ for each Higgs boson $H_i$ according to
\begin{equation}
X_i=X(H_i):\max \frac{Q_\MOD\left(X(H_i)\right)}{Q_\EXPEC\left(X(H_i)\right)}.
\label{eq:fullmodvsexp}
\end{equation} 
This is followed by a straightforward exclusion test on the individually most sensitive analysis applications
\begin{equation}
k_i=\frac{Q_\MOD(X_i)}{Q_\OBS(X_i)}.
\label{eq:fullmodvsobs}
\end{equation} 
The result of these tests contains more information than the single test of \HB\ \HBclassic\ (such as exclusion/non-exclusion by individual Higgs bosons), which is now made available to the user (see Sect.~\ref{Sec:Manual} for details). A combined \HB\ exclusion is also calculated, where the result is interpreted as model exclusion if $k_i>1$ for \emph{any} of the $k_i$. The combined (single-number) output is then calculated as
\begin{equation}
k_0=\max_i k_i,
\end{equation}
\begin{equation}
X_0=X_i : \max_i k_i .
\end{equation}

By the construction of the \HBfull\ method, it follows directly that the two methods are equivalent for models with a single Higgs boson. It is also clear that the \HBfull\ method can only give \emph{stronger} exclusion than the \HBclassic\ method. This is consistent with the fact that the exclusion of the \HBfull\ method will correspond to a limit at somewhat \emph{lower} statistical confidence level than $95\%$. Still, the deviation from the strict $95\%$~C.L.~should be minor in this approach compared to the alternative (naive) testing of all Higgs bosons versus all observed limits, since the number of Higgs bosons in a model in general is much smaller than the number of implemented experimental analyses. Furthermore, a non-negligible dilution of the $95\%$~C.L.~interpretation of the combined result is only expected in the case where more than one test $X_i$ leads to a ratio $k_i \approx 1$.

\begin{figure}
\centering
\includegraphics[width=0.45\columnwidth]{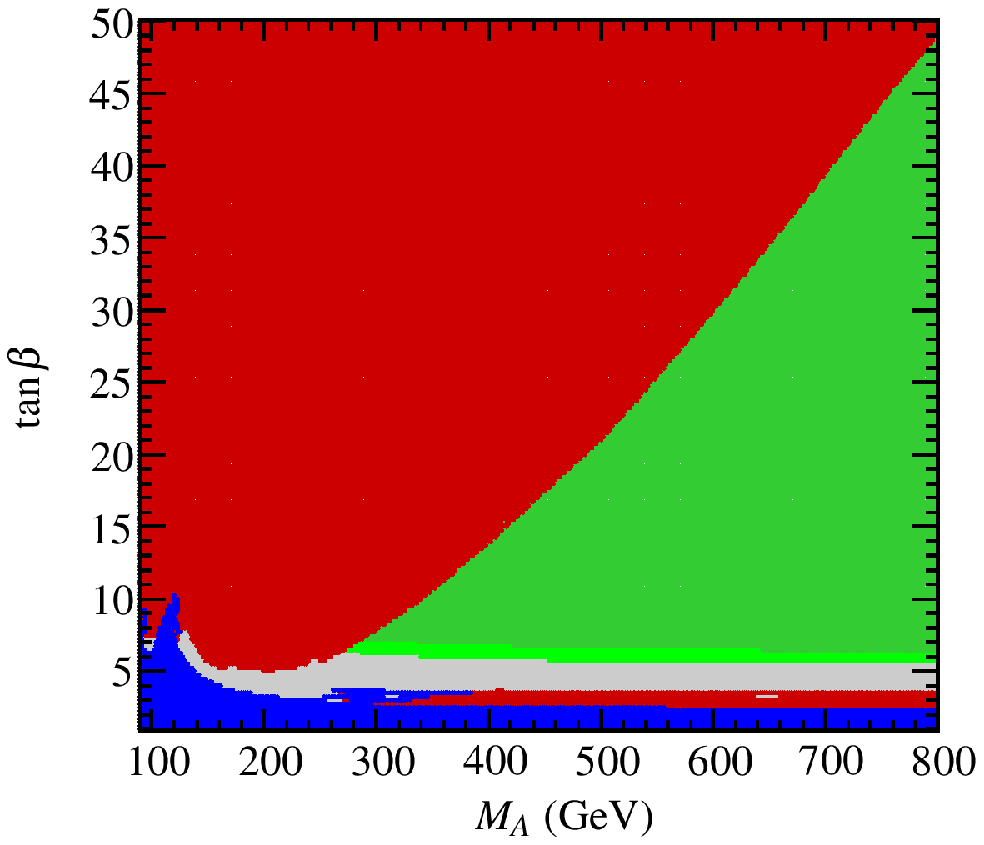}
\includegraphics[width=0.45\columnwidth]{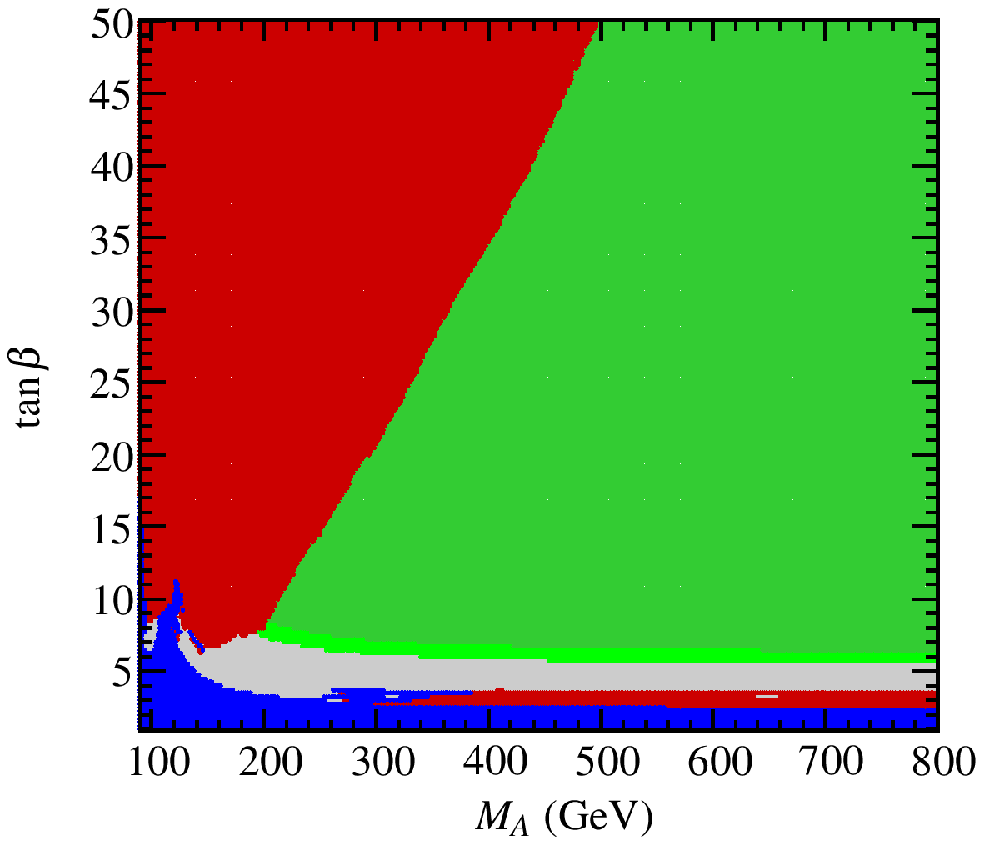}
\caption{Exclusion regions in the MSSM parameter space for the \mhmod\ benchmark scenario~\cite{Carena:2013qia}. Results from \HB\ \HBfull\ (left) are compared to the results from \HB\ \HBclassic\ (right). The colours show exclusion by the LHC (red), LEP (blue), and the favored region where $M_h=125.7\pm 2$~GeV (dark green), $M_h=125.7\pm 3$~GeV (light green).}
\label{fig:fullclassic}
\end{figure}

\begin{figure}[t!]
\centering
\includegraphics[width=0.32\columnwidth]{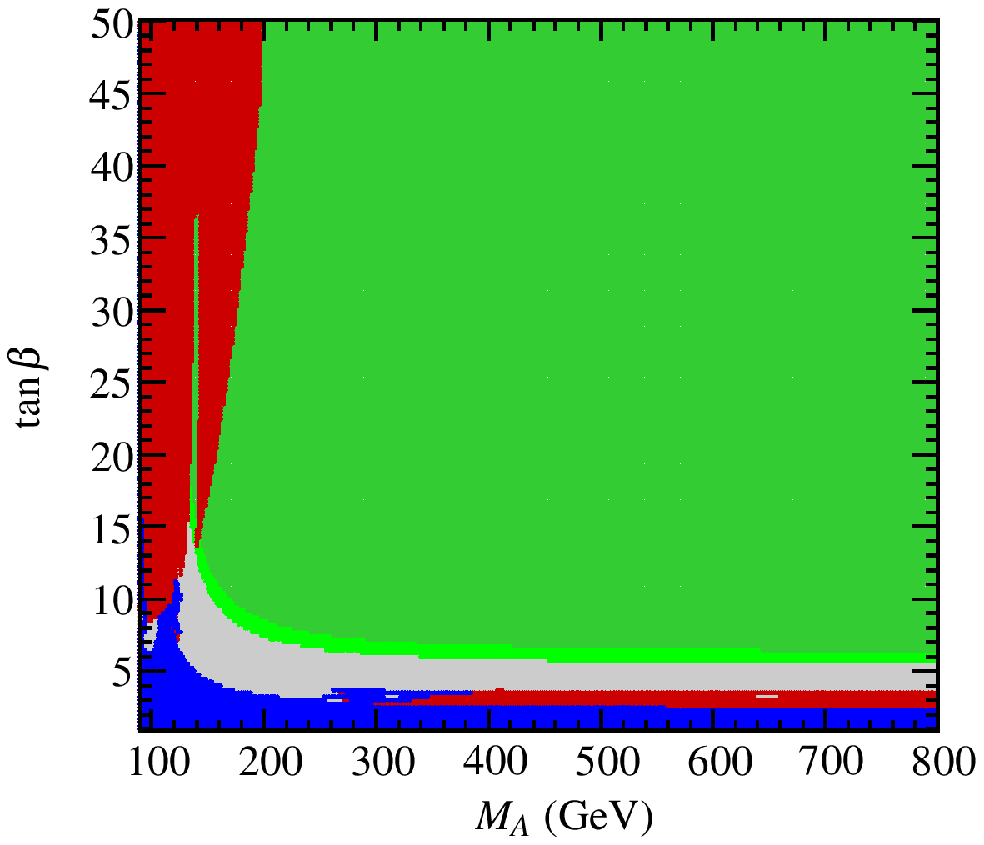}
\includegraphics[width=0.32\columnwidth]{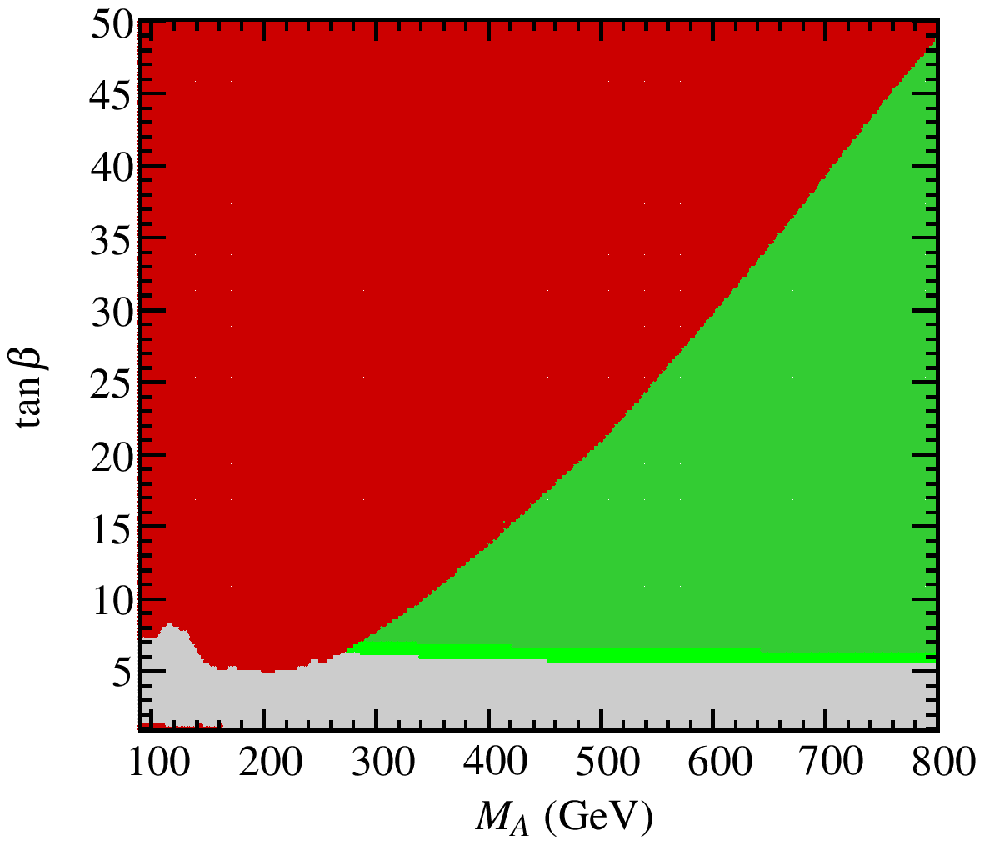}
\includegraphics[width=0.32\columnwidth]{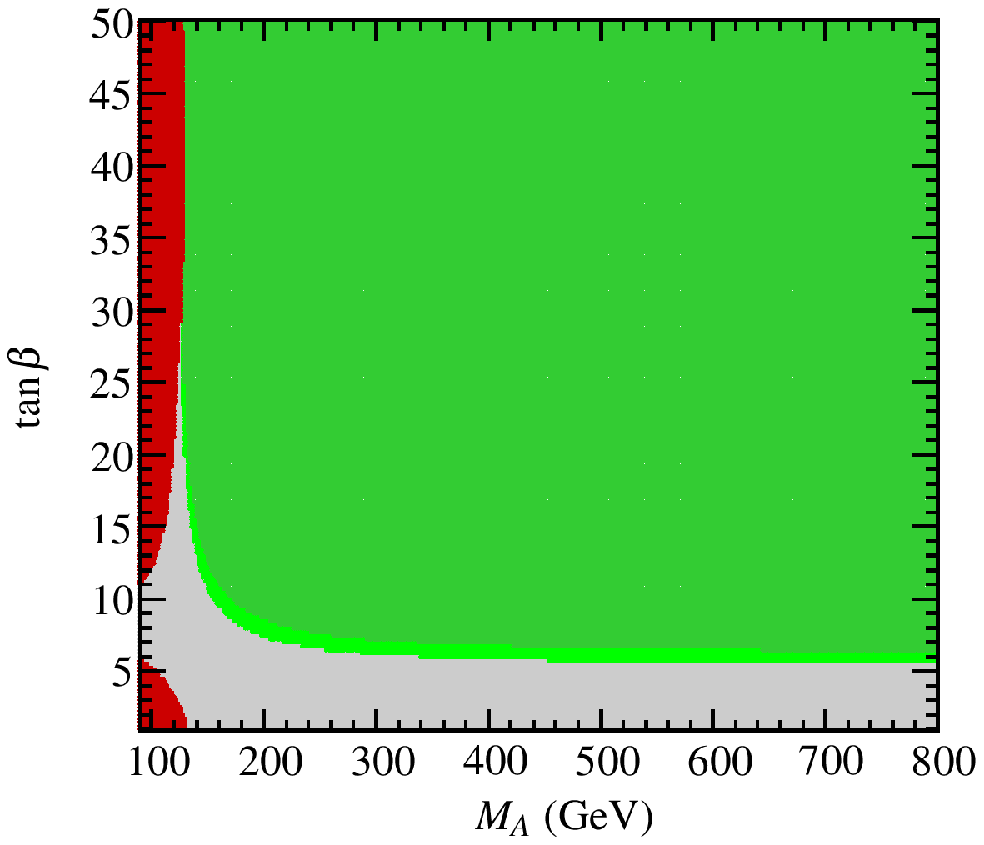}
\caption{Contribution to the \emph{full} \HB\ exclusion in the MSSM parameter space for the \mhmod\ benchmark scenario \cite{Carena:2013qia} from exclusion of the individual Higgs bosons: $h$ (left), $H/A$ (center), and $H^\pm$ (right). The colour coding is the same as in Fig.~\ref{fig:fullclassic}.}
\label{fig:individualH}
\end{figure}

To illustrate the difference between the \HBclassic\ and \HBfull\ methods of \HB, we show in Fig.~\ref{fig:fullclassic} three versions of the excluded regions in an MSSM benchmark scenario, the so-called \mhmod\ scenario \cite{Carena:2013qia}. The MSSM  has three neutral Higgs bosons ($h,H,A$), where in this scenario the $h$ boson can have a mass close to the LHC signal around $M_h\sim 125\gev$ (this region, considering a $2$ $(3)$~GeV total uncertainty on $M_h$ is indicated by dark (light) green colour in the figure). The exclusion bounds, as evaluated by \HB, are shown separately for LEP exclusion (blue) and the LHC (red). When evaluating the limits in this figure, a theory uncertainty of $3$~GeV is taken into account in the evaluation of the lightest Higgs mass, see Sect.~\ref{Sec:MassUnc} for details on how this is done. As can be seen from this figure, the \HBfull\ method gives the strongest exclusion, corresponding to the most accurate application of the existing limits in this scenario (as also used in \cite{Carena:2013qia}). The difference to the \HBclassic\ method can be seen in particular for high $M_A$ and high $\tan\beta$ (the decoupling regime). Here the applicability of the \HBclassic\ method is limited, since the globally most sensitive channel is a search for the lightest (SM-like) Higgs boson, which cannot be excluded when is mass its in the signal region, $M_h\simeq 125\gev$.
This is in contrast to the results in the \HBfull\ method, which can be further illustrated by looking at the contribution of individual Higgs bosons as shown in Fig.~\ref{fig:individualH} for the same MSSM example. The first panel shows the exclusion contributed by $h$. The narrow unexcluded region around $\MA=135$~GeV results from a particular channel ($pp\to VH$, $H\to b\bar{b}$) being the most sensitive. For this channel, the observed limit is not strong enough here to exclude the lightest Higgs. The second panel shows the exclusion for $H/A$. They are treated together, since their masses are close to degenerate over most of the parameter space. The dominant exclusion therefore comes from the same search channels and their signal rates are added. Finally, the last plot shows the exclusion from  $H^\pm$. The exclusion region presented for the \HBfull\ method in Fig.~\ref{fig:fullclassic} consists of the union of the three different exclusion regions shown here.
In the \HB\ distribution we provide an updated example program, {\tt HBwithFH}, which can be used to test MSSM parameter points for exclusion using either the \HBfull\ or the \HBclassic\ \HB\ methods.

\newpage
\section{Theoretical predictions}
\label{Sec:Input}
The theoretical model predictions, which are compared to the experimental data in the analysis applications, are computed from the user input. A detailed explanation of this input has been given for previous version of \HB\ \cite{arXiv:0811.4169,*arXiv:1102.1898}. For completeness, we include the full input specification here, describing both the original input and the updates in \HBv{4}. For the input of the required theory predictions to \HB, the user can choose between four different formats. In the command line version of the program, these are labelled by the variable {\tt whichinput}, which can take the values {\tt hadr} (hadronic), {\tt part} (partonic), {\tt effC} (effective couplings), or {\tt SLHA} (SUSY Les Houches Accord). The type of input required for each of these formats is summarized in \refta{table:instructions2}, and the following subsections contain more detailed information about each type of input. In particular, we describe the internal treatment of LHC cross sections (at $7$ and $8$ TeV). 
Technical details on the subroutines which should be used for the different input options can be found in Sect.~\ref{Sec:Manual}.
\begin{table}[h!]
\centering
\renewcommand{\arraystretch}{1.0}
\footnotesize
\begin{tabular}{ll}
\br
{\tt whichinput} & ({\tt character(LEN=4)}) \\
\mr
{\tt hadr} & Higgs masses, total decay widths, ratios of LEP cross sections, \\
           & ratios of Tevatron hadronic cross sections, \\
           & ratios of LHC hadronic cross sections, branching ratios.\\[1mm]
{\tt part} & Higgs masses, total decay widths, ratios of LEP cross sections, \\
           & mainly ratios of partonic hadron-collider cross sections, branching ratios.\\[1mm]
{\tt effC} & Higgs masses, total decay widths, \\
           & ratios of effective couplings squared, some branching ratios.\\[1mm]
{\tt SLHA} & Higgs masses, total decay widths, \\
           & ratios of effective couplings squared, branching ratios.\\
\br
\end{tabular}\\
\caption{ The possible values of the variable {\tt
    whichinput}, which indicates the format of the theoretical
  predictions provided by the user for the neutral Higgs sector.} 
  \label{table:instructions2}
\end{table}

\subsection{Hadronic input}
\label{section:hadr}
The hadronic input option {\tt whichinput=hadr} requires model input in the most general form and therefore contains the lowest degree of approximation. The complete input in this format involves the specification of:
\begin{enumerate}
\item[(1)] The masses for the neutral Higgs bosons, $h_k$ ($k=1,\ldots, \nHneut)$, and the (singly) charged Higgs bosons, $H_k^{\pm}$ $(k=1,\ldots, \nHplus)$, in the model.
\BEA
M_{h_k},\, M_{H_k^{\pm}},\non
\EEA
\item[(2)] their total decay widths,
\BEA
\Gamma_\TOT(h_k),\,\Gamma_\TOT(H_k^{\pm}),\non
\EEA
\item[(3)] whether the neutral Higgs bosons are $\cp$-even, $\cp$-odd or states of mixed $\cp$,\\

\item[(4)] the neutral Higgs branching ratios which have SM equivalents 
\BEA
&&\BR_\MOD(h_k\to \text{SM}),\non\\
&&\text{with SM = $\{s\bar{s}$, $c\bar{c}$, $b\bar{b}$, $\mu^+ \mu^-$,$\tau^+\tau^-$,
 $W^+W^-$, $ZZ$, $Z\gamma$, $\gamma\gamma$, $gg\}$,}\non
\EEA
\item[(5)] the neutral Higgs branching ratios without SM equivalents
\BEA
&&\BR_\MOD(h_k\to h_i h_j),\non \\
&&\BR_\MOD(h_k\to {\rm invisible}),\non
\EEA
\item[(6)] the charged Higgs branching ratios to SM particles 
\BEA
\BR(H_j^{+} \to {\text{SM}}),\:{\rm where}\non\:
\text{SM = $\{c\bar{s}$, $c\bar{b}$,  $\tau^+ \nu_{\tau}\}$,}\non
\EEA
\item[(7)] the top quark branching ratios
\BEA
&&\BR(t \to W^{+}b),\non\\
&&\BR(t \to H_j^{+}b),\non
\EEA
\item[(8)] normalized cross sections ratios $R_{\sigma}(P)$ (see below for the definition), for the LEP Higgs production processes
\BEA 
 & &  e^+e^- \to h_j Z              ,\non\\
 & &  e^+e^- \to b \bar{b} h_j      ,\non\\
 & &  e^+e^- \to \tau^+ \tau^- h_j  ,\non\\
 & &  e^+e^- \to h_i h_j            ,\non\\
 & &  e^+e^- \to H^+_j H^-_j        ,\non
\EEA
\item[(9)] normalized ratios $R_{\sigma}(P)$ (see below) of hadronic Higgs production cross sections at the Tevatron for the processes
\BEA 
 & &  p\bar{p} \to h_j   ,\non\\
 & &  p\bar{p} \to b h_j ,\non\\
 & &  p\bar{p} \to h_j W ,\non\\
 & &  p\bar{p} \to h_j Z ,\non\\
 & &  p\bar{p} \to h_j  {\rm \, via \, VBF}  ,\non\\
 & &  p\bar{p} \to t \bar{t} h_j             ,\non
\EEA
\item[(10)] and, finally, normalized ratios $R_{\sigma}(P)$ of hadronic Higgs
	production cross sections at the LHC (both for $\sqrt{s}=7$ and $8\tev$, given as separate inputs) for the processes 
\BEA 
 & &  pp \to h_j   ,\non\\
 & &  pp \to b h_j ,\non\\
 & &  pp \to h_j W ,\non\\
 & &  pp \to h_j Z ,\non\\
 & &  pp \to h_j  {\rm \, via \, VBF}  ,\non\\
 & &  pp \to t \bar{t} h_j             .\non
\EEA
\end{enumerate}
It is important to note that this corresponds to an exhaustive list of the possible input. In certain (most) cases only a subset of these inputs may be required.  For example, if the user is only interested in limits from the LHC, no LEP or Tevatron cross sections need to be given as input (the corresponding values can be set to zero). Likewise, for models with only neutral Higgs bosons, no input involving either the charged Higgs sector or top decays will be used. The meaning of most of these quantities should be pretty clear already from the notation; for those that require further clarifications we provide some more details below.

For input items (8), (9) and (10), the normalized cross section of a Higgs production process $P$ is simply defined by
\begin{equation} 
R_{\sigma}(P)=\frac{\sigma_{\rm model} (P)}{\sigma_{\rm ref} (P)}.
\label{eq:R-sigma}
\end{equation}
Where a SM equivalent exists, the reference cross section
$\sigma_{\rm ref} (P)$ for Higgs boson $h_k$ is
$\sigma_{\rm ref} (P)=\sigma_{\rm SM} (P)$, evaluated for $M^{\rm SM}_H=M_{h_k}$.
The only neutral Higgs production process considered in \HB\ up to now which does not have a SM equivalent 
is $e^+e^- \to h_j h_i$. In this case, we choose as the reference cross section a fictitious
production process for two scalar particles ($H'$, $H$) with masses
$M_{H'}=M_{h_j}$ and $M_H = M_{h_i}$ that proceeds via a virtual
$Z$ exchange with a standardized squared coupling constant $g_{H'HZ}^2 =e^2/(4 \sw^2 \cw^2)$,
where $e$ denotes the electromagnetic coupling constant, and $\sw$ ($\cw$) the sine (cosine) of the weak mixing angle, respectively.\footnote{The chosen reference cross section coincides with the MSSM cross
section (at tree level) for the process $e^+ e^- \to h A$, if the Higgs mixing-angle
dependent factor $\cos(\beta-\alpha)$ is divided out of the tree level
coupling. Therefore, $R_{\sigma}(e^+ e^- \to h A)$ is simply given by
$\cos^2(\beta-\alpha)$ in the MSSM (with real parameters) at tree level.}
Similarly, for the process $P=e^+e^- \to H^+_j H^-_j$, the reference cross section is
the cross section of the process $e^+e^- \to H^+ H^-$ 
in a two-Higgs-doublet model (e.g. the MSSM) at tree-level 
(i.e.~with $s$-channel $\gamma$ and $Z$ exchange). 
This reference cross section depends solely on
the mass of the charged Higgs boson and SM quantities.
As a consequence, the leading-order prediction in the MSSM is $R_{\sigma}(e^+e^- \to H^+_j H^-_j)=1$.

The branching ratio to ``invisible'', $\BR_\MOD(h_k\to {\rm invisible})$, is defined
as the branching ratio of a neutral Higgs boson into particles which are only infered in the detector by their contribution to the missing transverse energy. Examples of this includes the lightest neutralino in the MSSM~\cite{Eboli:2000ze,*Chang:2008cw}, inert scalars~\cite{LopezHonorez:2006gr,*Branco:2011iw,*Goudelis:2013uca}, or majorons in supersymmetric models with spontaneous breaking of $\mathcal{R}$-parity~\cite{Grossman:2002ry,*Hirsch:2004rw,*Ghosh:2011qc}.

The hadronic input is the most versatile, since it allows the user to provide the predictions in the form of the most precise calculations available. It is also the only input format which allows for studying e.g.~effects of parton distribution functions or hadronic uncertainties on the Higgs exclusion bounds. On the other hand, this input format is also the most demanding, and in order to make it more convenient for the user to correctly normalize the rate predictions, the
\HB\ library provides a series of Fortran functions which allow
the user to access the predictions of certain SM quantities, including 
the hadronic SM Higgs production cross sections, total decay width, and branching ratios as a
function of the Higgs mass. 

\subsection{Parton-level input}
\label{section:part}
The second possibility for specifying the \HB\ input (with {\tt whichinput=part}) is using ratios of \emph{partonic} cross sections as far as is possible. This input format is in many cases more convenient for the user to calculate than the hadronic option. It requires (at most) the following model predictions
\begin{enumerate}
\item[(1)-(8)] the same as for {\tt whichinput=hadr},

\item[(9)] normalized ratios $R_{\sigma}(P)$ (as defined by Eq.~\ref{eq:R-sigma}) of hadronic Higgs production cross sections at the Tevatron ($\sqrt{s}=1.96 \tev$) and the LHC (at $\sqrt{s}=7/8\tev$) for the neutral Higgs production processes
\BEA 
 & &  p\bar{p} \to h_j  {\rm \, via \, VBF}\non\\
 & &  p\bar{p} \to t \bar{t} h_j \non\\
 & &  p p \to h_j  {\rm \, via \, VBF}\non\\
 & &  p p \to t \bar{t} h_j \non
\EEA
\item[(10)] normalized ratios $R^{h_j+y}_{nm}$ (defined below) of parton-level cross sections for
 neutral Higgs production, which are assumed to be valid both at the Tevatron and the LHC, for the following processes
\BEA
& & gg,b\bar{b} \to h_j,\non\\
& & u\bar{d}, c\bar{s} \to h_j W^+, \non\\
& & \bar{u}d,\bar{c}s \to h_j W^-,\non\\
& & gg,q\bar{q} \to h_j Z\; (q=d,u,s,c,b), \non\\
& & bg \to h_j b.\non
\EEA
\end{enumerate}
The normalized cross section ratio $R^{h_j+y}_{nm}$ for a partonic
neutral Higgs production process, $nm\to h_j + y$, is defined by
\begin{equation}
R^{h_j+y}_{nm}=\frac{\hat{\sigma}^{\rm model}_{nm\to h_{j}+y} }{\hat{\sigma}^{\rm SM}_{nm\to H+y} }.
\label{eq:R-nm-partonic}
\end{equation}
It should be calculated for a parton-system center-of-mass energy squared $\hat{s}=\hat{s}_0$, where $\hat{s}_0$ denotes the partonic production threshold $\hat{s}_0=(M_{h_j}+M_y)^2$. For this approximation to be valid the dependence of the partonic cross section on $\hat{s}$ is required to be mild.  For the case of single Higgs boson production, $M_y=0$.

The partonic cross section ratios $R^{h_j+y}_{nm}$ can be a lot easier to calculate than hadronic cross section ratios for a wide range of models. In addition, it is often (at least to a good approximation) the case that 
\begin{equation*}
R^{h_j +W^+}_{nm}=R^{h_j +W^-}_{nm}=R^{h_j +Z}_{q\bar{q}}
\label{eq:gaugebosonpartratios}
\end{equation*}
for all $nm$. This reduces the number of partonic cross
section ratios which need to be provided by the user from $13$ to $5$.
An example of a model of this type is given by the MSSM with real parameters,
where the common ratio can be calculated approximately from the normalized (squared) effective coupling of the Higgs boson to a pair of $Z$ bosons, $(g^\MOD_{h_jZZ}/g^\SM_{HZZ})^2$. For a more complete discussion of the use of the efffective coupling approximation as input to \HB, see \refse{section:effC}.

Internally, \HB\ uses the approximate relations
 \begin{align}
\label{eq:R-sigma-approx}
R^{\mathrm{TEV}}_\sigma(P) & \simeq \sum_{\{n,m\}}
	R_{nm}^{h_j+y}
	\frac{
	  \sigma_\SM(p\bar p \to n m \to H+y)
	}{\sigma_\SM(p\bar p \to H+y)}\,,\\
	\label{eq:R-sigma-approx2}
R^{\mathrm{LHC}}_\sigma(P) & \simeq \sum_{\{n,m\}}
	R_{nm}^{h_j+y}
	\frac{
	  \sigma_\SM(p p \to n m \to H+y)
	}{\sigma_\SM(p p \to H+y)}\,,
\end{align}
to calculate the missing hadronic cross section ratios from the partonic cross section ratios. Here, $\sigma_\SM(p p \to n m \to H+y)$ denotes the contribution from the partonic initial state $nm$ to the total hadronic cross section for the process $p p \to H+y$ in the SM. The hadronic LHC ratios are evaluated separately for $7$ and $8\tev$ but, as already mentioned, using the same values for $R_{nm}^{h_j+y}$.

\begin{figure}[t]
\centering
\hspace{0.5em}
 \includegraphics[width=0.32\columnwidth]{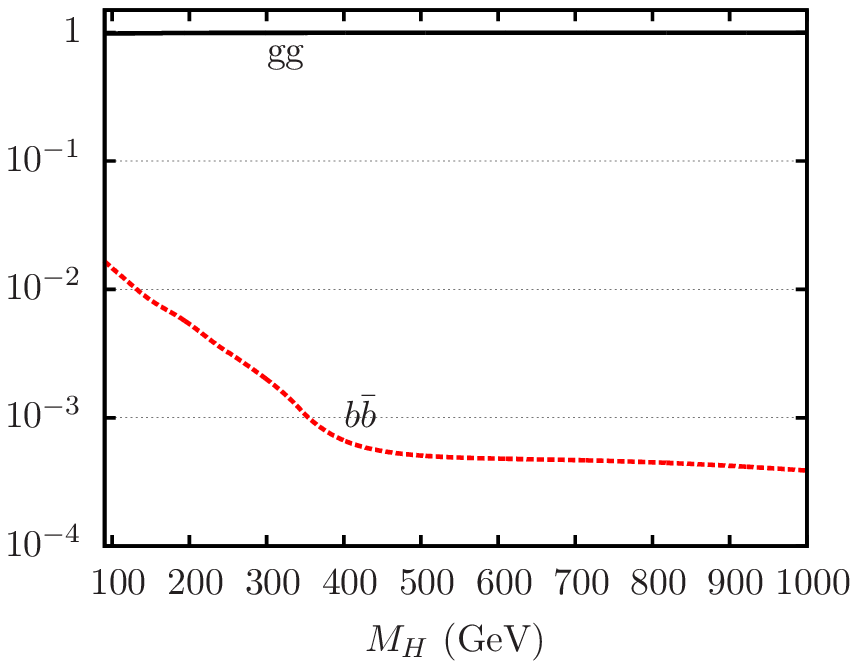}
  \includegraphics[width=0.32\columnwidth]{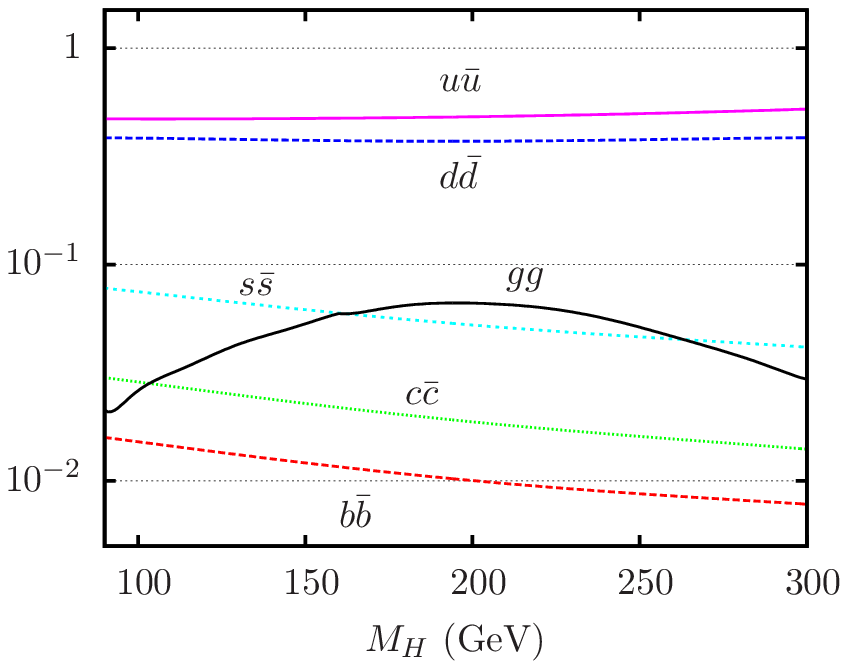}
  \includegraphics[width=0.32\columnwidth]{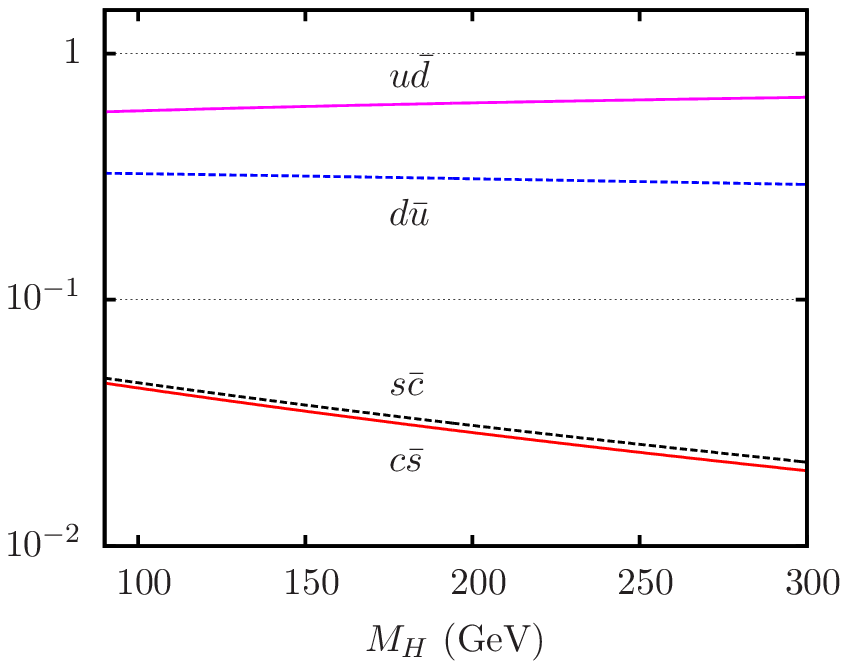}
    \caption{Relative SM contributions from different partonic subprocesses to the total hadronic cross sections for 
	$pp\to H$ (left), $pp\to HZ$ (center), and $pp\to HW^\pm$ (right) at $\sqrt{s}=8$~TeV.}
	\label{fig:CS-ratios} 
\end{figure}

The hadronic cross section ratios for the SM appearing in Eqs.~\eqref{eq:R-sigma-approx}, \eqref{eq:R-sigma-approx2} are provided within \HB. Further discussion of the applicability of this approximation, and
details of how the ratios $\sigma_\SM(p\bar p \to n m \to H+y)/\sigma_\SM(p\bar p \to H+y)$ 
are calculated for the Tevatron have been presented in \cite{arXiv:0811.4169}. 
In \HBv{4}, these cross section ratios are provided 
for the LHC with $\sqrt{s}=7\tev$ and $\sqrt{s}=8\tev$ based on 
the prediction  
of the LHC Higgs Cross Section Working Group \cite{Dittmaier:2011ti,*Dittmaier:2012vm,*Heinemeyer:2013tqa}
for the gluon fusion cross section
and the $HZ$ and $HW$ cross sections,
and the program {\tt bbh@nnlo} 1.3~\cite{hep-ph/0304035}.
Results on the relative contributions from different parton configurations (as indicated in the figure) to the total hadronic cross section
for single Higgs production, $HZ$ production, $HW^\pm$ production (at $\sqrt{s}=8\tev$) are shown in Fig.~\ref{fig:CS-ratios}.

\subsection{The effective coupling approximation}
\label{section:effC}
With the effective coupling option ({\tt whichinput=effC}), the user input is greatly simplified and 
reduced to a smaller number of quantities. From this input, physical predictions corresponding to input with the partonic option {\tt whichinput=part} are calculated.
The effective couplings involve specifying (at most)
\begin{enumerate}
\item[(1)-(2)] the same as for {\tt whichinput=hadr},
\item[(3)] normalized scalar and pseudoscalar (squared) effective Higgs couplings to fermions 
\BEA
\left(\frac{g^\MOD_{s,h_k(\text{OP})}}{g^\SM_{H(\text{OP})}}\right)^2,\,
\left(\frac{g^\MOD_{p,h_k(\text{OP})}}{g^\SM_{H(\text{OP})}}\right)^2,\,
\text{ with OP = $\{s\bar{s}, c\bar{c}, b\bar{b}, t\bar{t}, \mu^+ \mu^-, \tau^+\tau^-\}$},
\non
\EEA
\item[(4)] normalized (squared) effective Higgs couplings to bosons 
\BEA
\left(\frac{g^\MOD_{h_ih_jZ}}{g^{\REF}_{H'HZ}}\right)^2,\,
\left(\frac{g^\MOD_{h_k(\text{OP})}}{g^\SM_{H(\text{OP})}}\right)^2,\,
\text{ with OP = $\{W^+W^-, ZZ, Z\gamma, \gamma\gamma, gg, ggZ\}$},\non
\EEA 
\item[(5)] neutral Higgs branching ratios without SM equivalents, charged Higgs
	branching ratios to SM particles, and top quark branching ratios according to (5)--(7) of {\tt whichinput=hadr}.
\end{enumerate}
The scalar and pseudoscalar components of a Higgs coupling to a fermion pair are defined in the
usual way, via the Feynman rule for the coupling of a generic neutral Higgs
boson $h$ to fermions:
\begin{equation}
g_{h f \bar f}  = \mathrm{i}(g_s \text{\bf 1} + g_p \gamma_5),
\end{equation}
where $g_s$ and $g_p$ are real-valued scalar and pseudoscalar coupling
constants respectively, and $\text{\bf 1}$ and $\gamma_5$ are the usual
matrices in Dirac space.
A \cp-even scalar, like the SM Higgs boson, has $g_p=0$ and a \cp-odd scalar has $g_s=0$. 

Where these exist, the reference couplings are taken as the SM tree-level equivalents:
\begin{align}
\label{eq:gref-HZZ}
\left(g^\SM_{HZZ}\right)^2 & = 
	\left(\frac{e \MZ^2}{\sw \MW}\right)^2,\\
\label{eq:gref-HWW}
\left(g^\SM_{HWW}\right)^2 & = 
	\left(\frac{e\MW}{\sw}\right)^2,\\
\label{eq:gref-Hff}
\left(g^\SM_{Hf\bar f}\right)^2 & = 
	\left(\frac{e m_f}{2\sw\MW}\right)^2,
\end{align}
where $\MZ$ is the $Z$ boson mass, $\MW$ the $W$ boson mass, and $m_f$ the mass of fermion $f$.  The reference coupling $(g^{\REF}_{H'HZ})^2$, that does not have a SM equivalent, is defined as
\begin{equation}
(g^{\REF}_{H'HZ})^2= \frac{e^2}{4\sw^2 \cw^2}.
\label{eq:gref-HHZ}
\end{equation}

The effective couplings 
$(g^\MOD_{h_k \gamma\gamma}/g^\SM_{H \gamma\gamma})^2$ (and similarly for $\gamma Z$) are loop-induced. They can most conveniently be defined via
\begin{equation}
\left(\frac{g^\MOD_{h_k \gamma\gamma}}{g^\SM_{H \gamma\gamma}}\right)^2
	= \frac{\Gamma^\MOD_{h_k\to \gamma\gamma}(M_{h_k})}
		{\Gamma^\SM_{H\to \gamma\gamma}(M_H=M_{h_k})}.
\end{equation}
For the Higgs-gluon-gluon effective coupling, $(g^\MOD_{h_kgg}/g^\SM_{Hgg})^2$,  there is a choice of definition. It can be defined either via decay widths as
\begin{equation}
\left(\frac{g^\MOD_{h_k gg}}{g^\SM_{Hgg}}\right)^2
	=\frac{\Gamma^\MOD_{h_k\to gg}(M_{h_k})}
		{\Gamma^\SM_{H\to gg}(M_H=M_{h_k})},
\label{eq:defggheffCa}
\end{equation}
or via partonic cross sections:
\begin{equation}
\left(\frac{g^\MOD_{h_k gg}}{g^\SM_{Hgg}}\right)^2=R_{g g}^{h_k}.
\label{eq:defggheffCb}
\end{equation}
It has to be understood that both these definitions will involve approximations.
It is therefore only recommended to use the input of effective couplings when
both definitions result in similar values for 
$(g^\MOD_{h_kgg}/g^\SM_{Hgg})^2$. 
However, under certain circumstances, this
condition can be relaxed. For example, in a model in which the 
LEP searches for Higgs bosons decaying into hadrons
are not relevant, the effective $h_kgg$ coupling can be defined solely by \refeq{eq:defggheffCb}. Conversely, if for some reason the gluon fusion Higgs production mechanism is not relevant, the effective coupling can be defined solely by
\refeq{eq:defggheffCa}.

The calculation of LEP and Tevatron cross section ratios from the effective couplings has been described in \cite{arXiv:0801.0045,*hep-ph/0311123}. Here we shall focus on the extension of this procedure to LHC cross sections.
Partonic cross section ratios for the LHC are calculated as 
\begin{align}
\label{eq:effCparta}
R_{g g}^{h_k}&=
	\left(\frac{g^\MOD_{h_kgg}}{g^\SM_{Hgg}}\right)^2
\,,\\
R_{g g}^{h_kZ}&=\left(\frac{g^\MOD_{h_kggZ}}{g^\SM_{HggZ}}\right)^2\,,\\
\label{eq:effCpartb}
R_{b \bar b }^{ h_k}
	=
	R_{b g,\bar{b} g}^{h_kb,h_k \bar{b}} 
	&=\left(\frac{g^\MOD_{s,h_k b\bar b}}{g^\SM_{H b\bar b}}\right)^2+
\left(\frac{g^\MOD_{p,h_k b\bar b}}{g^\SM_{H b\bar b}}\right)^2
\,, \\
\label{eq:effCpartc}
R_{q \bar q' }^{ h_k W^+}
	= 
R_{q' \bar q }^{ h_k W^-}
	&=
	\left(\frac{g^\MOD_{h_kWW}}{g^\SM_{HWW}}\right)^2
\,,\\
\label{eq:effCpartd}
R_{q'' \bar q'' }^{ h_k Z}
	&=
	\left(\frac{g^\MOD_{h_kZZ}}{g^\SM_{HZZ}}\right)^2
\,,
\end{align}
where $(q,q')\in\{(u,d),(c,s)\}$ and $q''\in\{u,d,c,s,b\}$.
The normalized hadronic cross section ratio for $t \bar t$ production
together with a $\cp$-even Higgs boson for the LHC
is obtained using 
\begin{equation}
\label{eq:effCtthLHC}
R_{\sigma}(p p \to t \bar t  h^{\cp\mathrm{-even}}_k )
	=
	\left(\frac{g^\MOD_{s,h_k t \bar t}}{g^\SM_{H t \bar t}}\right)^2.
\end{equation}
For $h_k$ production via VBF, the normalized hadronic cross section ratio is calculated using the approximate relation
\begin{equation}
\label{eq:effCvbf}
R^{\text{LHC}}_{\sigma}(p p \to h_k \text{ via VBF})
	=
 	 R_{\text{VBF,LHC}}^{WW}\,
		\left(\frac{g^\MOD_{h_kWW}}{g^\SM_{HWW}}\right)^2
	+R_{\text{VBF,LHC}}^{ZZ}\,
		\left(\frac{g^\MOD_{h_kZZ}}{g^\SM_{HZZ}}\right)^2.
\end{equation}
The $M_H$ dependence of the relative fractions of VBF events induced from $WW$ and $ZZ$ fusion, denoted as $R_{\text{VBF,LHC}}^{WW}$ and $R_{\text{VBF,LHC}}^{ZZ}$, respectively, is mild and, for the Tevatron case, can be approximated by constant values. For the LHC case, we obtain these functions by fitting to SM results produced with
HAWK~1.1 \cite{arXiv:0707.0381,arXiv:0710.4749} for $p p$ collisions at $\sqrt{s}=7\tev$. Note that, for models where $(g^\MOD_{h_kWW}/g^\SM_{HWW})^2=(g^\MOD_{h_kZZ}/g^\SM_{HZZ})^2=(g^\MOD_{h_kVV}/g^\SM_{HVV})^2$ (which is often the case), Eq.~\eqref{eq:effCvbf} reduces simply to
\begin{equation}
R^{\text{LHC}}_{\sigma}(p p \to h_k \text{ via VBF})
	=
		\left(\frac{g^\MOD_{h_kVV}}{g^\SM_{HVV}}\right)^2,
\end{equation}
independent of $R_{\text{VBF,LHC}}^{WW}$ and $R_{\text{VBF,LHC}}^{ZZ}$.

The input scheme for decay widths and branching ratios is not affected by the extension to include LHC results in \HBv{4}, and the calculation from the effective couplings follows what has been published earlier \cite{arXiv:0811.4169,*arXiv:1102.1898}.\footnote{Since the calculation of branching ratios relies on the total Higgs width, we would like to point out that it is especially important to give accurate values for $\Gamma_\TOT(h_k)$ when using this input option.} The only difference compared to previous versions is that we have updated the SM reference values to agree with those recommended by the LHC Higgs cross section working group \cite{Dittmaier:2011ti,*Dittmaier:2012vm} for $\MH$ between $80\gev$ and $1\tev$. 

In order to constrain Higgs bosons with masses below $\sim 10 \gev$, the effective coupling input option is usually not appropriate because the implemented SM reference branching ratios are rather inaccurate for such low masses. It is therefore strongly recommended to use one of the other input formats and enter the branching ratios for such a light Higgs boson directly. It can also be relevant in this mass region to consider constraints from other colliders than LEP, which are not included in \HB.

\subsection{Input using the SUSY LesHouches Accord}
\label{sec:inputdescriptionslha}
For the convenience of users interested in supersymmetric
models, an input option using the SUSY Les Houches Accord (SLHA) \cite{arXiv:0801.0045,*hep-ph/0311123} is now offered.
This option (available by setting {\tt whichinput=SLHA}) uses the calculated decay information from the SLHA file and cross sections approximated through the effective couplings approach. It requires the following input to be read from an SLHA file:
\begin{enumerate}
\item The masses for the neutral Higgs bosons $h_k \; (k=1\ldots\nHneut)$ and singly charged Higgs bosons $H_j^{\pm} \; (j=1\ldots\nHplus)$,
\BEA
M_{h_k},\, M_{H_j^{\pm}},\non
\EEA
\item the Higgs total decay widths,
\BEA
\Gamma_\TOT(h_k),\,\Gamma_\TOT(H_k^{\pm}),\non
\EEA
\item neutral Higgs branching ratios with SM equivalents
\BEA
&&\BR_\MOD(h_k\to \text{SM}),\non\\
&&\text{with SM = $\{s\bar{s}$, $c\bar{c}$, $b\bar{b}$, $\mu^+ \mu^-$,$\tau^+\tau^-$,
 $W^+W^-$, $ZZ$, $Z\gamma$, $\gamma\gamma$, $gg\}$,}\non
\EEA
Note that the decay modes have to be specified as two-body decays, irrespectively of whether they are on- or off-shell.
\item the neutral Higgs branching ratios without SM equivalents
\BEA
&&\BR_\MOD(h_k\to h_i h_j),\non \\
&&\BR_\MOD(h_k\to {\rm invisible}),\non
\EEA
\item the charged Higgs branching ratios to SM particles 
\BEA
\BR(H_j^{+} \to {\text{SM}}),\:{\rm where}\:
\text{SM = $\{c\bar{s}$, $c\bar{b}$,  $\tau^+ \nu_{\tau}\}$,}\non
\EEA
\item the top quark branching ratios
\BEA
&&\BR(t \to W^{+}b),\non \\
&&\BR(t \to H_j^{+}b),\non
\EEA
\item normalized scalar and pseudoscalar (squared) effective Higgs couplings to fermions 
\BEA
\left(\frac{g^\MOD_{s,h_k(\text{OP})}}{g^\SM_{H(\text{OP})}}\right)^2,\,
\left(\frac{g^\MOD_{p,h_k(\text{OP})}}{g^\SM_{H(\text{OP})}}\right)^2,\,
\text{ with OP = $\{b\bar{b}, t\bar{t}, \tau^+\tau^-\}$},
\non
\EEA
\item normalized (squared) effective Higgs couplings to bosons 
\BEA
\left(\frac{g^\MOD_{h_ih_jZ}}{g^{\REF}_{H'HZ}}\right)^2,\,
\left(\frac{g^\MOD_{h_k(\text{OP})}}{g^\SM_{H(\text{OP})}}\right)^2,\,
\text{ with OP = $\{W^+W^-, ZZ, gg, ggZ\}$},\non
\EEA 

\end{enumerate}
In the SLHA input the effective couplings are only used to calculate the Higgs production cross section ratios (unlike the effective coupling option, where they are also used to calculate the branching ratios). The Higgs decay branching ratios are taken directly from the corresponding decay blocks in the SLHA file. In the case of incomplete input, Higgs masses which are not specified in the SLHA file will be set equal to minus one (such that these Higgs bosons are not tested against any limits), whereas any other input that is not specified will be set equal to zero.

Table \ref{table:SLHApossibleparticleshiggs} lists the PDG codes of the particles that can be considered 
as neutral Higgs bosons by \HB. The setting of {\tt nHzero} determines how many of these are used, starting from the top of Table~\ref{table:SLHApossibleparticleshiggs}. For example,
with {\tt nHzero=3}, the properties of particles with the PDG numbers 25, 35, and 36 are read from the SLHA file. Note that no $\cp$ properties are inferred from the PDG numbers of the neutral Higgs bosons.
The invisible Higgs branching ratios are obtained from the branching ratios of
Higgs bosons into a weakly-interacting lightest supersymmetric particle (LSP).
\HB\ finds the weakly-interacting candidate with the lowest mass (considering neutralinos and sneutrinos as  candidates) and confirms that this particle is indeed the LSP by comparing its mass against
the masses of the charged leptons, the lightest chargino, and the gluino. If the LSP is not neutral, the invisible Higgs branching ratio is set to zero.
\begin{table}[h]
\centering
\renewcommand{\arraystretch}{1.0}
\footnotesize
\begin{tabular}{ll}
\br
\multicolumn{2}{c}{Neutral Higgs bosons}\\
Common notation & PDG number\\
\mr
$h$ or $h_1$ &25 \\
$H$ or $h_2$&35 \\
$A$ or $h_3$ or $A_1$&36 \\
$h_3$&45 \\
$A_2$ &46 \\
\br
\end{tabular}\\
\caption{PDG particle codes for particles that represent neutral Higgs bosons. A number {\tt nHzero} of these will be considered by \HB, starting from the top row.
} 
\label{table:SLHApossibleparticleshiggs}
\end{table}

To specify the required effective couplings, as described by points (vii) and (viii) above, \HB\ requires two extra input blocks which are not part of the normal SLHA. An example of these blocks is shown in Table~\ref{fig:newSLHAblocks}. There are some cases when \HB\ is unable to use an SLHA file, including any of the following:
\begin{itemize}
\item The Block MODSEL indicates that there is $\mathcal{R}$-parity violation,
\item either Block SPINFO or DCINFO contains an entry with the label '4' (which is used to indicate an unphysical parameter point),
\item the number of neutral Higgs bosons, {\tt nHzero}, is greater than 5, 
\item the number of charged Higgs bosons, {\tt nHplus}, is greater than 1.
\end{itemize}
The settings for {\tt nHzero} and {\tt nHplus} are given as input, either as
arguments to the subroutine {\tt initialize\_HiggsBounds} or on the command
line, and are not read from the SLHA file. If \HB\ is unable to use an SLHA file (i.e.~if one of the situations listed above applies), it might still be possible to run \HB\ with one of the other input options. It is nevertheless recommended that the user investigates and understands the reason behind the SLHA failure before proceeding.

The supersymmetric spectrum calculator \texttt{SPheno}~\cite{Porod:2003um,*Porod:2011nf} can directly write the \HB\ SLHA input blocks to its output SLHA file.\footnote{When using this option it is necessary to let {\tt SPheno} give the Higgs branching ratios for (off-shell) two-body final states instead of three-body decays. This can be done by changing the \texttt{SPhenoInput} block entry \texttt{13} to a non-zero value.}  For \FH\ \cite{hep-ph/9812320,*hep-ph/9812472,*hep-ph/0611326, *Hahn:2009zz,hep-ph/0212020} we provide a stand-alone program, \texttt{HBSLHAinputblocksfromFH}, which creates the necessary SLHA blocks from the \FH\ output.
\begin{table}[h]
\renewcommand{\arraystretch}{1.0}
{\tt\scriptsize
\begin{tabular}{lllllll}
\br
\multicolumn{7}{l}{Block HiggsBoundsInputHiggsCouplingsFermions}\\
\# ScalarNormEffCoupSq & PseudoSNormEffCoupSq &    NP &    IP1 &    IP2 &    IP3 & \\
     1.0000001E+00		&        1.0000101E+00	   &   3  	&   25 &     5 &     5 &   \# h0-b-b eff. coupling\textasciicircum2\\
     1.0000002E+00  		&      1.0000102E+00        	  & 3   &     35     & 5 &     5&   \# HH-b-b eff. coupling\textasciicircum2\\
     1.0000003E+00 		&       1.0000103E+00 	 &      3 &    36    &  5&      5 &  \# A0-b-b eff. coupling\textasciicircum2\\
\multicolumn{7}{l}{\#} \\
     1.0000004E+00 		&        1.0000104E+00 	&       3  &   25 &     6   &   6&   \# h0-top-top eff. coupling\textasciicircum2\\
     1.0000005E+00       	&	 1.0000105E+00     	&   3   &  35    &  6    &  6  & \# HH-top-top eff. coupling\textasciicircum2\\
     1.0000006E+00  		&      1.0000106E+00      	&  3	&     36    &  6 &     6   &\# A0-top-top eff. coupling\textasciicircum2\\
\multicolumn{7}{l}{\#} \\
     1.0000007E+00 		&       1.0000107E+00	&        3  &   25  &   15 &    15  & \# h0-tau-tau eff. coupling\textasciicircum2\\
     1.0000008E+00      	&	  1.0000108E+00   	&     3    & 35   &  15  &   15  & \# HH-tau-tau eff. coupling\textasciicircum2\\
     1.0000009E+00 		&       1.0000109E+00    	&    3    & 36  &   15   &  15 &  \# A0-tau-tau eff. coupling\textasciicircum2\\
\multicolumn{7}{l}{\#}
\end{tabular}
\begin{tabular}{lllllll}
\multicolumn{7}{l}{Block HiggsBoundsInputHiggsCouplingsBosons}\\
    1.0000010E+00	&     3	&     25	&     24	&     24	& &   \# h0-W-W effective coupling\textasciicircum2\\
    1.0000011E+00	&     3	&     35	&     24	&     24	& &   \# HH-W-W effective coupling\textasciicircum2\\
    1.0000012E+00	&     3	&     36	&     24	&     24	& &   \# A0-W-W effective coupling\textasciicircum2\\
\multicolumn{7}{l}{\#}\\
     1.0000013E+00	&     3	&     25	&     23	&     23	& &   \# h0-Z-Z effective coupling\textasciicircum2\\
     1.0000014E+00	&     3	&     35	&     23	&     23	&  &  \# HH-Z-Z effective coupling\textasciicircum2\\
     1.0000015E+00	&     3	&     36	&     23	&     23	&  &  \# A0-Z-Z effective coupling\textasciicircum2\\
\multicolumn{7}{l}{\#}\\
     1.0000016E+00 	&    3		&     25	&     21	&     21	& &   \# h0-gluon-gluon effective coupling\textasciicircum2\\
     1.0000017E+00	&     3	&     35	&     21	&     21	& &   \# HH-gluon-gluon effective coupling\textasciicircum2\\
     1.0000018E+00	&     3	&     36	&     21	&     21	& &   \# A0-gluon-gluon effective coupling\textasciicircum2\\
\multicolumn{7}{l}{\#}\\
     1.0000019E+00	&     3	&     25	&     25	&     23	& &   \# h0-h0-Z effective coupling\textasciicircum2\\
     1.0000020E+00	&     3	&     25 	&    35	&     23	& &   \# h0-HH-Z effective coupling\textasciicircum2\\
     1.0000021E+00	&     3	&     25	&     36	&     23	& &   \# h0-A0-Z effective coupling\textasciicircum2\\
     1.0000022E+00	&     3	&     35	&     35	&     23	& &   \# HH-HH-Z effective coupling\textasciicircum2\\
     1.0000023E+00	&     3	&     35	&     36	&     23	& &   \# HH-A0-Z effective coupling\textasciicircum2\\
     1.0000024E+00	&     3	&     36	&     36	&     23	& &   \# A0-A0-Z effective coupling\textasciicircum2\\
\multicolumn{7}{l}{\#}\\
     1.0000025E+00 	&    4		&     25	&     21	&     21	&    23	&   \# h0-gluon-gluon-Z effective coupling \textasciicircum2\\
     1.0000026E+00	&     4	&     35	&     21	&     21	&    23	&   \# HH-gluon-gluon-Z effective coupling \textasciicircum2\\
     1.0000027E+00	&     4	&     36	&     21	&     21	&    23	&   \# A0-gluon-gluon-Z effective coupling \textasciicircum2\\
\br
\end{tabular}
}
\caption{Examples of the two new SLHA blocks which are required by \HB\ when using the SLHA input option.} 
\label{fig:newSLHAblocks}
\end{table}

\newpage
\section{New developments in \HBv{4}}
\label{Sec:New}

\subsection{Applying exclusion limits to arbitrary Higgs models}
\label{Sec:SMtest}
The aim of \HB\ is to apply limits derived in Higgs collider searches to models which have \textit{not} been directly investigated by the experimental analyses. These models can be \textit{arbitrary} in the sense that they may contain any number of neutral or (singly) charged Higgs bosons,\footnote{The total number of neutral or charged Higgs bosons that can be handled by the code is currently limited to $\le 9$ for practical (formatting) reasons.} or particles which behave like (elementary) Higgs bosons in Higgs collider searches. Examples of the latter include theories with composite Higgs bosons~\cite{Espinosa:2010vn} or dilatons~\cite{Bai:2009ms}. More specifically, the requirements on the theory in order for the results of \HB\ to be reliable are the following:

\begin{enumerate}
\item The narrow width approximation must be applicable, such that the predictions for each process can be factorized into Higgs production and decay.
\item The investigated model should not change the signature of the background processes considerably. Usually, new physics models which show strong deviations from the SM in the background processes of Higgs searches are not considered in the literature, since this would often put them in conflict with SM electroweak precision data~\cite{hep-ex/0509008,hep-ex/0612034}. Hence, they would most likely not be interesting for \HB\ anymore. The presence of such backgrounds would rather correspond to an opportunity for the discovery of physics beyond the SM in other areas.\label{assumption2}
\item The investigated model should not significantly change the kinematical distributions of the signal topology $X$ (e.g.~the $\eta$ and $p_T$ distributions of the final state particles) from that assumed in the corresponding analysis. For a more detailed discussion of this requirement, see \citere{arXiv:0811.4169,*arXiv:1102.1898}.\label{assumption3}
\end{enumerate}

The above requirements are typically sufficient to ensure the applicability of \textit{model-indepen\-dent} exclusion limits, \ie limits on a cross section of a certain Higgs signal topology, composed of \textit{one} production and \textit{one} decay process. If further model assumptions have been made in the experimental analysis, for instance on the $\cp$-properties of the Higgs boson or on the top quark branching ratios, \HB\ checks whether the investigated model fulfills them before applying the analysis.

The application of exclusion limits to arbitrary Higgs models becomes less trivial if the experimental analysis combines several Higgs signal topologies under the assumption of a specific model. This is the case for most of the Tevatron and LHC Higgs searches, where a SM Higgs boson is assumed. The exclusion limit is then set on a common signal scale factor for all considered SM Higgs topologies (also called \textit{signal strength modifier}), usually denoted by $\mu$ (sometimes also $\sigma/\sigma_\SM$ is used). For an analysis considering $i=1,\ldots, N$ signal topologies (each consisting of a production mode $P$, and a final state $F$), the prediction for this quantity can be computed for Higgs boson $h_k$ of the investigated model as
\begin{equation}
\mu = \frac{\sum_{i=1}^{N} \epsilon_i [\sigma_\mathrm{model} (P(h_k)) \times \BR_\mathrm{model}(h_k\to F)]_i}{\sum_{i=1}^{N} \epsilon_j [\sigma_\mathrm{SM} (P(H)) \times \BR_\mathrm{SM}(H\to F)]_j}.
\label{Eq:mu}
\end{equation}
The \textit{channel efficiencies}, $\epsilon_i$, are assumed to be the same for the model and the SM (see requirements (\ref{assumption2}) and (\ref{assumption3}) above). If these channel efficiencies were published together with the exclusion limit posed by an experimental analysis, the signal strength modifier $\mu$ could be computed for a given model without further assumptions. However, these efficiencies have so far been made publicly available only in a very few cases.\footnote{These efficiencies usually depend on the tested Higgs boson mass. Using a single number for $\epsilon_i$ therefore might appear to be a crude approximation. Nevertheless, for many searches, having access to this information even for one or a few values of the Higgs mass would already provide a better approximation of the full result than in the current situation.} In \HB\ we therefore neglect the channel efficiencies in Eq.~\eqref{Eq:mu}, leading to an unavoidable model-dependence of the resulting limit, since the calculation of $\mu$ via Eq.~\eqref{Eq:mu} with all $\epsilon_i \equiv 1$, is strictly speaking only valid if the model predictions for all signal topologies of the analysis contribute to the total signal rate in (approximately) the same proportions as in the SM.

In order to ensure that an analysis is applied only when this last requirement is fulfilled by the
model, \HB\ performs a SM-likeness test for every Higgs analysis performed under
SM assumptions. A test of this kind has been present in all versions of \HB\ \cite{arXiv:0811.4169,*arXiv:1102.1898}. However, this test was significantly improved in \HBv{3.8.0} onwards, as described in Ref.~\cite{Bechtle:2013gu}, and it is this improved version which we describe here.
Neglecting the channel efficiencies, the predicted signal strength modifier $\mu$ given in Eq.~\eqref{Eq:mu} can be obtained as $\mu \approx \sum_{i=1}^N \omega_i c_i$, where
\begin{equation}
c_i = \frac{[\sigma_\mathrm{model}(P(h_k)) \BR_\mathrm{model}(h_k\to F)]_i}{[\sigma_\mathrm{SM}(P(H)) \BR_\mathrm{SM}(H\to F)]_i}
\end{equation}
and
\begin{equation}
\omega_i = \frac{[\sigma_\mathrm{SM}(P(H)) \BR_\mathrm{SM}(H\to F)]_i}{\sum_{j=1}^N [\sigma_\mathrm{SM}(P(H)) \BR_\mathrm{SM}(H\to F)]_j}
\end{equation}
are the (SM normalized) channel signal strengths and the SM channel weights, respectively. The requirement that the signal topologies contribute in similar proportions to the total signal rate as in the SM is fulfilled if all channel signal strengths $c_i$ are similar to the total signal strength modifier $\mu$ (and thus similar to each other). A possible SM-likeness criterion would therefore be to require
\begin{equation}
\Delta \equiv \max_{i}~\left|\frac{\delta c_i}{\mu} \right| < \zeta
\label{Eq:SMtest1}
\end{equation}
with $\delta c_i = c_i - \mu$ and $\zeta \sim \mathcal{O}(\mathrm{few}~\%)$, \ie that the maximal relative deviation of the channel signal strength modifiers from the total signal strength modifier is less than a few percent. In fact, this criterion is very similar to what was used in earlier versions of \HB. However, this choice was found to be too restrictive in some cases, since it
may reject an analysis application which is actually justifiable, leading to overly conservative results. In particular, it is reasonable that channels contributing only a few percent to the total signal rate should be allowed to deviate more from their SM expectations, since their influence on $\mu$ is subdominant. We therefore introduce the SM channel weights $\omega_i$ in an improved SM-likeness test criterion,
\begin{equation}
\Delta \equiv \max_{i}~\omega_i \left|\frac{\delta c_i}{\mu} \right| < \zeta.
\label{Eq:SMtest}
\end{equation}
The default setting in \HB\ is $\zeta = 2\%$. This is a conservative choice, considering that the uncertainties on the rate predictions for individual channels (even in the SM) are generally larger. With the improved SM-likeness test, the maximal \textit{weighted} deviation of an individual signal strength modifier from the total signal strength modifier is required to be less than $2\%$. Models fulfilling this SM-likeness test for a SM analysis can be safely tested against its exclusion limit.

\begin{figure}
\centering
\subfigure[]{\scalebox{0.8}{\includegraphics{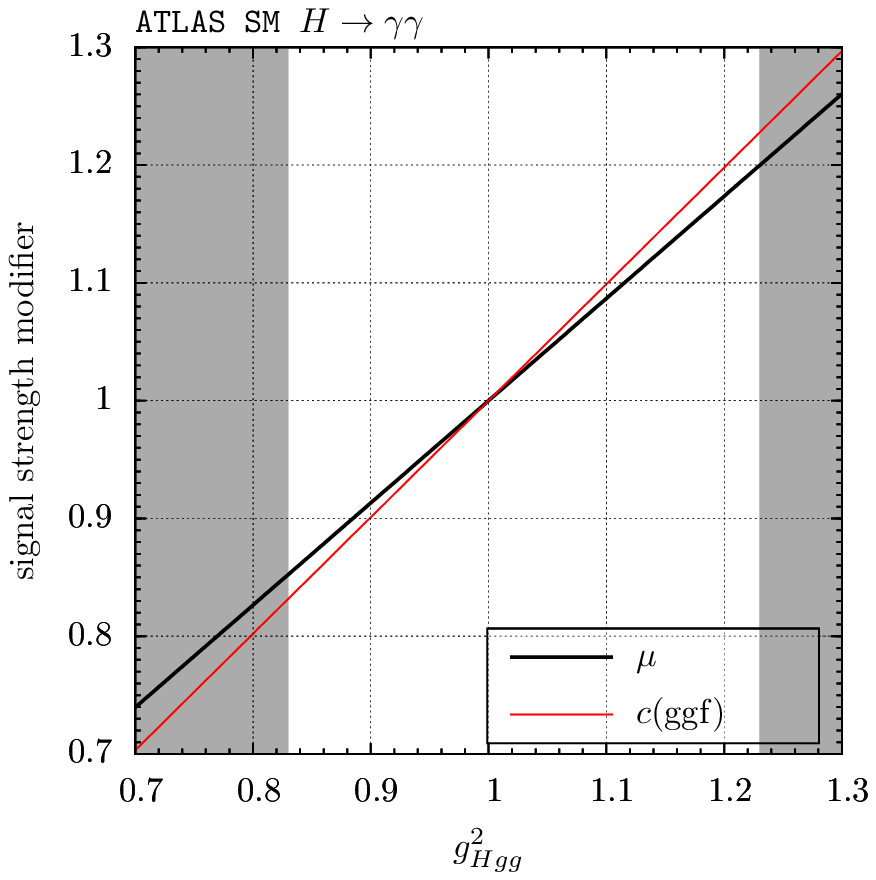}}}\hfill
\subfigure[]{\scalebox{0.8}{\includegraphics{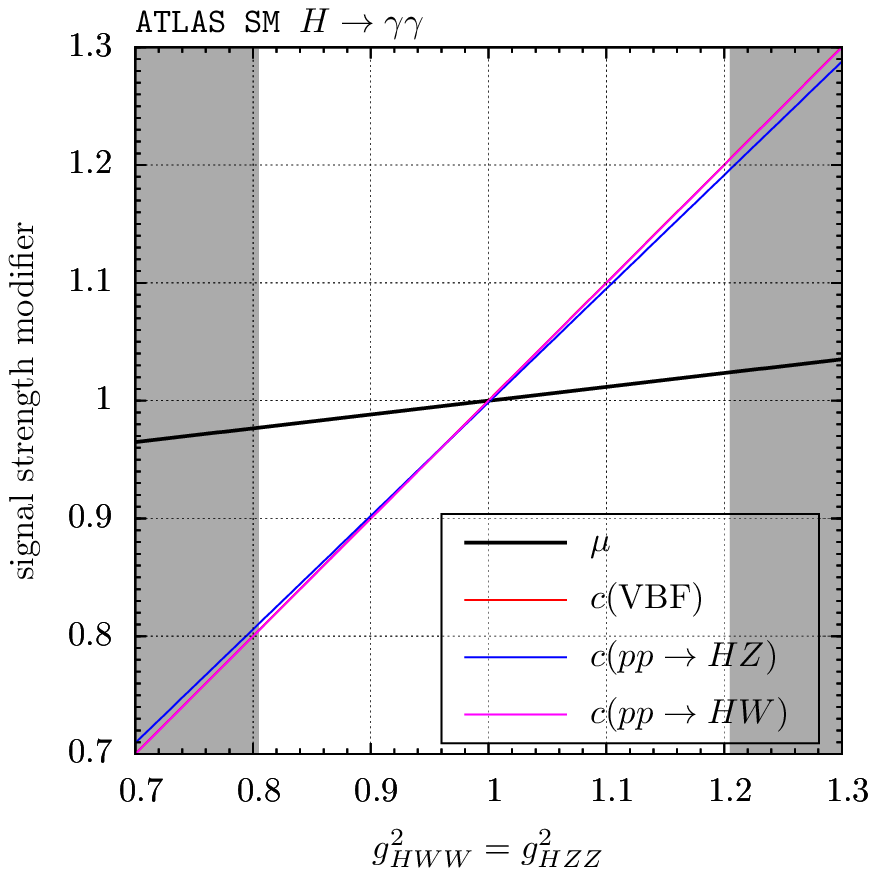}}}
\caption{Performance of the SM-likeness test. Total signal strength modifier $\mu$ and the relevant individual signal strength modifiers $c_i$ for the ATLAS $H\to\gamma\gamma$ search~\cite{1202.1414} with modified effective Higgs couplings (relative to the SM) $g^2_{Hgg}$ (left) and $g_{HVV}^2$ ($V=W,Z$) (right) for a Higgs boson with mass $m=125\gev$. The gray regions indicate the parameters for which the SM-likeness test fails.}
\label{Fig:SMtest}
\end{figure}

To illustrate the inclusion of the SM weights $\omega_i$ in the SM-likeness test criterion, we consider as an example the ATLAS $H\to \gamma\gamma$ search~\cite{1202.1414} and test a model with a single Higgs boson with mass $m=125\gev$. We depart from the SM by modifying either the squared effective Higgs coupling ratio to gluons (normalized to the SM), $g_{Hgg}^2$, or the coupling to vector bosons, $g_{HVV}^2$ ($V=W,Z$). All other effective Higgs couplings, in particular the $H\gamma\gamma$ coupling, are set to their SM values. At $m=125\gev$, the SM weights for the LHC at $\sqrt{s}=7\tev$ are
\begin{equation}
\omega \approx (87.7\%,~6.8\%,~3.2\%,~1.8\%,~0.5\%)
\end{equation}
for the production processes $(gg\to H,~\mathrm{VBF},~HW,~HZ,~Ht\bar{t})$.
In Fig.~\ref{Fig:SMtest} we show how the total signal strength modifier $\mu$ and the $c_i$ for the signal topologies are influenced by the modified effective Higgs couplings. Varying $g^2_{Hgg}$ influences only the $gg\to H$ (ggf) cross section. However, due to its large SM weight, $\omega_\mathrm{ggf} \approx 87.7\%$, the total signal strength modifier $\mu$ follows  $c (\mathrm{ggf})$ closely. The failure of the SM-likeness test at $g^2_{Hgg} = 0.835$ and $1.225$ is therefore eventually caused by the ggf signal topology, although the deviation $\delta c_i$ for the remaining signal topologies VBF, $HW$, $HZ$ and $Ht\bar{t}$ is much larger here. However, the SM weights of these channels are much smaller. The same effects can be seen when varying $g_{HVV}^2$ ($V=W,Z$). Now, the $c_i$ of the VBF, $HW$, $HZ$ signal topologies are affected by the modified effective coupling, but the total signal strength modifier $\mu$ is only slightly influenced due the small weight of these channels. Again, the deviation between $\mu$ and $c(\mathrm{ggf})$ eventually causes the SM-likeness test to fail. Due to the inclusion of the SM weights in Eq.~(\ref{Eq:SMtest}), subdominant signal topologies are allowed to deviate further from $\mu$.

\begin{figure}
\centering
\includegraphics[width=0.45\textwidth]{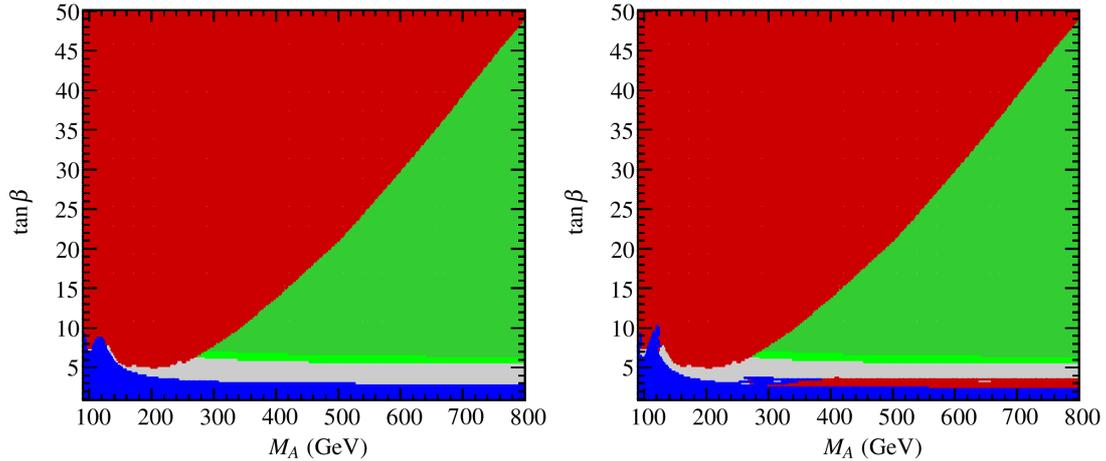}
\includegraphics[width=0.45\textwidth]{HB_mhmodp_dm3_mA_tanb}
\vspace{-0.5em}
\caption{Combined \HB\ exclusion in the the \mhmod\ benchmark scenario of the MSSM using a SM-likeness test without weights, Eq.~\eqref{Eq:SMtest1} (left), and using the new SM-likeness test with weights included, Eq.~\eqref{Eq:SMtest} (right). The colour coding is the same as in Fig.~\ref{fig:fullclassic}.}
\label{Fig:mhmax_SMtest}
\end{figure}

In comparison with the old SM-likeness test (which was used in \HB\ up to version \texttt{3.7.0}), the new criterion leads to a wider applicability of SM Higgs search results to other Higgs sectors, and thus to a significant improvement of the performance of \HB. This is shown in Fig.~\ref{Fig:mhmax_SMtest} for the \mhmod\ benchmark scenario of the MSSM~\cite{Carena:2013qia}. Without SM weights (left panel), the LHC exclusion approximately follows the results from the dedicated MSSM search for $H/A\to \tau\tau$ \cite{12050}, and no additional exclusion can be set. In particular there is \emph{no} LHC exclusion from the SM-like Higgs boson, $h$. With the full weighted criterion active (the default setting in \HBv{4}), the lightest MSSM Higgs boson can become sufficiently SM-like at large $\MA$ and small $\tan\beta$ for the combined SM Higgs searches of ATLAS and CMS to be applied, giving
additional areas of exclusion (right panel). Exclusion by individual Higgs bosons for the same scenario
can be seen in Fig.~\ref{fig:individualH}, which has also been produced using the weighted SM-likeness criterion.

\subsection{Including Higgs mass uncertainties}
\label{Sec:MassUnc}
In several theories with extended Higgs sectors, the Higgs boson masses are not free
parameters but can be predicted as a function of the other model parameters up to a certain theoretical accuracy.
This is the case, for example, in the MSSM where out of the five physical Higgs states typically only one mass, $M_A$ or $\MHp$, is used as an (on-shell) input parameter. The remaining Higgs masses then become predictions of the model, with a theoretical uncertainty that varies within the parameter space and with the sophistication of the theoretical prediction.

We have extended \HB\ to be able to take this type of theoretical uncertainty into account when evaluating the Higgs exclusion limits. For theories that have no Higgs mass uncertainties, or where they are negligibly small, this new feature can be left deactivated. In \HBv{4}, the Higgs mass uncertainties are taken into account approximately by varying each mass within a user-defined interval.\footnote{Changes to the relative rates induced by the Higgs mass variation is considered to be ``second-order'', and are therefore neglected in this procedure. This approximation is valid when the rate predictions vary slowly within the mass uncertainty interval, which sets an upper limit on the reasonable size of the mass uncertainties.} If the tested Higgs boson is unexcluded by the probed limit (in the normal \HB\ sense) for any mass in this interval, the tested parameter point of the model is regarded as being unexcluded. This leads to an overall conservative (weaker) result for the exclusion limit when uncertainties are included.

Technically, the number of mass points $N$ considered in the variation can be specified by a variable. The default setting is $N=3$ (this corresponds to testing the central mass value, $M_H$, and the two endpoints, $M_H\pm \Delta M_H$, of the specified uncertainty interval). When a sensitive limit varies rapidly with $M_H$, it is advisable to increase $N$ for best results. The mass variation is performed for each Higgs boson independently. In the \emph{classic} \HB\ method this variation is also simultaneous, which leads to a multi-dimensional computation grid with a complexity growing as $\mathcal{O}(N^{n_H})$, where $n_H$ is the number of Higgs bosons with a non-zero mass uncertainty.\footnote{To avoid unnecessary calculations, uncertainties smaller than the minimal mass spacing at which the experimental results are available (currently $0.1$ GeV for some channels) are not considered.} For the \emph{full} method, since the limit from each Higgs boson is already considered independently of the others, the complexity remains managable, i.e. $\mathcal{O}(n_H N )$. Nevertheless, the user is strongly encouraged to specify uncertainties only for those Higgs bosons where this is numerically relevant.

\begin{figure}
\centering
 \resizebox{0.5\columnwidth}{!}{\input{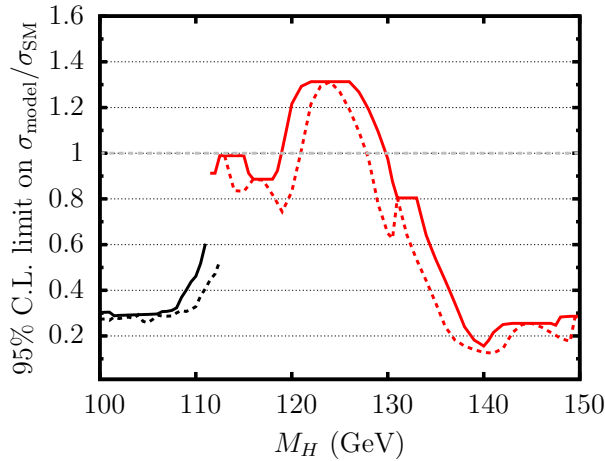}}
 \vspace{-0.5em}
\caption{Upper limits from \HB\ (at $95\%$ C.L.) on the relative signal strength versus the Higgs boson mass in the SM, which has zero theoretical uncertainty (dashed lines), and in a model with a SM-like Higgs boson with a theoretical mass uncertainty of $2$~GeV (solid lines). The two colours indicate mass ranges where the most sensitive limit is from either LEP (black) or the LHC (red).}
\label{fig:deltam_sm}
\end{figure}
The effects of a theoretical mass uncertainty on the resulting \HB\ limits are demonstrated in Fig.~\ref{fig:deltam_sm}, which shows the combined exclusion for a SM-like Higgs boson with $\Delta M_H=0$~GeV (solid lines), and similarly for a Higgs boson with SM-like couplings but a theory mass uncertainty of $\Delta M_H=2$~GeV (dashed lines). In this figure, the mass range excluded at $95\%$ C.L.~corresponds to where the limit on $\sigma_{\mathrm{model}}/\sigma_{\mathrm{SM}}<1$. Including the mass uncertainties can be seen here as a broadening of the allowed range for the Higgs mass prediction in the model by $\pm 2$ GeV around the signal region. It can also be seen that for a given mass point the resulting upper limit on the signal rate is always weaker or equal to the upper limit obtained without theoretical mass uncertainty. Including a theory mass uncertainty in \HB\ therefore produces overall more conservative limits, which is as expected.

\begin{figure}
\centering
\includegraphics[width=0.32\columnwidth]{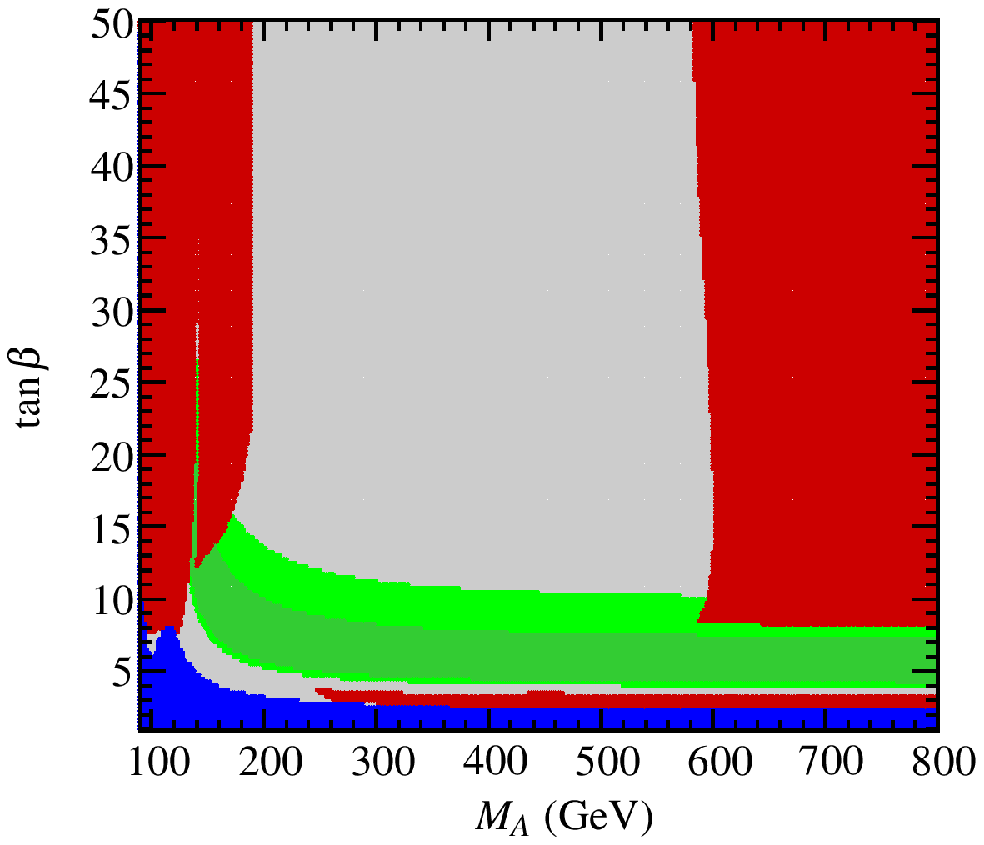}
\includegraphics[width=0.32\columnwidth]{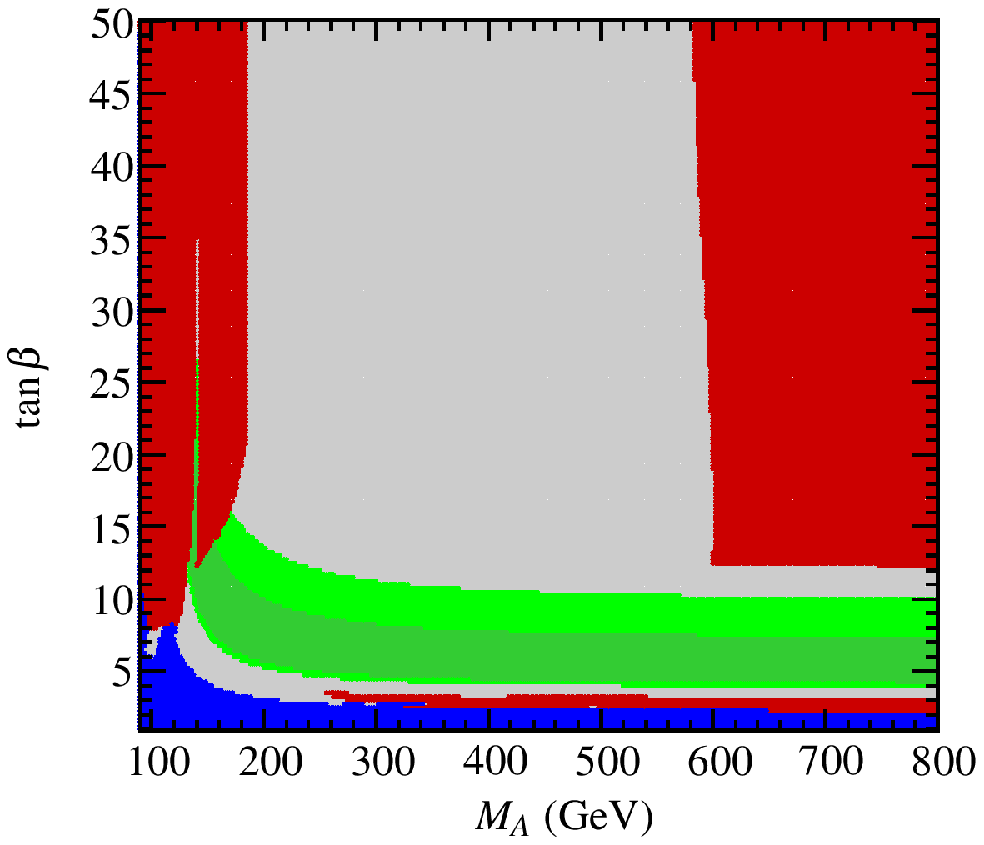}
\includegraphics[width=0.32\columnwidth]{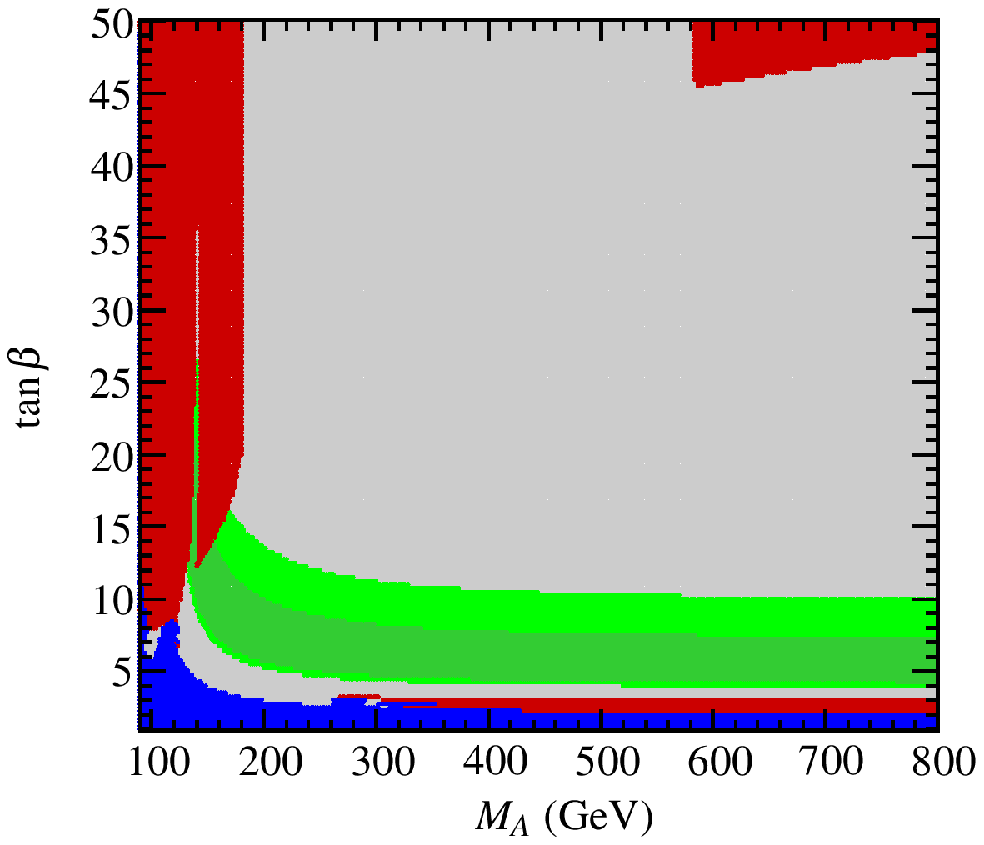}
\vspace{-0.5em}
\caption{Contribution from the lightest MSSM Higgs boson, $h$, to the \emph{full} \HB\ exclusion in the parameter space for the \mhmax\ benchmark scenario \cite{Carena:2013qia}. The results are shown for a theory mass uncertainty of $\Delta M_h=0$~GeV (left), $\Delta M_h=1$~GeV (center), and $\Delta M_h=2$~GeV (right). The colour coding is the same as used in Fig.~\ref{fig:fullclassic}.}
\label{fig:deltam_mssm}
\end{figure}

This point is further illustrated in Fig.~\ref{fig:deltam_mssm}, which shows the resulting limits from the light (SM-like) MSSM Higgs boson, $h$, when running \HB\ \HBfull\ for the \mhmax\ benchmark scenario \cite{Carena:2013qia}. Similar to above, the green band shows the region of parameter space (in this scenario) where $M_h=125.7\pm 2(3)\gev$. For large values of $\MA$ and $\tanb$, the \mhmax\ scenario gives rise to values of $\Mh$ that are \emph{too high} compared to the measured LHC signal. The predicted value for $\Mh$ increases with $\tan\beta$. $\Mh \gtrsim 128\gev$ is excluded when no theory uncertainty is applied (cf.~Fig.~\ref{fig:deltam_sm}).
 The three panels in Fig.~\ref{fig:deltam_mssm} show, from left to right, the results when using a mass uncertainty (resulting from the calculation of $M_h$ in the model \cite{hep-ph/0212020}) of $\Delta M_h=0$~GeV, $\Delta M_h=1$~GeV, and $\Delta M_h=2$~GeV. It can be seen that the exclusion at high $\MA$ from the limit on the lightest Higgs boson goes down to lower $\tanb$ values when $\Delta M_h$ is small. This illustrates the importance of taking Higgs mass uncertainties into account when interpreting exclusion limits (and compatibility with observed signals, see \cite{Bechtle:2013xfa}) in the MSSM and other scenarios for physics beyond the SM. 

\subsection{LEP $\chi^2$ extension}
\label{Sec:LEPchi2}
An unfortunate limitation of both the model-independent limits
implemented in \HB, as well as the model-dependent
search limits presented by the experiments, is that they are available
only at one fixed confidence interval, which is $95\%$~C.L. for all
searches implemented here. The result of an exclusion
provided by \HB\ based on these searches therefore has a
confidence limit of \emph{at least} $95\%$~C.L., and in many cases higher. 
However, the exact level of confidence at which a signal with the properties given to
\HB\ is either excluded or allowed, is generally
unknown.
\begin{figure}
  \begin{center}
    \subfigure[]{\includegraphics[width=0.47\textwidth,clip=]{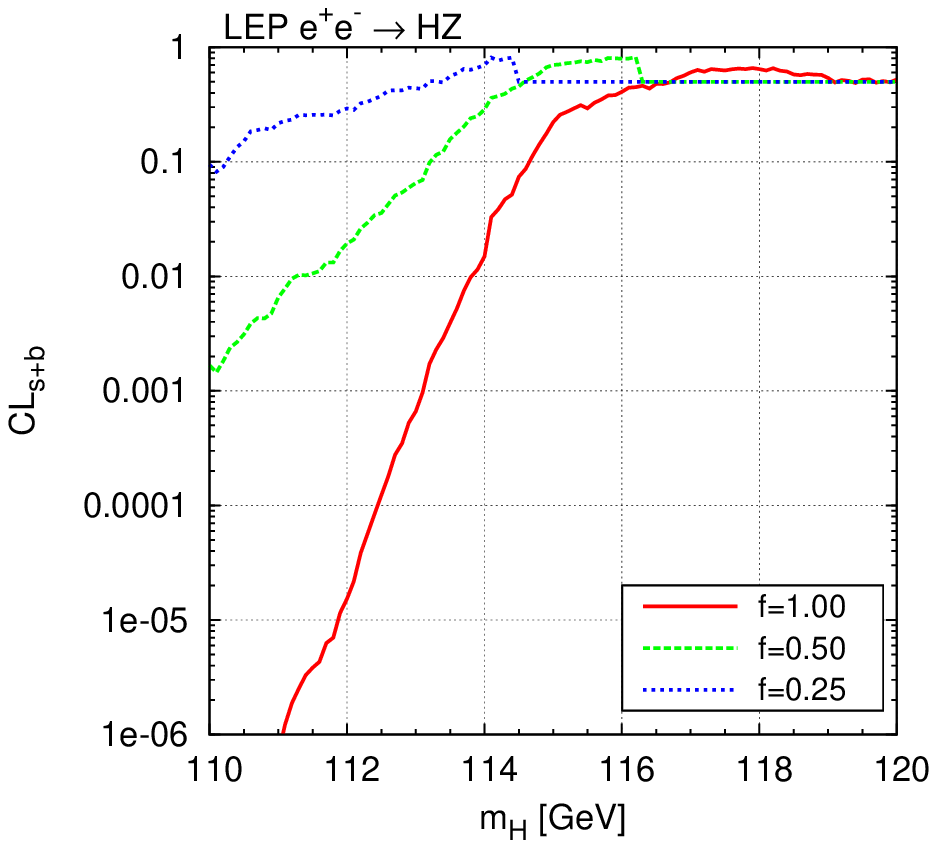}\label{subfig:LEPchi2Ext:CLhzsm}}
    \subfigure[]{\includegraphics[width=0.47\textwidth,clip=]{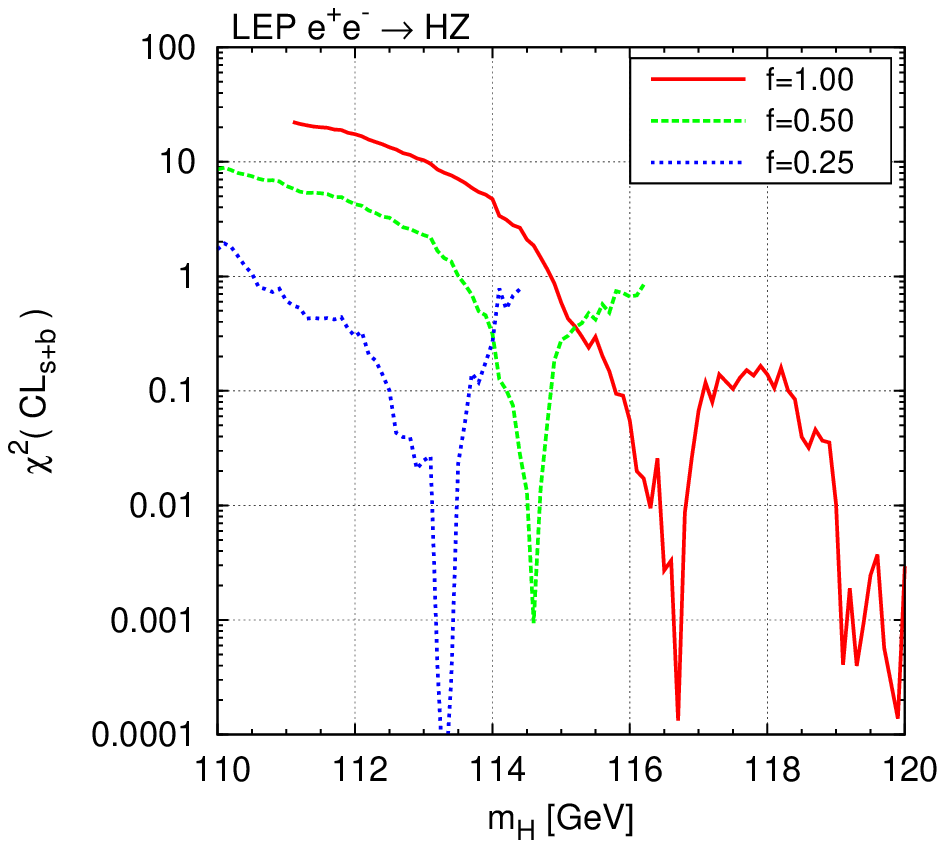}\label{subfig:LEPchi2Ext:chi2hzsm}}
    \subfigure[]{\includegraphics[width=0.47\textwidth,clip=]{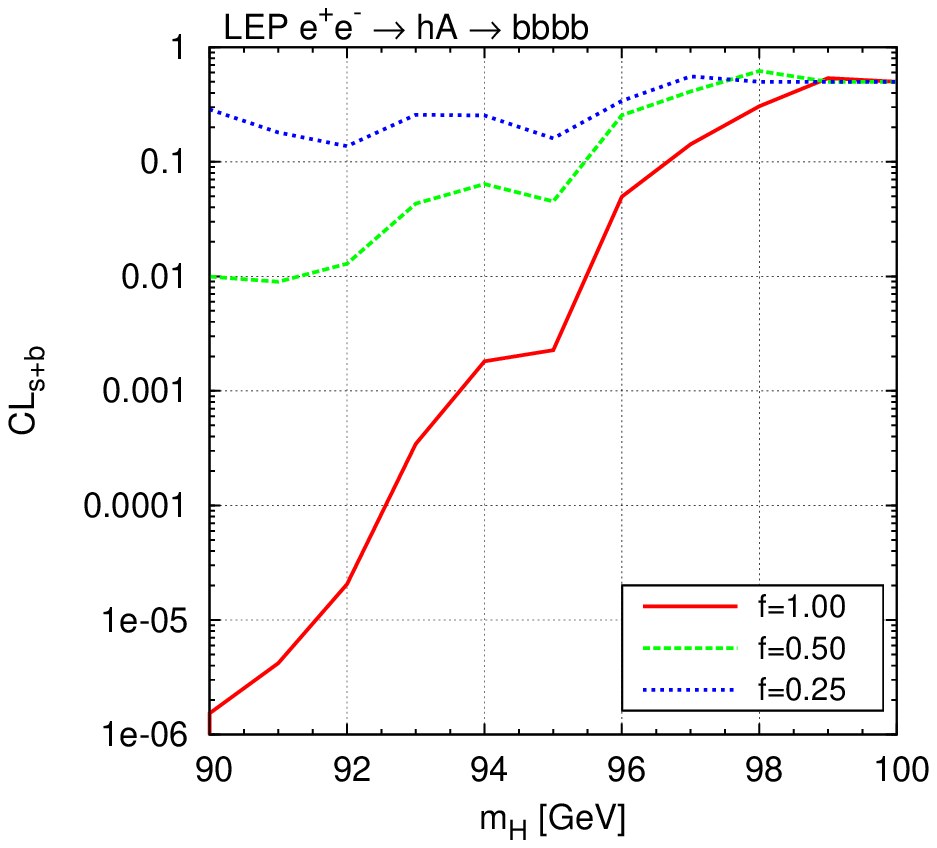}\label{subfig:LEPchi2Ext:CLhhbbbb}}
    \subfigure[]{\includegraphics[width=0.47\textwidth,clip=]{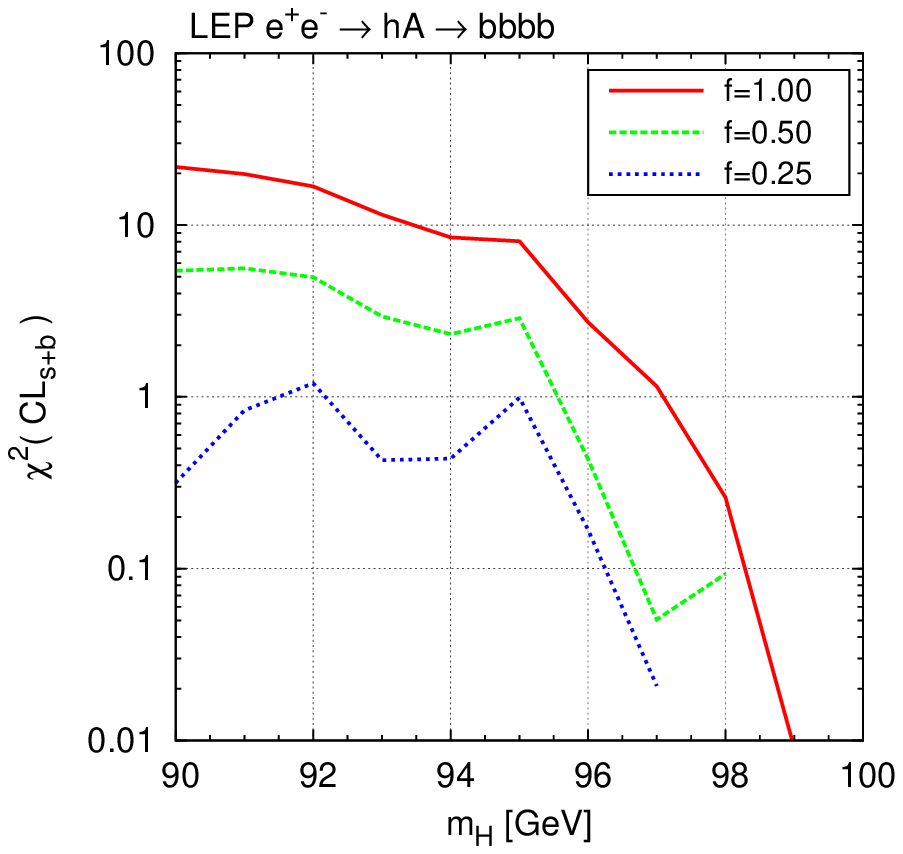}\label{subfig:LEPchi2Ext:chi2hhbbbb}}
    \vspace{-1em}
  \end{center}
  \caption{Examples for transferring the LEP $CL$ into a value for $\chi^2_{H}$ using three different values of the scale factor $f=(0.25, 0.5, 1.0)$.
    The upper row shows (a) the LEP $CL_{s+b}$ result for $e^+e^-\to hZ$ in the SM, and (b) the corresponding $\chi^2_H$ (note the logarithmic scale). In the lower row, similar results are shown for $e^+e^-\to hA\to
    b\bar{b}b\bar{b}$, with (c) again being the $CL_{s+b}$ values and (d) the result for $\chi^2_{H}$.}
  \label{fig:LEPchi2Ext}
\end{figure}

This has unfortunate consequences for the use of these limits in
applications like global fits (see e.g.~Ref.~\cite{Bechtle:2012zk,*Buchmueller:2012hv,*Bechtle:2012jw} for examples of such fits in the MSSM). There, a model point
where the predicted Higgs signal is excluded for example at $96\%$~C.L.,
i.e.~with a significance of slightly more than $2\,\sigma$, might
still be a very good fit if the other properties of the model point
in the global fit match the data well. However, the conventional \HB\ output only contains information about whether the parameter point is experimentally excluded at \emph{at least} $95\%$~C.L. and thus can only be treated as a ``hard cut'' on the validity of a parameter point.

In order to circumvent this problem, at least for the LEP Higgs
searches, the full information on $CL_{s+b}$ and $CL_s$ for all Higgs
mass combinations in the model-independent LEP searches
from~\cite{hep-ex/0602042} have been re-calculated for varying cross
sections~\cite{PIK}. These can be written as $\sigma_i=f_i\, \sigma_{i, \mathrm{ref}}$,
 where $\sigma_{i, \mathrm{ref}}$ is the reference cross section times branching fraction for search $i$, motivated by the SM
Higgs boson or the corresponding cross section for non-SM-Higgs bosons
(see~\cite{Schael:2006cr} for details), and $f_i$ is an arbitrary scaling parameter. A
logarithmic grid in the scaling parameters $f_i$ with $100$ points between
$10^{-3}$ and $1$ is used. Using an interpolation, the actual $CL$ can be
calculated for every Higgs production mode at LEP for every physically
allowed cross section. 

This $CL$ can then be transferred into a quantity whose properties
closely follow that of a $\chi^2$ function. This is achieved by assuming that
the distribution of $-2\ln
Q$~\cite{Schael:2006cr} is Gaussian in the asymptotic limit. Transferring the one-sided
$CL$ into the two-sided calculation of a $\chi^2$, the following formula
can then be used 
$$\chi^2_{H} = 2\,\mathrm{InvErf}^{2}(1-2CL_{s+b}).$$
The resulting $\chi^2_{H}$ can be used as a continuous expression of the
agreement between the result of the LEP Higgs boson searches and the
model predictions. Note that, in the case of a strong excess in one of
the searches, $\chi^2_{H}$ is not only large for models 
whose predicted cross section times branching fraction is above the
observed limit, but also for predictions much smaller than the
observed rate in data.

In cases when the predicted cross section is lower than the minimal (rescaled) value available in the table, the corresponding $\chi^2$ value is set to zero. When the predicted cross section exceeds the tabulated values, no reliable  $\chi^2$ value can be calculated, and the value $\chi^2=-999$ is returned to indicate that a problem has occured. This default behavior can be changed (by setting a flag in {\tt usefulbits.f90}) to use instead the $\chi^2$ value for the maximal (rescaled) cross section available for that combination of Higgs masses.

An example of the relation between the LEP $CL_{s+b}$ and
$\chi^2_{H}$, also for different values of $f$, is given in
Fig.~\ref{fig:LEPchi2Ext}. It
    can be seen that for $CL_{s+b}\approx 0.5$, indicating very good
    agreement of the signal plus background prediction with the data,
    fluctuations of $\chi^2_H$ around 0 are unavoidable, but
    numerically irrelevant. In addition, the possibility exists to
follow a prescription from~\cite{Ellis:2007fu} to include a mass
uncertainty into $\chi^2_H$ by folding the full $\chi^2$ distribution
with a gaussian $G_{\Delta M_H}$ with a mass uncertainty $\Delta M_H$
given by the user,\footnote{Note that this mass uncertainty can be specified completely independently from the uncertainties discussed in Sect.~\ref{Sec:MassUnc}.} instead of evaluating $\chi^2_H$ just at the given
$M_H$:
$$\chi^2_{H,\rm{bare}}(M_H,\Delta M_H)=-2\ln\left(\int_{-\infty}^{+\infty}e^{-\frac{1}{2}\chi^2_H(M')}\, G_{\Delta M_H}(M_H-M')\mathrm{d}M'\right).$$
Since the folding introduces small, but non-zero values
$\chi^2_{H,\mathrm{bare}}(M_H,\Delta M_H)$ for $M_H>116.4$~GeV, where
no sensitivity is expected for the SM-like Higgs search channels, the
final $\chi^2_H(M_H,\Delta M_H)$ is obtained by subtracting
$\chi^2_H(116.4\,\mathrm{GeV},\Delta M_H)$ from
$\chi^2_{H,\mathrm{bare}}(M_H,\Delta M_H)$ for $M_H\leq 116.4$~GeV, and by setting $\chi^2_H(M_H,\Delta
M_H)=0$ above. A similar procedure is adapted for non-SM like
searches, where the point of vanishing sensitivity is determined for
each search prior to the folding.
    
\begin{figure}[t]
  \centering
  \includegraphics[width=0.45\textwidth]{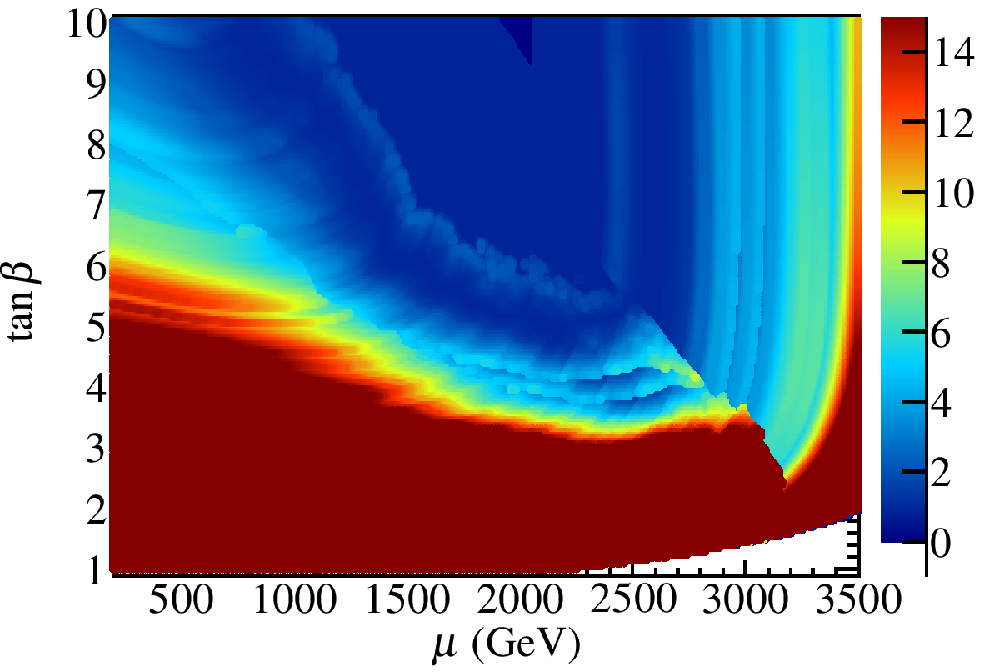}
  \includegraphics[width=0.45\textwidth]{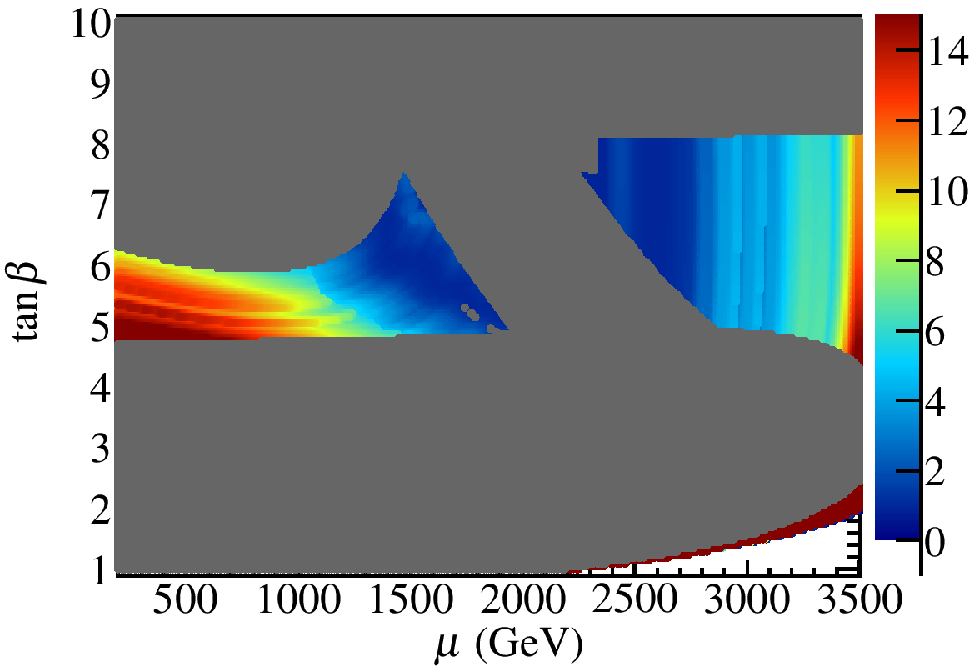}
  \caption{\HB\ results for the LEP $\chi^2$ (colours) in the low-$M_H$ scenario of the MSSM \cite{Carena:2013qia}. The LEP $\chi^2$ information is shown both on its own (left), and with with the combined LHC exclusion bounds in gray (right). The latest limit from ATLAS charged Higgs searches \cite{ATLAS2013090} is not applied here. These results, which are included from \HBv{4.1}, lead to exclusion over the whole parameter space of this benchmark scenario.}
  \label{fig:lowmh}
\end{figure}

    This implementation has already been used in
global fits of constrained SUSY models~\cite{Bechtle:2012zk}. A non-trivial example of how the LEP $\chi^2$ information can be applied is given in Fig.~\ref{fig:lowmh}. This figure shows the MSSM low-$M_H$ benchmark scenario \cite{Carena:2013qia}, where the heavier $\cp$-even Higgs boson is interpreted as the LHC signal around $M_H\sim 126$~GeV. In that case, the lightest Higgs boson, $h$, is usually below the SM LEP limit and has suppressed couplings to gauge bosons. This is reflected in the figure, where a sizeable $\chi^2$ penalty can be seen to result in parts of the parameter space, corresponding to regions of low $M_h$ (an uncertainty of $2\gev$ was used here) where the couplings to gauge bosons is such that the LEP Higgs searches are sensitive to the production of such a state. The sharp edge in the $\chi^2$ distribution in Fig.~\ref{fig:lowmh} is obtained at the boundary between two regions of parameter space where the $\chi^2$ contribution comes from the channels $e^+e^-\to hZ$, $h\to b\bar{b}$ and $e^+e^-\to hA \to 4b$, respectively. Using the LEP $\chi^2$ information together with the \HB\ exclusion at $95\%$~C.L. from Tevatron/LHC, Fig.~\ref{fig:lowmh} (right) gives the most complete information available from direct Higgs search limits.

Even after the discovery of a SM-like Higgs boson, Higgs boson exclusions still plays, and will continue to play, a vital role in fitting models of physics beyond the SM with an extended Higgs sector. It would therefore be to great advantage if the Tevatron
and LHC collaborations could follow the example of the LEP Higgs WG
and provide exclusion limits for varying values of $f\,\sigma_{\mathrm{ref}}$ in addition to the results that are presented at $95\%$~C.L..

\vspace{-2em}
\section{User Operating Instructions}
\label{Sec:Manual}
In this section we describe in detail the two main methods to use \HBv{4}: The command line version and the library of subroutines. There is also an online version that provides
quick access to all the functionality of \HB, without the need to install the code.
\newpage

\subsection{Installation}
\label{Sect:installation}
The \HB\ source code, the online version, and the documentation can all
be obtained at the URL\\[.3em]
\centerline{\url{http://higgsbounds.hepforge.org}}\\[.3em]
The \HBv{4} code is mostly written in Fortran 90 but includes also a few Fortran 2003 features. 
It has been tested with a variety of Fortran compilers, including the free GNU compiler\footnote{Version {\tt 4.2} (or higher) of {\tt gfortran} is required for full support of the Fortran 2003 features.} ({\tt gfortran}) which accompanies most Linux distributions.

Before compiling the \HB\ code, the user should first make changes to the {\tt configure} script to appropriately reflect the compiler and path settings on the user's system. The code can then be compiled by running
\cbox{./configure}
\centerline{\tt make}\\[.3em]
which creates the main \HB\ executable and the library of subroutines, {\tt libHB.a}.
Any program for which the \HB\ subroutines should be used
can be compiled and linked to the library by adding {\tt -L<HBpath> -lHB} to
the command line, for example, 
\cbox{gfortran myprog.f90 -o myprog -L<HBpath> -lHB}
where {\tt <HBpath>} is the location of the \HB\ library. The \HB\ subroutines make use of the Fortran file handles 10, 11, 44, 45 and 87, which means that users should avoid these file handles in programs calling the subroutines.

The default behavior of \HBv{4} is to use the \HBfull\ (new) method to generate combined exclusion. To set the \HBclassic\ method as the default, the user can modify the flag {\tt run\_HB\_classic} in the file {\tt usefulbits.f90} before compiling \HB. When running the subroutine version of \HB, it is also possible to access the results from both methods without changing the default behavior, see below.

The library of subroutines and the command-line version share a common set of features, which we will describe first. We will then give the proper operating instructions to use each of these \HB\ formats individually. 

\subsection{Common features: Input}
\label{subsec:Common features: Input}
Regardless of the operation mode, \HB\ requires five basic types of user input:
\begin{enumerate}
\item the number of neutral Higgs bosons in the model under study ({\tt nHzero}) 
\item the number of (singly, positively) charged Higgs bosons in the model ({\tt nHplus}) 
\item the set of experimental analyses which should be considered
  ({\tt whichanalyses}) 
\item the desired input format for the theoretical predictions ({\tt whichinput})
\item the theoretical predictions for the Higgs sector of the model (given as arrays)
\end{enumerate}
The variables {\tt nHzero} and {\tt nHplus} are currently both limited to the range $0$--$9$, but if necessary this could easily be extended in the future. The possible values for the choice of experimental analyses ({\tt whichanalyses}) are described in
\refta{table:instructions1}. 
\begin{table}[t]
\centering
\renewcommand{\arraystretch}{1.0}
\footnotesize
\begin{tabular}{cll}
\br
{\tt code} & {\tt whichanalyses} & ({\tt character(LEN=5)}) \\
\mr
1 & {\tt onlyL} & only LEP analyses\\
2 & {\tt onlyH} & only analyses from hadron colliders (Tevatron and LHC)\\
3 & {\tt LandH} & analyses from both lepton and hadron colliders (LEP, Tevatron and LHC)\\
4 & {\tt onlyP} & only published analyses (defined as analyses with an arXiv number)\\
\br
\end{tabular}
\caption{Possible settings for the variable {\tt whichanalyses}, which determines the subset of experimental analyses
  to be considered by \HB.
\label{table:instructions1}
} 
\end{table}

\HB\ expects the theoretical input to be provided in one of three
formats labelled by the variable {\tt whichinput}. These formats are
described in detail in \refse{Sec:Input}, and their required input is briefly
 summarized in \refta{table:instructions2}. 
In Appendix~\ref{app:input_tables} (Tables~\ref{table:instructions3}--\ref{table:instructions3b}) we assign names and list the full contents of all the possible input arrays for the theory predictions. These names will be used below to describe the input requirements of each version of \HB\ individually. 
\subsection{Common features: Output}
\HB\ provides the user with four types of basic output:

\begin{enumerate}
\item whether or not the model under study is excluded by Higgs searches at the 95\% C.L. ({\tt HBresult}) 
\item the reference number of the 
analysis application ($X_0$)
with the highest statistical sensitivity ({\tt chan})  
\item the number of Higgs bosons that contributed to the theoretical rate
  for the corresponding process ({\tt ncombined}) 
\item the ratio $k_0=Q_{\rm model}/Q_{\rm obs}$ for the process with highest statistical sensitivity ({\tt obsratio}). 
\end{enumerate}

As discussed in Section~\ref{Sec:General}, the extended \HB\ algorithm now offers similar quantities to be calculated individually for each Higgs boson in the model. When making use of the \HBfull\ method, the corresponding output quantities are promoted to arrays of length $n+1$, where $n$ is the total number of (neutral and charged) Higgs bosons in the model. The combined result (contained in element $0$ of these arrays) from this extended test can also be used in analogy to the result of \HB\ \emph{classic}. When several Higgs bosons exclude the same point through different searches, the values for {\tt chan}, {\tt obsratio}, and {\tt ncombined} in the combined result refers to the channel giving the strongest exclusion.

\refta{table:instructions4} shows how to interpret the possible values of {\tt HBresult}
and {\tt obsratio} (one entry in the case of arrays), which are complementary. 
When using either the library of subroutines or the command-line version, the keys associating the
reference numbers (as given by {\tt chan}) with the analysis applications is written in human-readable format in the file {\tt   Key.dat}. In the online version, this information appears directly on the screen.  
When the SLHA option is used for input, the \HB\ results can be added to SLHA files in the form of a
new block, called {\tt HiggsBoundsResults}. An example of this block is shown in \refta{tab:newSLHAblocksOutput}.  
\begin{table}[h]
\vspace{-0.5em}
\centering
\renewcommand{\arraystretch}{1.0}
\footnotesize
\begin{tabular}{ccl}
\br
{\tt HBresult (int)} &{\tt obsratio (double)}&\\
\mr
           0 &   $\ge 1.0$   & parameter point is excluded at $95\%$ C.L.\\
           1 &   $<1.0$      & parameter point is not excluded at $95\%$ C.L.\\
          -1 &   $\le 0.0$   & invalid parameter set\\
\br
\end{tabular}
\caption{The possible values of the output variables {\tt
    HBresult} and {\tt obsratio}, which indicate whether a parameter
  point has been excluded at the 95\% C.L. by the experimental results
  under consideration.} 
\label{table:instructions4}
\vspace{-0.5em}
\end{table} 

\begin{table}[h]
\centering
\renewcommand{\arraystretch}{1.0}
{\tt\footnotesize
\begin{tabular}{llll}
\br
\multicolumn{4}{l}{Block HiggsBoundsResults}\\
\multicolumn{4}{l}{\# HBresult  : result flag (1: allowed, 0: excluded, -1: unphysical)}\\
\multicolumn{4}{l}{\# chan      : most sensitive channel (see below). chan=0 if no channel applies}\\
\multicolumn{4}{l}{\# obsratio  : ratio [sig x BR]\_model/[sig x BR]\_limit (< 1: allowed, > 1: excluded)}\\
\multicolumn{4}{l}{\# ncomb     : number of Higgs bosons combined for most sensitive channel}\\
\multicolumn{4}{l}{\#}\\
0	&    4.0.0	&     ||LandH||	&            \# HB version, HB setting whichanalyses\\
\multicolumn{4}{l}{\#}\\
\multicolumn{4}{l}{\#CHANNELTYPE 1: channel with the highest statistical sensitivity}\\
    1	&      1	&       1	&                  \# chan \\
    1	&      2	&       0	&                  \# HBresult      \\
    1  &    3		&   23.53108	&               \# obsratio  \\
    1	&      4	&       1	&                  \# ncombined \\
    1	&      5	&  \multicolumn{2}{l}{||(ee)->(h1)Z->(b b-bar)Z   (hep-ex/0602042, table 14b (LEP))|| }\\
\br
\end{tabular}
}
\caption{Example of the output SLHA Block \texttt{HiggsBoundsResults}. Strings appearing in the output are wrapped with `$||$'.} 
\label{tab:newSLHAblocksOutput}
\end{table}

\subsection{Library of subroutines}
In this section we list all the user subroutines available through the \HB\ library.
\subroutine{initialize\_HiggsBounds}{(\textit{int}~nHzero, \textit{int}~nHplus, \textit{char(5)}~whichanalyses)}

In each run, this subroutine must be called before any other subroutine of the \HB\ package, and it must be called only once. It performs preparatory operations such as initialization of arrays and reading in the tables of experimental data.
If the neutral Higgs sector should not be tested with 
\HB, the user should set {\tt nHzero=0}. Similarly, if the user
does not wish to test the charged Higgs sector, set {\tt nHplus=0}. The possible settings for {\tt whichanalyses} are shown in \refta{table:instructions1}.
\subroutine{initialize\_HiggsBounds\_int}{(\textit{int}~nHzero, \textit{int}~nHplus, \textit{int}~whichanalyses)}

This is an alternative version of {\tt initialize\_HiggsBounds}, which takes an integer for the third argument instead of  a string constant. This code specifies the set of experimental data that is used by \HB\ according to the first column of \refta{table:instructions1}.
\subroutine{Higgs}{Bounds\_neutral\_input\_effC(\textit{double(n)}~Mh, \textit{double(n)}~MhGammaTot, \textit{double(n)}~g2hjss\_s, \textit{double(n)}~g2hjss\_p, \textit{double(n)}~g2hjcc\_s,  \textit{double(n)}~g2hjcc\_p, \textit{double(n)}~g2hjbb\_s, \textit{double(n)}~g2hjbb\_p, \textit{double(n)}~g2hjtoptop\_s, \textit{double(n)}~g2hjtoptop\_p, \textit{double(n)}~g2hjmumu\_s, \textit{double(n)}~g2hjmumu\_p, \textit{double(n)}~g2hjtautau\_s, \textit{double(n)}~g2hjtautau\_p, \textit{double(n)}~g2hjWW, \textit{double(n)}~g2hjZZ, \textit{double(n)}~g2hjZga, \textit{double(n)}~g2hjgaga, \textit{double(n)}~g2hjgg, \textit{double(n)}~g2hjggZ, \textit{double(n,n)}~g2hjhiZ, \textit{double(n)}~BR\_hjinvisible, \textit{double(n,n)}~BR\_hjhihi)}

This subroutine sets the model input for the neutral Higgs sector using the effective couplings ({\tt whichinput=effC}), as defined in Sect.~\ref{section:effC}. Using this method also excludes the use of either parton-level or hadron-level input. The meaning of the input arrays (all of length $n={\tt nHzero}>0$) is summarized in Appendix \ref{app:input_tables}, \refta{table:instructions3}. If any of the effective couplings are deemed to be irrelevant, the corresponding array may be filled with zeros. However, this needs to be exercised with some caution for quantities relevant in a SM-like Higgs search (as most of the limits reported from the LHC are). It is possible that setting certain couplings  artificially to zero could lead to the model failing the SM-likeness test, cf. Sect.~\ref{Sec:SMtest}.
\subroutine{Higgs}{Bounds\_neutral\_input\_part(\textit{double(n)}~Mh, \textit{double(n)}~GammaTot, \textit{int(n)}~CP, \textit{double(n)}~lep\_hjZ\_ratio, \textit{double(n)}~CS\_lep\_bbhj\_ratio, \textit{double(n)}~CS\_lep\_tautauhj\_ratio, \textit{double(n,n)}~CS\_lep\_hjhi\_ratio, \textit{double(n)}~CS\_gg\_hj\_ratio, \textit{double(n)} CS\_bb\_hj\_ratio, \textit{double(n)}~CS\_bg\_hjb\_ratio, \textit{double(n)}~CS\_ud\_hjWp\_ratio, \textit{double(n)}~CS\_cs\_hjWp\_ratio, \textit{double(n)}~CS\_ud\_hjWm\_ratio, \textit{double(n)}~CS\_cs\_hjWm\_ratio, \textit{double(n)} CS\_gg\_hjZ\_ratio, \textit{double(n)}~CS\_dd\_hjZ\_ratio, \textit{double(n)} CS\_uu\_hjZ\_ratio, \textit{double(n)}~CS\_ss\_hjZ\_ratio, \textit{double(n)}~CS\_cc\_hjZ\_ratio, \textit{double(n)}~CS\_bb\_hjZ\_ratio, \textit{double(n)}~CS\_tev\_vbf\_ratio, \textit{double(n)}~CS\_tev\_tthj\_ratio, \textit{double(n)} CS\_lhc7\_vbf\_ratio, \textit{double(n)} CS\_lhc7\_tthj\_ratio, \textit{double(n)}~CS\_lhc8\_vbf\_ratio, \textit{double(n)}~CS\_lhc8\_tthj\_ratio, \textit{double(n)}~BR\_hjss, \textit{double(n)}~BR\_hjcc, \textit{double(n)}~BR\_hjbb, \textit{double(n)}~BR\_hjmumu, \textit{double(n)}~BR\_hjtautau, \textit{double(n)}~BR\_hjWW, \textit{double(n)}~BR\_hjZZ, \textit{double(n)}~BR\_hjZga, \textit{double(n)}~BR\_hjgaga, \textit{double(n)}~BR\_hjgg, \textit{double(n)}~BR\_hjinvisible, \textit{double(n,n)}~BR\_hjhihi)}

This routine is used to set the input for the neutral Higgs sector using parton-level cross sections ({\tt whichinput=part}), as defined in Sect.~\ref{section:part}. Using this method excludes the simultaneous use of effective couplings or hadron-level input. The meaning of the input arrays (of length $n={\tt nHzero}>0$) are summarized in more detail in Appendix~\ref{app:input_tables}, \refta{table:xsections} (cross sections) and \refta{table:BR} (branching ratios). As for the effective coupling case, quantities which are not required by any channel that has a competitive sensitivity can be set to zero to simplify the input (and the same caveats about searches for SM-like Higgs bosons apply).
\subroutine{Higgs}{Bounds\_neutral\_input\_hadr(\textit{double(n)}~Mh, \textit{double(n)}~GammaTot, \textit{double(n)}~CP,                  
 \textit{double(n)}~CS\_lep\_hjZ\_ratio, \textit{double(n)}~CS\_lep\_bbhj\_ratio,
 \textit{double(n)}~CS\_lep\_tautauhj\_ratio, \textit{double(n,n)}~CS\_lep\_hjhi\_ratio,           
 \textit{double(n)}~CS\_tev\_hj\_ratio,  \textit{double(n)}~CS\_tev\_hjb\_ratio,      
 \textit{double(n)}~CS\_tev\_hjW\_ratio, \textit{double(n)}~CS\_tev\_hjZ\_ratio,      
 \textit{double(n)}~CS\_tev\_vbf\_ratio, \textit{double(n)}~CS\_tev\_tthj\_ratio,  
 \textit{double(n)}~CS\_lhc7\_hj\_ratio, \textit{double(n)}~CS\_lhc7\_hjb\_ratio,      
 \textit{double(n)}~CS\_lhc7\_hjW\_ratio, \textit{double(n)}~CS\_lhc7\_hjZ\_ratio,      
 \textit{double(n)}~CS\_lhc7\_vbf\_ratio, \textit{double(n)}~CS\_lhc7\_tthj\_ratio,                       
 \textit{double(n)}~CS\_lhc8\_hj\_ratio,  \textit{double(n)}~CS\_lhc8\_hjb\_ratio,      
 \textit{double(n)}~CS\_lhc8\_hjW\_ratio, \textit{double(n)}~CS\_lhc8\_hjZ\_ratio,      
 \textit{double(n)}~CS\_lhc8\_vbf\_ratio, \textit{double(n)}~CS\_lhc8\_tthj\_ratio,                       
 \textit{double(n)}~BR\_hjss, \textit{double(n)}~BR\_hjcc, \textit{double(n)}BR\_hjbb,
 \textit{double(n)}~BR\_hjmumu, \textit{double(n)}~BR\_hjtautau, \textit{double(n)}~BR\_hjWW,
 \textit{double(n)}~BR\_hjZZ, \textit{double(n)}~BR\_hjZga, \textit{double(n)}~BR\_hjgaga,
 \textit{double(n)}~BR\_hjgg, \textit{double(n)}~BR\_hjinvisible, \textit{double(n,n)}~BR\_hjhihi)}

This subroutine sets the input for the neutral Higgs sector using hadron-level cross sections ({\tt whichinput=hadr}), as defined in Sect.~\ref{section:hadr}. Using this method excludes the use of effective couplings or parton-level input. The names for the input arrays (of length $n={\tt nHzero}>0$) are described in Appendix~\ref{app:input_tables}, \refta{table:xsections} (cross sections) and \refta{table:BR} (branching ratios). Similarly to the effective coupling case, quantities which are not required by any channel that has a competitive sensitivity can be set to zero to simplify the input (and the same caveats about searches for SM-like Higgs bosons apply).
\subroutine{Higgs}{Bounds\_charged\_input(\textit{double(k)}~MHplus, \textit{double(k)}~GammaTot,
 \textit{double(k)}~CS\_lep\_HpjHmi\_ratio, \textit{double(k)}~BR\_tWpb, \textit{double(k)}~BR\_tHpjb, 
 \textit{double(k)}~BR\_Hpjcs, \textit{double(k)}~BR\_Hpjcb, \textit{double(k)}~BR\_Hptaunu)}

The subroutine {\tt HiggsBounds\_charged\_input} gives the charged Higgs
sector input to \HB.  The use of this subroutine is only required if
$k={\tt nHplus}$ is non-zero (recall that {\tt nHplus} is set in subroutine {\tt
initialize\_HiggsBounds}). Currently, only results from searches for \emph{light} charged Higgs bosons ($M_{H^\pm}< m_t$) are available. Once results from heavy charged Higgs searches are presented, this interface will be extended with input of the necessary cross sections.
The names used for the input arrays are described in Appendix~\ref{app:input_tables}, \refta{table:BR}.
\subroutine{HiggsBounds\_input\_SLHA(}{\textit{char(:)}~SLHAfilename)}

This subroutine can be used for supersymmetric theories as an alternative to the other routines which provide model input to \HB. It is called with a string-type variable, {\tt SLHAfilename}, which gives the name of an SLHA file (full path should be included if not in the current working directory). The model predictions are then read in from this file, which should contain the two \HB-specific blocks as described in Sect.~\ref{sec:inputdescriptionslha}. Furthermore, it will set the mass uncertainties of the neutral and charged Higgs bosons according to the values given in the SLHA block DMASS (if available).
\subroutine{HiggsBounds\_set\_mass\_uncertainties(}{\textit{double(n)}~dMhneut, \textit{double(k)}~dMhch)}

This subroutine allows the user to specify theory mass uncertainties for the neutral and charged Higgs bosons of the model. The implementation and use of these uncertainties when the exclusion limits are evaluated is discussed in detail in Sect.~\ref{Sec:MassUnc}. The default is for all the uncertainties to be zero. The treatment of mass uncertainties in the limit setting is invoked automatically by setting any of them to a non-zero value.
The routine takes two arrays as arguments: {\tt dMhneut(n)} (of length $n=1\ldots${\tt nHzero}), which specifies the (absolute) uncertainties for the neutral Higgs boson masses in GeV, and {\tt dMhch(k)} ($k=1\ldots${\tt nHplus}) which does the same for the charged Higgs bosons. If either ${\tt nHzero}=0$ or ${\tt nHplus}=0$, the corresponding uncertainty array will not be used (and can therefore be set to arbitrary values). 
\subroutine{run\_HiggsBounds(}{\textit{double}~HBresult, \textit{int}~chan, \textit{double}~obsratio, \textit{int}~ncombined)}

After initializating and setting the model input using one of the methods discussed above, this subroutine is called to perform the main part of the {\tt HiggsBounds} calculations. The results from the run is given as output. The combined result, {\tt HBresult}, is reported according to the description in~\refta{table:instructions4}. The channel with the highest exclusion power is identified by its code, {\tt chan} (the channel codes are written to the file {\tt Key.dat}), and the corresponding ratio of the model prediction to the observed limit in this channel is given by {\tt obsratio}. Finally, the number of Higgs bosons combined in this prediction is {\tt ncombined}. The default behavior of this subroutine (which can be controlled by setting a flag in {\tt usefulbits.f90}) is to use the \emph{full} exclusion method of \HB, rather than the \emph{classic} method employed in previous versions.
\subroutine{run\_HiggsBounds\_classic(}{\textit{double}~HBresult, \textit{int}~chan, \textit{double}~obsratio, \textit{int}~ncombined)}

This subroutine can be used to run \HB\ directly in the \emph{classic} mode, without changing any flag. As discussed in Sect.~\ref{Sec:General}, the \HBclassic\  method tests for exclusion using only the globally most sensitive analysis (considering all the Higgs boson). This corresponds to the behavior of \HB\ prior to version 4. The output variables have the same definitions as for {\tt run\_HiggsBounds}.
\subroutine{run\_HiggsBounds\_full(}{\textit{double(N)}~HBresult, \textit{int(N)}~chan, \textit{double(N)}~obsratio, \textit{int(N)}~ncombined)}

This subroutine runs \HB\ in the \emph{full} mode. This is similar to the default behavior of {\tt run\_HiggsBounds}, but with the important difference that when running this subroutine the results from each individual Higgs boson can be accessed in the output. Each of the output variables is therefore an array (with elements $n=0\ldots$ {\tt nHzero}+{\tt nHplus}), where element $0$ contains the combined result (the same as obtained from {\tt run\_HiggsBounds}) and the remaining entries hold the individual results: first the entries for the neutral Higgs bosons, followed by the results for the charged Higgs bosons.
\subroutine{run\_HiggsBounds\_single(}{\textit{int}~h, \textit{double}~HBresult, \textit{int}~chan, \textit{double}~obsratio, \textit{int}~ncombined)}

This subroutine produces the results for a single Higgs boson, which should be indexed by {\tt h}. The indexing is such that the neutral Higgs bosons correspond to ${\tt h}=1\ldots {\tt nHzero}$, followed by the charged Higgs bosons of the model, ${\tt h}={\tt nHzero}+1\ldots {\tt nHzero+nHplus}$. To get the results for more than one individual Higgs boson, it is recommended to instead use the subroutine {\tt run\_HiggsBounds\_full} for better performance.
\subroutine{ HiggsBounds\_SLHA\_output()}{}

When using the SLHA input, the subroutine {\tt HiggsBounds\_SLHA\_output} can be called after using (any of the different) {\tt run\_HiggsBounds} routines in order to write the block {\tt HiggsBoundsResults} to the SLHA file. The results are written in terms of the combined exclusion, see Table~\ref{tab:newSLHAblocksOutput} for an example.
\subroutine{finish\_HiggsBounds()}{}

The subroutine {\tt finish\_HiggsBounds} should be called once at the end of the program, after all other \HB\ subroutines. This deallocates the allocatable arrays used within \HB.
 
\subsection{Command-line version}
\label{sect:command}
When using \HB\ from the command line, the run options are specified in the program call, which should be of the form 
\cbox{./HiggsBounds <whichanalyses> <whichinput> <nHzero> <nHplus> <prefix>}
The variable {\tt <prefix>} is a string which is added to the front of
input and output file names. It may include directory names or other
information identifying the run files. If \verb|whichinput=SLHA|, {\tt <prefix>} should contain the full name of the SLHA file to use, including the path if it is not in the current working directory. When running \HB\ from the command line, the program behaviour (\HBfull/\HBclassic) is determined by a flag specified in the file {\tt usefulbits.f90} (the same as for the subroutine {\tt run\_HiggsBounds}). The default setting is that the \HBfull\ method is used.

\subsubsection*{Input file format}
The arrays containing the theoretical model predictions are read in from text files, with each value given in a separate column (separated by whitespace). The contents of each input file is described in \refta{table:contentsoffiles1} and \refta{table:contentsoffiles2}. Note that all these files will not be necessary at the same time. This will be specified below.
Each row in the input files starts with a line number, ${\tt k}$, which identifies predictions belonging to the
same parameter point in different files. The input files must not contain any comments or
blank lines. Care should be taken with the order of the array elements in the
files. The elements of a one-dimensional array, e.g. {\tt Mh} for ${\tt nHzero}=3$, is given in
the order 
\begin{quote}
\centering
{\tt Mh(1), Mh(2), Mh(3)}.
\end{quote}

\begin{table}[!t]
\centering
\renewcommand{\arraystretch}{1.0}
\footnotesize
\begin{tabular}{ll}
\br
Data file name & Contents\\
\mr
{\tt MH\_GammaTot.dat}                & {\tt k},          {\tt Mh, MhGammaTot}                                 \\[1mm]
{\tt MHplus\_GammaTot.dat}    & {\tt k},          {\tt Mhplus, MhplusGammaTot}                         \\[1mm]  
{\tt MHall\_uncertainties.dat} (optional) & {\tt k}, {\tt dMh}, {\tt dMhplus} \\[1mm]  
{\tt CP\_values.dat}           & {\tt k}, {\tt $\cp$\_value} \\[1mm]
{\tt effC.dat}               & {\tt k},          {\tt g2hjss\_s,g2hjss\_p,g2hjcc\_s,g2hjcc\_p}                       \\
                                      &\phantom{{\tt k},} {\tt g2hjbb\_s,g2hjbb\_p,g2hjtoptop\_s,g2hjtoptop\_p}          \\
                                       &\phantom{{\tt k},} {\tt g2hjmumu\_s,g2hjmumu\_p,g2hjtautau\_s,g2hjtautau\_p,}          \\
                                      &\phantom{{\tt k},} {\tt g2hjWW,g2hjZZ,g2hjZga,g2hjgaga,g2hjgg,g2hjggZ}               \\                                      &\phantom{{\tt k},} {some elements of {\tt g2hjhiZ} (lower left triangle - see example)}            \\
{\tt BR\_H\_OP.dat}                      & {\tt k},          {\tt BR\_hjss,} \\ 
                                      &\phantom{{\tt k},} {\tt BR\_hjcc,BR\_hjbb,BR\_hjmumu,BR\_hjtautau,}                  \\
                                      &\phantom{{\tt k},} {\tt BR\_hjWW,BR\_hjZZ,BR\_hjZga},                               \\
                                      &\phantom{{\tt k},} {\tt BR\_hjgaga,BR\_hjgg}               \\[1mm] 
{\tt BR\_H\_NP.dat}                     & {\tt k},   {\tt BR\_hjinvisible},       some elements of {\tt BR\_hjhihi} (see example)                               \\[1mm] 
{\tt BR\_t.dat}                & {\tt k}, {\tt BR\_tWpb}, {\tt BR\_tHpb}                              \\[1mm]      
{\tt BR\_Hplus.dat}            & {\tt k}, {\tt BR\_Hpcs, BR\_Hpcb, BR\_Hptaunu  }                              \\[1mm] 
{\tt additional.dat} (optional)        & {\tt k},  .. . \\
\br
\end{tabular}
\caption{File names and data format for the contents of \HB\ input files (part I).
  The right column shows the order of the input data arrays 
  within one row of the input file ({\tt k} is the line number). For the order of elements
  within the arrays, see the text for details.}
  \label{table:contentsoffiles1}
\end{table}
The correspondence between the array elements and the physical input quantities is clarified in Appendix B.
Not all of the elements of the two dimensional arrays are
required. From the array {\tt g2hjhiZ} only the lower left triangle (including the diagonal) is
required (and similarly for {\tt lepCS\_hjhi\_ratio} below),
since they are symmetric matrices. From the general matrix  
\begin{align}
\begin{pmatrix}
{\tt g2hjhiZ(1,1)} & \Gray{\tt g2hjhiZ(1,2)} & \Gray{\tt g2hjhiZ(1,3)}\non\\
{\tt g2hjhiZ(2,1)} &      {\tt g2hjhiZ(2,2)} & \Gray{\tt g2hjhiZ(2,3)}\non\\
{\tt g2hjhiZ(3,1)} &      {\tt g2hjhiZ(3,2)} &      {\tt g2hjhiZ(3,3)}\non 
\end{pmatrix},
\end{align}
the required elements should be written in the input file using the order 
\begin{quote}
{\tt g2hjhiZ(1,1)}, {\tt g2hjhiZ(2,1)}, {\tt g2hjhiZ(2,2)}, 
{\tt g2hjhiZ(3,1)}, \\* {\tt g2hjhiZ(3,2)}, {\tt g2hjhiZ(3,3)} .
\end{quote}

The branching ratios for the Higgs decays to lighter Higgs bosons, $h_j \to h_i h_i$, are given via the matrix {\tt BR\_hjhihi(j,i)}:
\begin{align}
\begin{pmatrix} 
\Gray{\tt BR\_hjhihi(1,1)} & {\tt BR\_hjhihi(1,2)}     & {\tt BR\_hjhihi(1,3)} 
                                                                         \non\\
{\tt BR\_hjhihi(2,1)}      & \Gray{\tt BR\_hjhihi(2,2)} & {\tt BR\_hjhihi(2,3)} 
                                                                          \non\\
{\tt BR\_hjhihi(3,1)}      & {\tt BR\_hjhihi(3,2)}      & 
                                                \Gray{\tt BR\_hjhihi(3,3)}\non
\end{pmatrix}
\end{align}
Here, only the off-diagonal components are required since the diagonal elements are not physical quantities. The required 
elements should be given in the order
\begin{quote}
{\tt BR\_hjhihi(1,2)}, {\tt BR\_hjhihi(1,3)}, {\tt BR\_hjhihi(2,1)},
{\tt BR\_hjhihi(2,3)}, \\* {\tt BR\_hjhihi(3,1)}, {\tt BR\_hjhihi(3,2)}.
\end{quote}
\begin{table}[h]
\centering
\renewcommand{\arraystretch}{1.0}
\footnotesize
\begin{tabular}{ll}
\br
Data file name & Contents\\
\mr
{\tt LEP\_HZ\_CS\_ratios.dat}         & {\tt k},          {\tt CS\_lep\_hjZ\_ratio}                            \\[1mm]  
{\tt LEP\_H\_ff\_CS\_ratios.dat}& {\tt k},          {\tt CS\_lep\_bbhj\_ratio}, {\tt CS\_lep\_tautauhj\_ratio}                        \\[1mm]     
{\tt LEP\_2H\_CS\_ratios.dat}         & {\tt k},          some elements of {\tt CS\_lep\_hjhi\_ratio} (see example)\\[1mm]
{\tt LEP\_HpHm\_CS\_ratios.dat}& {\tt k},          {\tt CS\_lep\_HpjHmj\_ratio}  \\
{\tt TEVLHC\_H\_0jet\_partCS\_ratios.dat}& {\tt k},          {\tt CS\_gg\_hj\_ratio}, {\tt CS\_bb\_hj\_ratio}  \\[1mm]   
{\tt TEVLHC\_H\_1jet\_partCS\_ratios.dat}& {\tt k},           {\tt CS\_bg\_hjb\_ratio}                         \\[1mm]  
{\tt TEVLHC\_HW\_partCS\_ratios.dat}     & {\tt k},          {\tt CS\_ud\_hjWp\_ratio}, {\tt CS\_cs\_hjWp\_ratio},             \\  
                                      &\phantom{{\tt k},} {\tt CS\_ud\_hjWm\_ratio}, {\tt CS\_cs\_hjWm\_ratio}              \\[1mm]  
{\tt TEVLHC\_HZ\_partCS\_ratios.dat}     & {\tt k},        {\tt CS\_gg\_hjZ\_ratio}, {\tt CS\_dd\_hjZ\_ratio}, {\tt CS\_uu\_hjZ\_ratio,}              \\ 
                                      &\phantom{{\tt k},} {\tt CS\_ss\_hjZ\_ratio}, {\tt CS\_cc\_hjZ\_ratio,} {\tt CS\_bb\_hjZ\_ratio}             \\[1mm]
{\tt TEV\_H\_vbf\_hadCS\_ratios.dat}  & {\tt k},          {\tt CS\_tev\_vbf\_ratio}                        \\[1mm] 
{\tt TEV\_H\_tt\_hadCS\_ratios.dat}   & {\tt k},          {\tt CS\_tev\_tthj\_ratio}                        \\[1mm] 
{\tt TEV\_1H\_hadCS\_ratios.dat}      & {\tt k},          {\tt CS\_tev\_hj\_ratio}, {\tt CS\_tev\_hjb\_ratio},              \\
                                      &\phantom{{\tt k},} {\tt CS\_tev\_hjW\_ratio}, {\tt CS\_tev\_hjZ\_ratio,}              \\
                                      &\phantom{{\tt k},} {\tt CS\_tev\_vbf\_ratio,} {\tt CS\_tev\_tthj\_ratio}\\[1mm] 
{\tt LHC7\_H\_vbf\_hadCS\_ratios.dat}  & {\tt k},          {\tt CS\_lhc7\_vbf\_ratio}                        \\[1mm] 
{\tt LHC7\_H\_tt\_hadCS\_ratios.dat}   & {\tt k},          {\tt CS\_lhc7\_tthj\_ratio}                        \\[1mm] 
{\tt LHC7\_1H\_hadCS\_ratios.dat}      & {\tt k},          {\tt CS\_lhc7\_hj\_ratio}, {\tt CS\_lhc7\_hjb\_ratio},              \\
                                      &\phantom{{\tt k},} {\tt CS\_lhc7\_hjW\_ratio}, {\tt CS\_lhc7\_hjZ\_ratio,}              \\
                                      &\phantom{{\tt k},} {\tt CS\_lhc7\_vbf\_ratio}, {\tt CS\_lhc7\_tthj\_ratio}\\[1mm] 
{\tt LHC8\_H\_vbf\_hadCS\_ratios.dat}  &{\tt k},          {\tt CS\_lhc8\_vbf\_ratio}                      \\[1mm] 
{\tt LHC8\_H\_tt\_hadCS\_ratios.dat}   &{\tt k},          {\tt CS\_lhc8\_tthj\_ratio}                         \\[1mm] 
{\tt LHC8\_1H\_hadCS\_ratios.dat}      &{\tt k},          {\tt CS\_lhc8\_hj\_ratio}, {\tt CS\_lhc8\_hjb\_ratio},              \\
                                      & \phantom{{\tt k},} {\tt CS\_lhc8\_hjW\_ratio}, {\tt CS\_lhc8\_hjZ\_ratio,}             \\
                                      &\phantom{{\tt k},} {\tt CS\_lhc8\_vbf\_ratio,} {\tt CS\_lhc8\_tthj\_ratio}                \\
\br
\end{tabular}\\
\caption{File names and data format for the contents of \HB\ input files (part II).
  The right column shows the order of the input data arrays 
  within one row of the input file ({\tt k} is the line number). For the order of elements
  within the arrays, see the text for details.
  Note that several arrays appear in two different input files.  These files are never
  required simultaneously in one run of \HB.}
  \label{table:contentsoffiles2}
\end{table}

\begin{table}
\centering
\renewcommand{\arraystretch}{1.0}
\footnotesize
\begin{tabular}{lll}
\br
{\tt whichinput = part}           		&{\tt hadr}  &{\tt effC}\\
\mr
{\tt MH\_GammaTot.dat}               	 &{\tt MH\_GammaTot.dat}     & {\tt MH\_GammaTot.dat}\\ 
{\tt MHplus\_GammaTot.dat}            	&{\tt MHplus\_GammaTot.dat} & {\tt MHplus\_GammaTot.dat}\\
{\tt MHall\_uncertainties.dat (*)}        &{\tt MHall\_uncertainties.dat (*)} & {\tt MHall\_uncertainties.dat (*)}\\
{\tt BR\_H\_NP.dat}  		                 &{\tt BR\_H\_NP.dat}        &{\tt effC.dat}  \\
{\tt BR\_H\_OP.dat}            		       & {\tt BR\_H\_OP.dat}       & {\tt BR\_H\_NP.dat} \\  
{\tt BR\_t.dat}                      			 &{\tt BR\_t.dat}            &{\tt BR\_t.dat}\\  
{\tt BR\_Hplus.dat}                   		&{\tt BR\_Hplus.dat}        &{\tt BR\_Hplus.dat}\\
{\tt LEP\_HZ\_CS\_ratios.dat}         	& {\tt LEP\_HZ\_CS\_ratios.dat}    & {\tt LEP\_HpHm\_CS\_ratios.dat}\\
{\tt LEP\_H\_ff\_CS\_ratios.dat}       	 &{\tt LEP\_H\_ff\_CS\_ratios.dat}& {\tt additional.dat (*)}\\
{\tt LEP\_2H\_CS\_ratios.dat}         	&{\tt LEP\_2H\_CS\_ratios.dat}     & \\
{\tt LEP\_HpHm\_CS\_ratios.dat}         &{\tt LEP\_HpHm\_CS\_ratios.dat} &\\
{\tt TEVLHC\_H\_0jet\_partCS\_ratios.dat}& {\tt TEV\_1H\_hadCS\_ratios.dat} & \\   
{\tt TEVLHC\_H\_1jet\_partCS\_ratios.dat}& {\tt LHC7\_1H\_hadCS\_ratios.dat} &\\  
{\tt TEVLHC\_HW\_partCS\_ratios.dat}     &{\tt LHC8\_1H\_hadCS\_ratios.dat} &\\  
{\tt TEVLHC\_HZ\_partCS\_ratios.dat}     &{\tt CP\_values.dat} &\\  
{\tt TEV\_H\_vbf\_hadCS\_ratios.dat}         & {\tt additional.dat (*)}&\\  
{\tt TEV\_H\_tt\_hadCS\_ratios.dat}         && \\
{\tt LHC7\_H\_vbf\_hadCS\_ratios.dat}         && \\
{\tt LHC7\_H\_tt\_hadCS\_ratios.dat}         && \\
{\tt LHC8\_H\_vbf\_hadCS\_ratios.dat}         && \\
{\tt LHC8\_H\_tt\_hadCS\_ratios.dat}         && \\
{\tt CP\_values.dat}                  && \\
{\tt additional.dat (*)}                  && \\
\br
\end{tabular}\\
\caption{\label{table:instructions5}List of possible input files for each
  setting of {\tt whichinput}. Optional files are marked with (*). The files required can also depend on the setting of {\tt whichanalyses}, see~\refta{table:instructions6}.} 
\end{table}
The file {\tt MHall\_uncertainties.dat} is optional. If it is provided, \HB\ will automatically include the theoretical Higgs mass uncertainties given in the file. This has been described in more detail in Sect.~\ref{Sec:MassUnc}.
The file {\tt additional.dat} is also listed as optional. If this file is included, it can
have any number of columns greater than $1$ (as for the previous files,
the first entry on each line should still be the line number). This input is particularly useful to keep track of variables which are not required by \HB, but which are helpful when plotting the results from a parameter scan. For example, in the case of the MSSM, {\tt additional.dat} could be used to store the values of $\tan \beta$.

As in the subroutine version, the command line version of \HB\ expects a subset of the total list of possible input arrays, which depends on the chosen setting of {\tt whichinput}. The maximal list of files used for
each value of {\tt whichinput} is given in \refta{table:instructions5}.  
Furthermore, some of the arrays will not be relevant for some of the choices for {\tt whichanalyses}. The command line
version of \HB\ will consider the list of input files
appropriate to the settings of {\tt whichinput} and {\tt whichanalyses}, and then attempt to
read only those input files where at least one of the arrays contained in the file will be
used. \refta{table:instructions6} contains a list of which input files are
actually relevant for each setting of {\tt whichanalyses}. Finally, the model predictions for the neutral and charged Higgs sectors are independently specified in different input files (except for the common optional input file {\tt MHall\_uncertainties.dat}). Therefore, the files {\tt Mhplus\_GammaTot.dat}, {\tt LEP\_HpHm\_CS\_ratios.dat}, {\tt BR\_t.dat} and {\tt BR\_Hplus.dat} are only required if the user wants to test the charged Higgs sector ({\tt nHplus > 0}). On the other hand, if the user is only interested in the constraints from charged Higgs boson searches, it is sufficient to give only these files while setting {\tt nHzero = 0}.
As for the subroutine version, if the user does not require processes
involving a particular branching ratio or cross section ratio to be
checked by \HB, that particular array can be filled with
zeros. 

For supersymmetric models, one possible way of generating \HB\ input files is to use the model building tool \texttt{SARAH}~\cite{Staub:2009bi,*Staub:2010jh,*Staub:2011dp} in conjunction with the spectrum generator \texttt{SPheno}~\cite{Porod:2003um,*Porod:2011nf}. These codes can directly write out the \HB\ input files required for the effective coupling approximation.
\begin{table}[t]
\centering
\renewcommand{\arraystretch}{1.0}
\footnotesize
\begin{tabular}{lllll}
\br
Data file name & \multicolumn{4}{l}{Setting of {\tt whichanalyses}}     \\
     & \multicolumn{4}{l}{which this file is relevant to}  \\
\mr
                                          &LandH & onlyL& onlyH & onlyP\\
\mr
{\tt MH\_GammaTot.dat}                    &y     & y     & y     &y       \\
{\tt MHplus\_GammaTot.dat}                &y     & y     & y     &y       \\
{\tt MHall\_uncertainties.dat} (optional)	& y	& y	& y 	& y	\\
{\tt effC.dat}		          &y     & y     & y     &y       \\   
{\tt LEP\_HZ\_CS\_ratios.dat}             &y     & y     &       &y      \\  
{\tt LEP\_H\_ff\_CS\_ratios.dat}          &y     & y     &       &y       \\  
{\tt LEP\_2H\_CS\_ratios.dat}             &y     & y     &       &y       \\
{\tt LEP\_HpHm\_CS\_ratios.dat}           &y     & y     &       &y      \\
{\tt TEVLHC\_H\_0jet\_partCS\_ratios.dat} &y     &       & y     &y      \\   
{\tt TEVLHC\_H\_1jet\_partCS\_ratios.dat} &y     &       & y     &y      \\  
{\tt TEVLHC\_HW\_partCS\_ratios.dat}      &y     &       & y     &y      \\  
{\tt TEVLHC\_HZ\_partCS\_ratios.dat}      &y     &       & y     &y      \\ 
{\tt TEV\_H\_vbf\_hadCS\_ratios.dat}      &y     &       & y     &y      \\ 
{\tt TEV\_H\_tt\_hadCS\_ratios.dat}       &y     &       & y     &y      \\
{\tt TEV\_1H\_hadCS\_ratios.dat}   &y     &       & y     &y      \\ 
{\tt LHC7\_H\_vbf\_hadCS\_ratios.dat}      &y     &       & y     &y      \\ 
{\tt LHC7\_H\_tt\_hadCS\_ratios.dat}       &y     &       & y     &y      \\
{\tt LHC7\_1H\_hadCS\_ratios.dat}   &y     &       & y     &y      \\ 
{\tt BR\_H\_OP.dat}                       &y     & y     & y     &y      \\   
{\tt BR\_H\_NP.dat}                       &y     & y     & y     &y     \\
{\tt BR\_t.dat}			          &y     &       & y     &y     \\
{\tt BR\_Hplus.dat}		          &y     & y     & y     &y    \\
{\tt CP\_values.dat}		          &y     & y     & y     &y    \\
{\tt additional.dat}  (optional)          &y     & y     & y     &y   \\
\br
\end{tabular}\\
\caption{List of input files relevant to each setting of {\tt whichanalyses} (marked by 'y'). The required files also depend on the settings of {\tt whichinput}, {\tt nHzero} and {\tt nHplus}; see \refta{table:instructions5} and
  the text for details.} 
  \label{table:instructions6}
\end{table}

\subsubsection*{Output file format}
When the command line version of \HB\ is used with
{\tt whichinput=hadr}, {\tt part} or {\tt effC}, the output
is written to the file {\tt
  <prefix>HiggsBounds\_results.dat}. A sample of this output is shown in
\reffi{tab:sampleoutput}. The key to the process numbering is written
to the file {\tt <prefix>Key.dat}. 
\begin{figure}
\renewcommand{\arraystretch}{1.0}
{\tt\scriptsize
\begin{tabular}{lll}
\br
\multicolumn{3}{l}{\# generated with HiggsBounds version 4.0.0beta on 15.03.2013 at 13:51} \\
\multicolumn{3}{l}{\# settings: LandH, part}\\
\multicolumn{3}{l}{\#}\\
\multicolumn{3}{l}{\# column abbreviations}\\
\# &   n          & : line id of input \\
\# &  Mh(i)   &  : Neutral Higgs boson masses in GeV\\
\# &  Mhplus(i) & : Charged Higgs boson masses in GeV\\
\# &  HBresult  & : scenario allowed flag (1: allowed, 0: excluded, -1: unphysical)\\
\# &  chan       & : most sensitive channel (see below). chan=0 if no channel applies\\
\# &  obsratio  & : ratio [sig x BR]\_model/[sig x BR]\_limit (<1: allowed, >1: excluded)\\
\# &  ncomb     & : number of Higgs bosons combined in most sensitive channel\\
\# &  additional & : optional additional data stored in <prefix>additional.dat (e.g. tan beta)\\
\multicolumn{3}{l}{\#}\\
\multicolumn{3}{l}{\# channel numbers used in this file}
\end{tabular}
\begin{tabular}{lrl}
\# &          14 & : (e e)->(h2)Z->(gamma gamma)Z   (LHWG Note 2002-02) \\
\# &         15  & : (e e)->(h3)Z->(gamma gamma)Z   (LHWG Note 2002-02)\\
\# &         233 & : (p p)->h2/VBF->Z Z-> l l q q where h2 is SM-like ([hep-ex] arXiv:1202.1416 (CMS))\\
\# &        350 & : (p p)->h2->tau tau (CMS-PAS-HIG-12-050)\\
\multicolumn{3}{l}{\# (for full list of processes, see Key.dat)}\\
\multicolumn{3}{l}{\#}
\end{tabular}
\begin{tabular}{rrrrrrrrrr}
\#cols:  n &           Mh(1) &          Mh(2)   &        Mh(3) &      Mhplus(1) &  HBresult &  chan  &   obsratio &    ncomb &   additional(1)\\
\multicolumn{10}{l}{\#}\\
             1 &     359.121  &        159.618 &        337.305 &        71.6423   &      0  & 233  &   212.358   &      1  &   0.592460    \\
             2 &    83.8032   &       49.2839  &       220.782  &       357.500    &     0  &  14  &   277167.    &     1   &    0.00000    \\
             3 &    85.8826   &      249.094   &      179.329   &      238.330     &    0 &  350  &   106712.  &       1   &    0.00000    \\
             4 &    127.520   &      164.372   &      55.4257   &      322.887     &    0  &  15   &  15141.7  &       1    &   0.294416E-01\\
             \vdots &\vdots &\vdots &\vdots &\vdots &\vdots &\vdots &\vdots &\vdots &\vdots \\
\br
\end{tabular}
}
\caption{Sample output file of the type {\tt
    <prefix>HiggsBounds\_Results.dat} obtained by running \HB\ from the command line. The results for the first four parameter points are shown.}
    \label{tab:sampleoutput} 
\end{figure}
When the command-line version of \HB\ is used with
{\tt whichinput=SLHA}, the \HB\ results are added to the SLHA file in the form of the block {\tt HiggsBoundsResults}, see \refta{tab:newSLHAblocksOutput}. It should be noted that it is not efficient to use the command-line version of \HB\ with SLHA input for large parameter scans, since the experimental data tables must be read in again for each SLHA file. 
If this is a concern, a better option is to use the \HB\ subroutines to create a program which can be called from the command line. An example program, {\tt HBwithSLHA}, demonstrating this is included.

\subsubsection*{Example input files}
\label{Sect:Examples}
The \HB\ package includes a full set of sample input files for
the case $n_H=3$, $n_{H^+}=1$, contained in the folder {\tt example\_data}. 
Each filename is
prefixed with {\tt HB\_randomtest50points\_}. 
To run the command-line version of 
\HB\ with these files as input, use, for example,
\cbox{./HiggsBounds LandH effC 3 1 'example\_data/HB\_randomtest50points\_'}
where the values of {\tt whichanalyses} and {\tt whichinput} can be varied
as desired.
The setting ${\tt nHplus}=0$ can also be used if
the user does not wish 
to test the charged Higgs sector,
\cbox{./HiggsBounds LandH effC 3 0 'example\_data/HB\_randomtest50points\_'}

\subsection{Example programs}

We provide a number of example programs which demonstrate the different features of the
\HB\ subroutines. These are available in the subfolder \texttt{/example\_programs/} of the main installation directory.
After the \HB\ library ({\tt libHB.a}) has been compiled
(using {\tt ./configure; make} as described previously), each of the examples can be compiled and with the command
\cbox{make <program name>}
More generally, any program linking with the \HB\ libraries can be compiled (assuming {\tt gfortran} is used) as follows
\cbox{gfortran example\_program.F -o example\_program -L<HBpath> -lHB}
The following example programs are provided with \HBv{4}:
\begin{itemize}
\item {\tt SM\_vs\_4thGen}\\
This first example compares the Higgs exclusion limits in the SM to those in a model with a heavy fourth fermion generation. 
The program demonstrates the use of the \HB\ functions for the SM branching ratios and total decay widths to calculate the Higgs decay widths and the effective normalized squared couplings.  This information is then used as input, and for each mass point \HB\ is called once with the SM input and once for the 4th generation model. The results are written to two separate output files.

\item {\tt HBwithFH}\\
This program demonstrate the
use of the subroutine version of {\tt Higgs\-Bounds} to test exclusion of MSSM model points. The theory predictions are provided by linking to the publicly available MSSM Higgs spectrum calculator \FH~\cite{hep-ph/9812320,*hep-ph/9812472,*hep-ph/0611326,*Hahn:2009zz,hep-ph/0212020}. The model parameters should be specified in the source file (see the code for details). The results are written directly to the screen.

\item {\tt HBwithCPsuperH}\\
This example is similar to the {\tt HBwithFH} example above, but uses the spectrum generator {\tt CPSuperH} \cite{hep-ph/0307377,*arXiv:0712.2360,*Lee:2012wa} for the theory predictions instead of \FH. As above, the model parameters should be specified directly in the source file and the results are written directly to the screen.

\item {\tt HBwithFH\_dm}\\
This is an updated version of the {\tt HBwithFH} example, which demonstrates the use of several new features in \HBv{4}. It makes it use of  {\tt FeynHiggs}-calculated Higgs mass uncertainties when evaluating the exclusion limits, and the output is provided both as a combined result and in terms of exclusion information from the individual MSSM Higgs bosons. Also in this case the model parameters should be specified directly in the source file, and the output is written to the screen.

\item {\tt HBwithSLHA}\\
When using this example, the user can provide input in the SLHA format with one or more input files. The program run settings are fixed as \verb|<whichanalyses>=LandH|, \verb|<nHzero>=3| and \verb|<nHplus>=1|, but this can be changed in the code. The set of SLHA files to be used as input should be named \verb|<stem>.i| where \verb|i|$=1\ldots$\verb|npoints|. The program can be called from the command line as:
\cbox{./example\_programs/HBwithSLHA <npoints> <stem>}
The block \verb|HiggsBoundsResults| will be added to each SLHA file. In
addition, the \HB\ results (\verb|i|, \verb|HBresult|,
\verb|chan| ,\verb|obsratio|, \verb|ncombined|) are collected in a summary output
file called \verb|<stem>-fromHB|. As in the command-line version, the \HB\ results are obtained using either the \HBfull\ (default) or \HBclassic\ method, following the setting of the corresponding flag in {\tt usefulbits.f90}.
\end{itemize}
In addition to the programs listed here, there are two more example codes specifically for the LEP $\chi^2$ extension. These are discussed in the next section.

\subsection{Installing and using the LEP $\boldmath{\chi^2}$ extension}
The usage of the LEP $\chi^2$ information is restricted to the subroutine version of \HB. In order to enable this feature, the user first needs to download a separate package containing the binary files with the relevant experimental information from the URL\\[.3em]
\centerline{\url{http://higgsbounds.hepforge.org}.}\\[.3em]
These files are contained in the tarball \texttt{csboutput\_trans\_binary.tar.gz}, which should be extracted to a user-defined directory \texttt{<clsbtablesdir>} (not exceeding 80 characters), such that the following file structure is obtained:
\cbox{<clsbtablesdir>/csboutput\_trans\_binary/*.binary}
A convenient choice for \texttt{<clsbtablesdir>} might be the \HB\ main directory. In the next step, \texttt{<clsbtablesdir>} has to be specified in the script \texttt{configure-with-chisq}, in addition to the usual compiler settings, cf. Sect.~\ref{Sect:installation}. Then, the \HB\ library can be built with:
\cbox{./configure-with-chisq}
\centerline{\tt make}\\[.3em]
After a successful compilation, new subroutines for the LEP $\chi^2$ extension are available. These are described in the following.
\subroutine{initialize\_HiggsBounds\_chisqtables(}{)}

This subroutine initializes the new arrays and tables needed for the LEP $\chi^2$ extension. It reads in all the relevant experimental information from the binary files installed in \texttt{<clsbtablesdir>}.
\subroutine{HB\_calc\_stats(}{\textit{double}~theory\_uncertainty, \textit{double}~chisq\_withouttheory,  \textit{double}~chisq\_withtheory, \textit{int}~channel)}

This routine is run to calculate the LEP $\chi^2$ value. The user can specify a theoretical mass uncertainty (in GeV), \texttt{theory\_uncertainty}. Note that this value is only used here, and not in ``standard'' mass uncertainties for the limits (which can be different). The resulting $\chi^2$ value is reported both including (\texttt{chisq\_withtheory}) and without (\texttt{chisq\_withouttheory}) this Higgs mass uncertainty. The \texttt{channel} code for the experimental analysis from which the $\chi^2$ value is derived is also given. This subroutine requires a preceeding call to the subroutine \texttt{run\_HiggsBounds\_classic}, in order to determine the most sensitive analyses for the model. Therefore, the usage of the LEP $\chi^2$ extension always requires a simultaneous run of the standard \HB\ program.
\subroutine{finish\_HiggsBounds\_chisqtables(}{)}

This deallocates the new arrays and tables, and should be called at the end of a run.

\subsubsection*{Usage}
The typical sequence of subroutine calls when using the LEP $\chi^2$ extension is the following:
{\footnotesize
\begin{verbatim}

  call initialize_HiggsBounds_chisqtables
  call initialize_HiggsBounds(nH,nHplus,whichanalyses)

  ... one of the input subroutines... 

  call run_HiggsBounds_classic(HBresult,chan,obsratio,ncombined)
  call HB_calc_stats(theory_uncertainty,chisq_withouttheory,chisq_withtheory,chan2)
  call finish_HiggsBounds_chisqtables
  call finish_HiggsBounds

\end{verbatim}
}
Note that the LEP $\chi^2$ functionality requires a classic \HB\ run to determine the most sensitive channel. For a consistent combination of LEP $\chi^2$ extension with Tevatron and LHC limits we recommend to initialize the LEP $\chi^2$ functionality with the option \texttt{whichanalyses="onlyL"} and performing a separate \HB\ run to consider the hadronic collider limits, i.e. \texttt{whichanalyses="onlyH"}.

Two example programs are provided for the LEP $\chi^2$ extension. They are called \texttt{HBchisq} and \texttt{HBchisqwithSLHA}, respectively, and are both contained in the directory \texttt{example\_programs} of the main \HB\ directory. After setting up \HB\ to use the LEP $\chi^2$ extensions, these examples can be compiled with
\cbox{make HBchisq}
The first example, \texttt{HBchisq}, simply scans over the SM Higgs boson mass within the range $M_H \in [100,~120]$~GeV and evaluates the LEP exclusion $\chi^2$ value. The second program, \texttt{HBchisqwithSLHA}, runs \HB\ on a set of $n$ SLHA files, named \texttt{<SLHA-filename>.i} with \texttt{i}$=1\ldots n$. It is called as
\cbox{/HBchisqwithSLHA <number of files> <SLHA-filename>}
The results, including the LEP $\chi^2$ values, are  printed for all parameter points to the new file {\tt <SLHA-filename>-fromHB}.

\section{Conclusions}
We have presented \HBv{4}, an extension of the \HB\ program
which can be used to study exclusion bounds on arbitrary Higgs sectors
using experimental results from LEP, the Tevatron and the LHC.
It includes the latest LHC results presented in 2013, many of which are based on the full 8 TeV dataset. 

We briefly reviewed the options for user input, including a new option (in the case of a
supersymmetric Higgs sector) which allows an SLHA file to be used as input.
Several improvements and updates of the code have been presented. This
includes in particular an improved SM-likeness test that now takes into
account the relative weight of the contributing channels, and in this
way substantially enlarges the parameter space in which SM analyses can
be applied. We have included the option of a theoretical Higgs mass
uncertainty, which can be relevant, e.g., in the MSSM. Taking the theory
uncertainties into account conservatively broadens the range of
non-excluded Higgs mass values. Concerning the LEP limits, we include an option to obtain the full $\chi^2$ information, i.e.\ not
``only'' a hard 95\%~C.L.\ cut. This is particularly useful
for fits in the Higgs sector. 

In view of the discovery of a Higgs signal at the LHC at $\sim 125.5 \gev$
we have included the option to test {\em every} Higgs boson in the model
under consideration {\em individually}. In this way we slightly deviate
from the pure 95\%~C.L.\ exclusion limit, but we ensure that models do
not falsely pass the {\tt HiggsBounds} test because the spectrum
contains one (SM-like) Higgs boson at a mass of $\sim 125.5 \gev$. 

{\tt HiggsBounds} can now readily be used together with its new sister code,
{\tt HiggsSignals}~\cite{Bechtle:2013xfa}. {\tt HiggsSignals} performs a $\chi^2$ evaluation of the compatibility between the predictions of arbitrary Higgs sectors to measured signal rates. This includes in particular the possibility to test the model predictions against the observed signal at $\sim 125.5 \gev$, but also future, hypothetical, signals of extended Higgs sectors. A combined analysis using both codes exploits all the
public information on the Higgs signal and the Higgs exclusion bounds
obtained at LEP, the Tevatron and the LHC. 

\section*{Acknowledgements}
We thank Nazila Mahmoudi and Lisa Zeune for helpful discussions and feedback on the code. We also thank Werner Porod and Florian Staub for interesting discussions and their effort to efficiently interface \texttt{SARAH} and \texttt{SPheno} with \texttt{HiggsBounds}. This work has been supported by the Helmholtz Alliance ``Physics at the Terascale'' and the Collaborative Research Center SFB676 of the DFG, ``Particles, Strings, and the early Universe". The work of S.H.\ was supported in part by CICYT (grant FPA 2010--22163-C02-01) and by the Spanish MICINN's Consolider-Ingenio 2010 Program under grant MultiDark CSD2009-00064. The work of T.S. was partially funded by the Bonn-Cologne-Graduate-School. O.S.\ is supported by the Swedish Research Council (VR) through the Oskar Klein centre.

\appendix
\clearpage
\newpage
\section{Experimental Data}
\subsection{Anlyses included in \HBv{4}}
\label{Appendix:Data}
\setcounter{figure}{10}
\renewcommand\thefigure{\arabic{figure}} 
The intention is to keep \HB\ continuously updated with the latest experimental results as they become available. Older results, which have been surpassed in sensitivity by newer analyses, are removed. After compiling \HB, the user can run the command 
\cbox{./AllAnalyses}
to print an up-to-date list of the implemented experimental results to the screen. 
Data from the following experimental analyses is included in \HBv{4}:
\begin{table}[h!]
\centering
\begin{tabular}{ll}
LEP Experiments & \cite{hep-ex/0602042,hep-ex/0111010,hep-ex/0107031,hep-ex/0107032,hep-ex/0206022,LHWGnotes,hep-ex/0401022,hep-ex/0404012,hep-ex/0410017,hep-ex/0501033} \\
CDF Collaboration & \cite{CDFnotes,arXiv:0809.3930,arXiv:0906.5613,arXiv:0906.1014,arXiv:0907.1269,arXiv:1001.4468} \\
D{\O} Collaboration & \cite{D0notes,arXiv:0806.0611,arXiv:0901.1887,arXiv:0905.3381,arXiv:0908.1811,arXiv:1001.4481,arXiv:1008.3564,arXiv:1011.1931,arXiv:1012.0874,arXiv:1106.4555,arXiv:1106.4885,arXiv:1107.1268} \\
ATLAS Collaboration & \cite{1202.1414,ATLASnotes,1107.5003,1108.5064,1109.3357,1109.3615,1202.1408,1202.1415,1204.2760,1207.7214}  \\
CMS Collaboration & \cite{12050,CMSnotes,1104.1619,1202.1416,1202.1488,1202.1997,1202.3478,1202.4083} \\
\end{tabular}
\end{table}
\vspace{-2em}
\subsection{Additions in \HBv{4.1}}
\begin{figure}[b!]
\vspace{-0.5em}
\centering
\includegraphics[width=0.42\columnwidth]{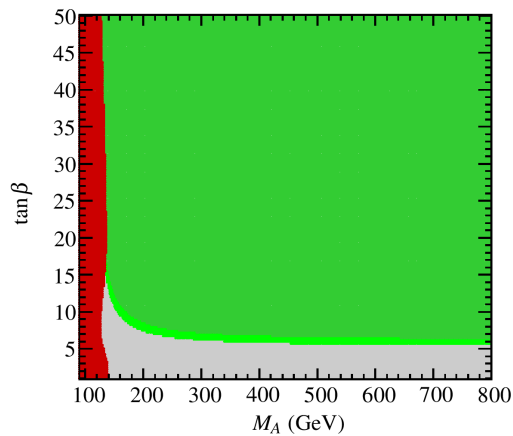}
\vspace{-0.5em}
\caption{Updated exclusion region from charged Higgs boson searches in the \mhmod\ scenario using \HBv{4.1}. This figure should be compared to Fig.~\ref{fig:individualH} (right).}
\label{fig:HB41Hp}

\end{figure}

In addition to the analyses listed above, \HBv{4.1} contains the results from the following experimental analyses:
\begin{table}[h!]
\vspace{-0.5em}
\centering
\begin{tabular}{ll}
ATLAS Collaboration & \cite{ATLASnotesin4.1,ATLAS2013090}  \\
CMS Collaboration & \cite{CMSnotesin4.1,Chatrchyan:2013vaa} \\
\end{tabular}
\end{table}
\newline
In particular the updated ATLAS results from light charged Higgs searches \cite{ATLAS2013090} are interesting for constraining the MSSM (and other models with multiple Higgs doublets) in the region $\MHp < 160$~GeV. In Fig.~\ref{fig:HB41Hp} we show an updated version of the results from charged Higgs exclusion in the \mhmod\ scenario presented in Fig.~\ref{fig:individualH}(c). The new limit excludes small values of $\MA$ for all $\tan\beta$. The same limit also excludes the whole parameter space of the MSSM low-$M_H$ scenario displayed in Fig.~\ref{fig:lowmh}.

\clearpage
\newpage
\section{Input arrays}
\label{app:input_tables}
This appendix contains tables which define the names of the input arrays the user must give as arguments to the various subroutines and/or data files used to specify the model predictions. The physics definitions for the different input quantities (given in the right-hand columns of these tables) follow~\refse{Sec:Input}.
\begin{table}[h]
\footnotesize
\centering
\begin{tabular}{llll}
\br
Input array &  \multicolumn{3}{l}{Quantity}\\
\mr
{\tt Mh(nHzero)}         &  $M_{h_i}$ &               \multicolumn{2}{l}{(GeV)}\\
{\tt MhGammaTot(nHzero)} &  $\Gamma_{\rm tot}(h_i)$ & \multicolumn{2}{l}{(GeV)} \\
\mr
{\tt Mhplus(nHplus)}         &  $M_{H_i^{\pm}}$ &               \multicolumn{2}{l}{(GeV)}\\
{\tt MhplusGammaTot(nHplus)} &  $\Gamma_{\rm tot}(H_i^{\pm})$ & \multicolumn{2}{l}{(GeV)} \\
\mr
{\tt g2hjss\_s(nHzero)}      & $(g^{\rm model}_{s,h_j{(\text{OP})}}/
                                    g^{\rm SM}_{H{(\text{OP})}})^2$,& OP =& $s\bar{s}$     \\
{\tt g2hjcc\_s(nHzero)}      & & & $c\bar{c}$     \\
{\tt g2hjbb\_s(nHzero)}      & & & $b\bar{b}$     \\
{\tt g2hjtoptop\_s(nHzero)}      & & & $t\bar{t}$     \\
{\tt g2hjmumu\_s(nHzero)}  & & & $\mu^+\mu^-$ \\
{\tt g2hjtautau\_s(nHzero)}  & & & $\tau^+\tau^-$ \\[1mm]

{\tt g2hjss\_p(nHzero)}      & $(g^{\rm model}_{p,h_j{(\text{OP})}}/
                                    g^{\rm SM}_{H{(\text{OP})}})^2$,& OP =& $s\bar{s}$     \\
{\tt g2hjcc\_p(nHzero)}      & & & $c\bar{c}$     \\
{\tt g2hjbb\_p(nHzero)}      & & & $b\bar{b}$     \\
{\tt g2hjtoptop\_p(nHzero)}      & & & $t\bar{t}$     \\
{\tt g2hjmumu\_p(nHzero)} & & & $\mu^+\mu^-$ \\
{\tt g2hjtautau\_p(nHzero)} & & & $\tau^+\tau^-$ \\
\mr
{\tt g2hjWW(nHzero)}      &$(g^{\rm model}_{h_j{(\text{OP})}}
                                    /g^{\rm SM}_{H{(\text{OP})}})^2$,  
				& OP = & $W^+W^-$           \\
{\tt g2hjZZ(nHzero)}      & & & $ZZ$           \\
{\tt g2hjZga(nHzero)}     & & & $Z\gamma$      \\
{\tt g2hjgaga(nHzero)}    & & & $\gamma\gamma$ \\
{\tt g2hjgg(nHzero)}      & & & $gg$           \\
{\tt g2hjggZ(nHzero)}     & & & $ggZ$           \\
\mr
{\tt g2hjhiZ(nHzero,nHzero)} & $(g^{\rm model}_{h_jh_iZ}
                                   g^{\rm ref}_{HH^{\prime}Z})^2$ &&\\
\br
\end{tabular}\\

\caption{Input arrays for model predictions of masses, total widths, and 
	 effective normalized squared couplings 
   recognized by \HB.
 The size of  each array is given in brackets in the first column. See \refse{Sec:Input} for the definition of the quantities
 in the second column.} 
 \label{table:instructions3}
\end{table}

\begin{table}[!h]
\footnotesize
\centering
\begin{tabular}{llll}
\br
Input array & \multicolumn{3}{l}{Quantity}\\
\mr 
{\tt CS\_lep\_hjZ\_ratio(nHzero)}             & $R_{\sigma}(P)$,
                                          & $P$ =
                                            & $e^+e^- \to h_j Z$   \\
{\tt CS\_lep\_bbhj\_ratio(nHzero,nHzero)}         & & & $e^+e^- \to b \bar{b} h_j$ \\
{\tt CS\_lep\_tautauhj\_ratio(nHzero,nHzero)}         & & & $e^+e^- \to \tau^+ \tau^- h_j$ \\
{\tt CS\_lep\_hjhi\_ratio(nHzero,nHzero)}         & & & $e^+e^- \to h_j h_i$ \\
\mr
{\tt CS\_lep\_HpjHmj\_ratio(nHzero)} & & & $e^+e^- \to H^+_j H^-_j$ \\
\mr
{\tt CS\_tev\_hj\_ratio(nHzero)}           & & & $p\bar{p} \to h_j$   \\
{\tt CS\_tev\_hjb\_ratio(nHzero)}          & & & $p\bar{p} \to b h_j$ \\
{\tt CS\_tev\_hjW\_ratio(nHzero)}          & & & $p\bar{p} \to h_j W$ \\
{\tt CS\_tev\_hjZ\_ratio(nHzero)}          & & & $p\bar{p} \to h_j Z$\\
{\tt CS\_tev\_vbf\_ratio(nHzero)}          & & & $p\bar{p} \to h_j 
                                                 {\rm \, via \, VBF}$\\
{\tt CS\_tev\_tthj\_ratio(nHzero)}  & & & $p\bar{p} \to t \bar{t} h_j$ \\
\mr
{\tt CS\_lhc7\_hj\_ratio(nHzero)}           & & & $pp \to h_j$  at $7\tev$ \\
{\tt CS\_lhc7\_hjb\_ratio(nHzero)}          & & & $pp \to b h_j$  at $7\tev$ \\
{\tt CS\_lhc7\_hjW\_ratio(nHzero)}          & & & $pp \to h_j W$  at $7\tev$ \\
{\tt CS\_lhc7\_hjZ\_ratio(nHzero)}          & & & $pp \to h_j Z$  at $7\tev$\\
{\tt CS\_lhc7\_vbf\_ratio(nHzero)}          & & & $pp \to h_j 
                                                 {\rm \, via \, VBF}$  at $7\tev$\\
{\tt CS\_lhc7\_tthj\_ratio(nHzero)}  & & & $pp \to t \bar{t} h_j$  at $7\tev$ \\
\mr
{\tt CS\_lhc8\_hj\_ratio(nHzero)}           & & & $pp \to h_j$  at $8\tev$ \\
{\tt CS\_lhc8\_hjb\_ratio(nHzero)}          & & & $pp \to b h_j$  at $8\tev$ \\
{\tt CS\_lhc8\_hjW\_ratio(nHzero)}          & & & $pp \to h_j W$  at $8\tev$ \\
{\tt CS\_lhc8\_hjZ\_ratio(nHzero)}          & & & $pp \to h_j Z$  at $8\tev$\\
{\tt CS\_lhc8\_vbf\_ratio(nHzero)}          & & & $pp \to h_j 
                                                 {\rm \, via \, VBF}$  at $8\tev$\\
{\tt CS\_lhc8\_tthj\_ratio(nHzero)}  & & & $pp \to t \bar{t} h_j$  at $8\tev$ \\
\br
\end{tabular}\\
\caption{\label{table:xsections}Input arrays 
	for model predictions for cross section ratios
	recognized by \HB.
  The size of
  each array is given in brackets in the first column. The LEP $e^+e^-$ and  hadronic Tevatron/LHC cross section ratios $R_{\sigma}(P)$ are defined in
\refeq{eq:R-sigma}.
\vspace*{-1em}} 
\end{table}

\begin{table}[t]
\footnotesize
\centering
\begin{tabular}{llll}
\br
Input array &  \multicolumn{3}{l}{Quantity}\\
\mr
{\tt BR\_hjss(nHzero)}       & BR($h_j \to {\text{OP}}$),& OP = &$s\bar{s}$     \\
{\tt BR\_hjcc(nHzero)}    & &                               &$c\bar{c}$     \\
{\tt BR\_hjbb(nHzero)}   & &                                &$b\bar{b}$     \\
{\tt BR\_hjmumu(nHzero)}& &                                 &$\mu^+\mu^-$ \\
{\tt BR\_hjtautau(nHzero)}& &                               &$\tau^+\tau^-$ \\
{\tt BR\_hjWW(nHzero)}    & &                               &$W^+W^-$           \\
{\tt BR\_hjZZ(nHzero)}    & &                               &$ZZ$           \\
{\tt BR\_hjZga(nHzero)}   & &                               &$Z\gamma$      \\
{\tt BR\_hjgaga(nHzero)}  & &                               &$\gamma\gamma$ \\
{\tt BR\_hjgg(nHzero)}    & &                               &$gg$           \\
\mr
{\tt BR\_hjinvisible(nHzero)}    &  \multicolumn{3}{l}{BR($h_j\to {\rm invisible}$)}  \\
{\tt BR\_hjhihi(nHzero,nHzero)} & \multicolumn{3}{l}{BR($h_j\to h_ih_i$)} \\
\mr
{\tt BR\_tWpb}& BR($t \to W^{+}b$)&    \\
{\tt BR\_tHpjb(nHplus)}& BR($t \to H_j^{+}b$)&    \\
\mr
{\tt BR\_Hpjcs(nHplus)}& BR($H_j^{+} \to {\text{OP}}$),& OP = &$c\bar{s}$     \\
{\tt BR\_Hpjcb(nHplus)}& & &$c\bar{b}$     \\
{\tt BR\_Hpjtaunu(nHplus)}& &  &$\tau^+ \bar\nu_{\tau}$     \\
\br
\end{tabular}\\
\caption{\label{table:BR}Input arrays for model predictions for branching ratios
    recognized by \HB.
  The size of
  each array is given in brackets in the first column. See \refse{Sec:Input} for the description of the notation used
  in the second column.   The elements of {\tt BR\_hjhihi} are ordered such that 
  {\tt BR\_hjhihi(j,i)}$ = \BR(h_j\to h_ih_i)$. }
\end{table}

\begin{table}[t]
\footnotesize
\centering
\begin{tabular}{llll}
\br
Input array & \multicolumn{3}{l}{Quantity}\\
\mr 
{\tt CS\_gg\_hj\_ratio(nHzero)}           & $R^{h_j}_{nm}$,
                                          & $nm$ = 
                                            & $gg$ \\
{\tt CS\_bb\_hj\_ratio(nHzero)}           & & & $b\bar{b}$   \\
\mr
{\tt CS\_ud\_hjWp\_ratio(nHzero)}         & $R^{h_j +W^+}_{nm}$,
                                          & $nm$ = 
                                            & $u\bar{d}$ \\
{\tt CS\_cs\_hjWp\_ratio(nHzero)}         & & & $c\bar{s}$ \\
\mr
{\tt CS\_ud\_hjWm\_ratio(nHzero)}         &$R^{h_j +W^-}_{nm}$
                                          & $nm$ =
                                            &  $d\bar{u}$ \\
{\tt CS\_cs\_hjWm\_ratio(nHzero)}         & & &  $s\bar{c}$ \\
\mr
{\tt CS\_gg\_hjZ\_ratio(nHzero)}          &$R^{h_j +Z}_{nm}$
                                          & $nm$ =                
                                            & $gg$ \\
{\tt CS\_dd\_hjZ\_ratio(nHzero)}          & & & $d\bar{d}$ \\
{\tt CS\_uu\_hjZ\_ratio(nHzero)}          & & & $u\bar{u}$ \\
{\tt CS\_ss\_hjZ\_ratio(nHzero)}          & & & $s\bar{s}$ \\
{\tt CS\_cc\_hjZ\_ratio(nHzero)}          & & & $c\bar{c}$ \\
{\tt CS\_bb\_hjZ\_ratio(nHzero)}          & & & $b\bar{b}$ \\
\mr
{\tt CS\_bg\_hjb\_ratio(nHzero)}          &$R^{h_j +b,h_j +\bar{b}}_{nm}$
                                          & $nm$ = 
                                            &  $bg,\bar{b}g$ \\
\br
\end{tabular}\\
\caption{\label{table:instructions3b}Input arrays 
	for model predictions of partonic cross section ratios
	recognized by \HB. The size of each array is given in brackets in the first column. The partonic cross section ratios $R^{h_j+y}_{nm}$ are defined in \refeq{eq:R-nm-partonic}.
\vspace*{5mm}} 
\end{table}

\clearpage
\newpage
\section*{References}

\bibliographystyle{JHEP}

\bibliography{HB4}

\end{document}